\documentclass{aa_old} 
 \usepackage[varg]{txfonts}
\usepackage{bbm}
\usepackage{fixmath}
\usepackage{graphicx}
\usepackage{caption}
\usepackage{subcaption}
\usepackage{booktabs}
\usepackage[table]{xcolor}
\usepackage{siunitx}
\usepackage{natbib,twoopt}
\usepackage[hyphenbreaks]{breakurl}
\usepackage[breaklinks]{hyperref}
\bibpunct{(}{)}{;}{a}{}{,} 
\usepackage{orcidlink}
\usepackage{placeins}
\definecolor{cobalt}{rgb}{0.06, 0.2, 0.65}
\hypersetup{
  colorlinks,
  citecolor=cobalt,
  linkcolor=[rgb]{0.8, 0.2, 1.0},
  urlcolor=cobalt,
}
\makeatletter
  \newcommandtwoopt{\citeads}[3][][]{\href{http://adsabs.harvard.edu/abs/#3}%
    {\def\hyper@linkstart##1##2{}%
     \let\hyper@linkend\@empty\citealp[#1][#2]{#3}}}
  \newcommandtwoopt{\citepads}[3][][]{\href{http://adsabs.harvard.edu/abs/#3}%
    {\def\hyper@linkstart##1##2{}%
     \let\hyper@linkend\@empty\citep[#1][#2]{#3}}}
  \newcommandtwoopt{\citetads}[3][][]{\href{http://adsabs.harvard.edu/abs/#3}%
    {\def\hyper@linkstart##1##2{}%
     \let\hyper@linkend\@empty\citet[#1][#2]{#3}}}
  \newcommandtwoopt{\citeyearads}[3][][]%
    {\href{http://adsabs.harvard.edu/abs/#3}
    {\def\hyper@linkstart##1##2{}%
     \let\hyper@linkend\@empty\citeyear[#1][#2]{#3}}}
\makeatother

\newcommand{\bmu}{\boldsymbol{\mu}}
\newcommand{\bsigma}{\boldsymbol{\sigma}}
\newcommand{\bz}{\boldsymbol{z}}
\newcommand{\bx}{\boldsymbol{x}}

\newcommand{\be}{\boldsymbol{e}}

\newcommand{\bcep}{$\beta\,$Cep\xspace}
\newcommand{\gdor}{$\gamma\,$Dor\xspace}
\newcommand{\dsct}{$\delta\,$Sct\xspace}
\newcommand{\rr}{RR~Lyr\xspace}
\newcommand{\rrab}{RR~Lyr~ab\xspace}
\newcommand{\rrc}{RR~Lyr~c\xspace}
\newcommand{\rrd}{RR~Lyr~d\xspace}
\newcommand{\dcep}{$\delta\,$Cep\xspace}
\newcommand{\tiicep}{t2~Cep\xspace}

\def\gaia{\textit{Gaia}\xspace}
\def\g{$G$\xspace}
\def\bp{$G_{\rm BP}$\xspace}
\def\rp{$G_{\rm RP}$\xspace}
\def\bprp{\mbox{$G_{\rm BP}-G_{\rm RP}$}\xspace}

\begin{document} 

\title{Learning novel representations of variable sources from multi-modal \gaia data via autoencoders}
\titlerunning{Learning representations of \gaia variable sources via autoencoders}
\authorrunning{P. Huijse et al.}

\author{
P. Huijse\orcidlink{0000-0003-3541-1697}\inst{1, 2}
\and
J. De Ridder\orcidlink{0000-0001-6726-2863}\inst{1}
\and
L. Eyer\orcidlink{0000-0002-0182-8040}\inst{3}
\and
L. Rimoldini \orcidlink{0000-0002-0306-585X} \inst{4}
\and 
B. Holl \orcidlink{0000-0001-6220-3266}\inst{3,4}
\and
N. Chornay \orcidlink{0000-0002-8767-3907}\inst{4}
\and
J. Roquette \orcidlink{0000-0002-0709-4703}\inst{3}
\and 
K. Nienartowicz \orcidlink{0000-0001-5415-0547}\inst{4,5}
\and 
G. Jevardat de Fombelle \orcidlink{0000-0001-6166-8221}\inst{4}
\and
D. J. Fritzewski \orcidlink{0000-0002-2275-3877}\inst{1}
\and
A. Kemp \orcidlink{0000-0003-2059-5841}\inst{1}
\and
V. Vanlaer \orcidlink{0000-0003-4923-6199
}\inst{1} 
\and
M. Vanrespaille \orcidlink{0000-0002-0420-2473} \inst{1}
\and 
H. Wang \orcidlink{0009-0004-1774-7167}\inst{1}
\and
M.I. Carnerero \orcidlink{0000-0001-5843-5515}\inst{6}
\and
C.M. Raiteri \orcidlink{0000-0003-1784-2784}\inst{6}
\and 
G. Marton \orcidlink{0000-0002-1326-1686}\inst{7,8}
\and 
M. Madar\'{a}sz \orcidlink{0009-0005-4037-5506} \inst{7,8}
\and 
G. Clementini \orcidlink{0000-0001-9206-9723}\inst{9}
\and
P. Gavras \orcidlink{0000-0002-4383-4836}\inst{10}
\and 
C. Aerts\orcidlink{0000-0003-1822-7126}\inst{1,11,12}
}
\institute{
Institute of Astronomy, KU Leuven, Celestijnenlaan 200D, B-3001 Leuven, Belgium\\
\email{pablo.huijse@kuleuven.be}
\and
Millennium Institute of Astrophysics, Nuncio Monse\~nor Sotero Sanz 100, Of. 104, Providencia, Santiago, Chile
\and
Department of Astronomy, University of Geneva, Chemin Pegasi 51, 1290 Versoix, Switzerland
\and
Department of Astronomy, University of Geneva, Chemin d’Ecogia 16, 1290 Versoix, Switzerland
\and
Sednai S\`arl, Geneva, Switzerland
\and
INAF - Osservatorio Astrofisico di Torino, Via Osservatorio 20, I-10025 Pino Torinese, Italy
\and
Konkoly Observatory, HUN-REN Research Centre for Astronomy and Earth Sciences, Konkoly Thege 15-17, 1121 Budapest, Hungary
\and
CSFK, MTA Centre of Excellence, Konkoly Thege 15-17, 1121, Budapest, Hungary
\and
INAF - Osservatorio di Astrofisica e Scienza dello Spazio di Bologna, Via Piero Gobetti 93/3, Bologna 40129, Italy
\and
Starion for European Space Agency, Camino bajo del Castillo, s/n, Urbanizacion Villafranca del Castillo, Villanueva de la Ca{\~n}ada, 28692 Madrid, Spain
\and
Department of Astrophysics, IMAPP, Radboud University Nĳmegen, PO Box 9010, 6500 GL Nĳmegen, The Netherlands
\and
Max Planck Institute for Astronomy, Koenigstuhl 17, 69117 Heidelberg, Germany
}

\date{Received ...; accepted ...}

\abstract
{\gaia Data Release 3 (DR3) published for the first time epoch photometry, BP/RP (XP) low-resolution mean spectra, and supervised classification results for millions of variable sources. This extensive dataset offers a unique opportunity to study their variability by combining multiple \gaia data products.}
{In preparation for DR4, we propose and evaluate a machine learning methodology capable of ingesting multiple \gaia data products to achieve an unsupervised classification of stellar and quasar variability.}
{A dataset of 4 million \gaia DR3 sources is used to train three variational autoencoders (VAE), which are artificial neural networks (ANNs) designed for data compression and generation. One VAE is trained on \gaia XP low-resolution spectra, another on a novel approach based on the distribution of magnitude differences in the \gaia \g band, and the third on folded \gaia \g band light curves. Each \gaia source is compressed into 15 numbers, representing the coordinates in a 15-dimensional latent space generated by combining the outputs of these three models.}
{The learned latent representation produced by the ANN effectively distinguishes between the main variability classes present in \gaia DR3, as demonstrated through both supervised and unsupervised classification analysis of the latent space. The results highlight a strong synergy between light curves and low-resolution spectral data, emphasising the benefits of combining the different \gaia data products. A two-dimensional projection of the latent variables reveals numerous overdensities, most of which strongly correlate with astrophysical properties, showing the potential of this latent space for astrophysical discovery.}
{We show that the properties of our novel latent representation make it highly valuable for variability analysis tasks, including classification, clustering and outlier detection.}

\maketitle

\section{Introduction}

Since its launch in 2013, the European Space Agency’s \gaia mission \citep{gaia2016mission} has been surveying more than a billion astronomical sources, charting their positions, luminosities, temperatures, and compositions. The \gaia Data Release 3 (DR3) catalogue \citep{gaia2023dr3}, published in June 2022, marks a significant milestone in the mission, providing the results of 34 months of observations processed by the \gaia Data Processing and Analysis Consortium (DPAC). \gaia DR3 represents a substantial improvement over previous data releases, offering an increase not only in the quantity and quality of the data but also in the variety of data products available for the community. 

\gaia DR3 has been particularly relevant for variability analysis  \citep{eyer2023vari}, offering photometric time series in three bands for 11.7 million sources and low-resolution spectra \citep{carrasco2021xpcalibration, deangeli2023xp} for more than 200 million sources. In addition, the variability processing and analysis coordination unit at the \gaia DPAC (DPAC-CU7) classified variable sources into 25 classes using supervised machine learning (ML) techniques \citep{rimoldini2023gaia}.
The published variability classifications for DR3 were generated by combining the outputs of several ML models trained with two supervised ML algorithms: Random Forest \citep{breiman2001random} and XGBoost \citep{chen2016xgboost}. 
These classifiers based their predictions on approximately $30$ attributes (features) derived from \gaia's astrometry and epoch photometry data, mainly time-series statistics, periodicity indicators, and colour information. The models were trained on a subset of \gaia sources with known ground-truth variability classes, selected from the crossmatch of 152 catalogues from the literature \citep{gavras2023gaia}.

Although a supervised classification of sources into a set of predefined classes is a crucial prerequisite for follow-up astrophysical analysis, it does not allow the discovery of new groups or subgroups of variable sources. Such (sub)groups may have remained unnoticed before because the relevant parameter space was too sparsely populated to note them, but may now become visible with the huge number of sources that \gaia is observing. Finding an over- or underdensity in a parameter space is, however, computationally daunting. On one hand, the sheer size of the \gaia archive implies that any discovery method must be able to cope with hundreds of millions of sources. On the other hand, \gaia provides a large diversity of data products, making the parameter space also high-dimensional.

In preparation for Data Release 4 (DR4), this work presents an unsupervised ML method that combines multiple \gaia products in a novel way. Unlike supervised methods, it does not require predefined variability classes, nor does it rely on classification labels available in the literature, thereby avoiding potential biases arising from the selection and availability of training labels. By not imposing a predefined set of categories, unsupervised methods are particularly well-suited to objective data exploration and hunting for novelties or anomalies. However, as will be shown, this flexibility also introduces significant challenges in evaluation and interpretation compared to supervised approaches. 

A second important feature of our method is that it is able to exploit multiple \gaia data products as input, namely the epoch photometry and the low-resolution BP/RP (XP) mean spectra. These data products (sometimes referred to as \textit{modalities}) are fed to Artificial Neural Networks (ANNs) to generate a compressed set of features. This falls within the field of Self-supervised Representation Learning \citep[SSRL; e.g.,][]{bengio2013representation, gui2024sslsurvey, donoso2023astromer}, a collection of ANN-based methodologies to discover numerical representations from complex data in order to facilitate downstream tasks, such as classification, clustering, and anomaly detection. Like traditional unsupervised methods, SSRL techniques do not rely on manually annotated labels. Instead, they generate supervisory signals directly from the input data. Two prominent SSRL approaches are autoencoders \citep[AEs; e.g.,][]{naul2018recurrent,jamal2020neural} and contrastive learning \citep[CL; e.g.][]{liu2021self, huertas2023brief}. In particular, AEs learn representations by compressing the data and then minimising the error between the original and an approximation reconstructed from its compressed version. These representations are referred to as latent variables, as they encapsulate the underlying features that summarise the observed variables. Multimodal SSRL \citep{guo2019deep, radford2021clip} involves training AE or CL models from examples that span multiple modalities, often captured by different sensors. In astronomy, for instance, different modalities may consist of photometric and spectroscopic data \citep{angeloudimultimodal}. Here, we explore the synergies between \gaia DR3 epoch photometry and XP mean spectra in the context of variability analysis.

We expect the results presented here to serve as a proof of concept for a future implementation using \gaia DR4, which is scheduled for publication in 2026. DR4 will double the number of measurements compared to previous data releases, and more importantly, it will provide epoch photometry for all observed sources rather than just those classified as variables in DR3. Additionally, DR4 will include new data products, such as time-series data for XP spectra, among others\footnote{\url{https://www.cosmos.esa.int/web/gaia/release/}}. ML models capable of exploiting these rich data products will be essential for the automatic analysis of variable sources.

This article is organised as follows: 
Section~\ref{sec:related} places our method in the context of related work in the literature. Section~\ref{sec:datasets} describes the datasets used to train the proposed models. Section~\ref{sec:models} details the theoretical foundations and implementation details of the models for each data modality. Section~\ref{sec:latent_variables} explains how the dimensionality of the latent variable vector is set. 
The quality of our learned representation is evaluated through outlier detection, classification and clustering tasks, and these results are presented in Sections~\ref{sec:outliers}, \ref{sec:classification} and \ref{sec:clustering}, respectively. Section~\ref{sec:astrophysical} provides an astrophysical interpretation of the learned representations. Finally, our conclusions are summarised in Section~\ref{sec:conclusions}.

\section{State-of-the-art of representation learning}
\label{sec:related}

AEs have previously been proposed to analyse \gaia data products. \citet{laroche2023closing} introduced a Variational Autoencoder \citep[VAE;][]{kingma2014} that takes normalised XP mean spectra coefficients as input and compresses them into six latent variables. The authors demonstrated that these latent variables can predict stellar parameters, such as effective temperature ($T_{\rm eff}$), surface gravity, and iron abundance ([Fe/H]), with greater accuracy than a purely supervised model. This highlights the benefits of leveraging large quantities of unlabelled data. \citet{martinez2022deep} trained a VAE to generate light curves of periodic variable stars using crossmatched data from OGLE-III \citep{udalski2008ogleiii} and \gaia DR2. Specifically, the model was trained to reconstruct OGLE's folded light curves conditioned on \gaia's physical parameters, e.g. $T_{\rm eff}$, stellar radius and luminosity. They achieved this by training a regressor that predicts the VAE's latent variables from the physical parameters, obtaining a physically enhanced AE that acts as an efficient variable star simulator. In our work, the main focus is on the self-supervised learning of the latent variables, rather than on the data generation aspects of VAE, although the latter are used to help interpret the results.

Multimodal learning leveraging multiple \gaia data products has also been explored in previous studies. \citet{guiglion2024beyond} proposed a supervised ANN to predict stellar parameters and chemical abundances by combining \gaia Radial Velocity Spectrometer (RVS) spectra \citep{sartoretti2023rvs}, XP mean spectra, mean photometric magnitudes, and parallaxes. Their dataset included 841,300 \gaia DR3 sources, of which 44,780 had stellar labels derived from APOGEE DR17 \citep{abdurrouf2022sdssdr17}. Training the model on these combined modalities enabled the authors to resolve and trace the [$\alpha$/M]–[M/H] bimodality in the Milky Way for the first time, demonstrating the power of integrating diverse data sources. Using the same dataset and similar architectures, \citet{buck2024deep} applied CL to train an ANN capable of projecting RVS and XP spectra into a shared latent space. They showed that the resulting latent space is highly structured and achieved good performance in both spectra generation and stellar label prediction. Here, we choose to combine XP mean spectra with epoch photometry data and exclude mean RVS spectra, as the latter are only available for bright sources and have a small intersection with public epoch photometry data.

Other recent examples of large ANN models trained using multimodal astronomical data include \citet{parker2024astroclip, zhang2024maven, rizhko2024self}. These studies leverage Contrastive Language-Image Pre-Training \citep[CLIP;][]{radford2021clip}, a technique where different modalities are projected into latent vectors using ANNs and aligned by minimising their distance in the latent space. \citet{parker2024astroclip} focused on extragalactic objects, training their model on 197,632 pairs of optical spectra and images from DESI \citep{desi2024desi}. In contrast, \citet{zhang2024maven} focused on supernovae and proposed a model that takes multiband difference photometry and spectra as input. The model was initially trained on 500,000 simulated events and subsequently fine-tuned using 4702 data pairs from the ZTF \citep{bellm2019ztf} and the Spectral Energy Distribution Machine \citep{blagorodnova2018sedm}. 
Although extragalactic sources are also considered in this work, our focus is on general variability classification applied to \gaia DR3 public data, which includes multiple classes of variable stars. 

In this context, the study by \citet{rizhko2024self} is more closely related to ours, as it considered ten stellar variability classes, including subtypes of eclipsing binaries, RR Lyrae and $\delta$ Scuti stars. Their model was trained on 21,708 sources, integrating light curves from the ASAS-SN variable star catalogue \citep{jayasinghe2019asassnvs}, spectra from the LAMOST \citep{cui2012lamost}, and additional features such as galactic coordinates, magnitudes from multiple surveys and parallaxes from \gaia EDR3. The models for each modality are jointly trained using CLIP and then fine-tuned to solve a classification task. Fine-tuning refers to the process of adapting a pre-trained model to a new task by continuing training on labelled data. The results achieved by \citet{rizhko2024self} show that self-supervised pre-training is beneficial in scenarios with limited labelled data and that the resulting latent space is useful not only for classification but also for outlier detection and similarity searches. 

In this work, we propose an AE-based model that generates a latent space where variability classes are well organised without relying on simulated data or fine-tuning via labels. The model is trained using more than 4 million sources with \gaia DR3 epoch photometry and low-resolution spectral data covering a wide range of variable phenomena. We leave the inclusion of additional data products, such as \gaia astrometric information or data from other surveys, for future work.

\section{Datasets}
\label{sec:datasets}

\subsection{Training data}
\label{sec:train_data}

To construct the dataset used to train the ANN, we started from the 11,754,237 \gaia DR3 sources that have publicly available epoch photometry data. This dataset comprises 10,509,536 sources identified as variable \citep{eyer2023vari} and another 1,244,319 sources with time series from the \gaia Andromeda Photometric Survey \citep{evans2023gaps}. We further filtered this set by requiring the sources to have XP mean spectra coefficients available \citep{deangeli2023xp}. This filtering step removed 5,381,570 sources, predominantly faint ones, with 82.2\% and 21.3\% having a \g band median magnitude above 18 and 20, respectively. Finally, we excluded sources with fewer than 40 observations (sparser time series yield less reliable results), yielding a total of 4,136,544 sources for training and evaluating the proposed models. This limit in data points follows \citet{deridder2023gaia} and ensures the reliability of the learned representations. Only 60,295 sources have RVS spectra available within the dataset.Appendix~\ref{sec:supporting_material} provides the ADQL query to retrieve the source identifiers (ids) of the training set.

Using the source ids, we obtain light curves and XP mean spectra through the \gaia datalink service via the interface provided by the astroquery Python package \citep{ginsburg2019astroquery}. \gaia DR3 provides epoch photometry data in the form of \g, \bp, and \rp light curves, which are obtained from the DR3 \texttt{epoch\_photometry} table as a collection of arrays. Observation times and magnitudes for the \g band are extracted from the \texttt{g\_transit\_time} and \texttt{g\_transit\_mag} arrays, respectively. Magnitude uncertainties are not provided but can be estimated using the \texttt{g\_transit\_flux\_over\_error} array. Observations flagged as spurious or doubtful by DPAC-CU7 are removed from the light curves by applying the \texttt{variability\_flag\_g\_reject}\footnote{This flag is the result of the operators described in Section~10.2.3 of the Gaia DR3 documentation \citep{rimoldini2022gaiadr3manual}.} array.

\gaia's blue (BP) and red (RP) on-board spectrophotometers enable the collection of low-resolution epoch spectra. \citet{carrasco2021xpcalibration} details the process by which this data is internally calibrated and time-averaged to produce the XP mean spectra published in \gaia DR3. These spectra are provided through the \texttt{xp\_continuous\_mean\_spectrum} table as two 55-dimensional real-valued arrays per source, \texttt{bp\_coefficients} and \texttt{rp\_coefficients}, representing the coefficients of a Hermite-polynomial basis for BP and RP, respectively. This continuous representation can be converted into an internally calibrated flux vs pseudo-wavelength format using the GaiaXPy Python package \citep{ruz-mieres2024}, which is the format used in this work.

\subsection{Data used for evaluation purposes}
\label{sec:test_data}

A subset of 38,740 sources within the training set have reliable variability class information. This subset corresponds to a selection from the cross-match by \citet{gavras2023gaia} after applying the procedures described in Section~3.1.2 and Appendix~A of \citet{rimoldini2023gaia}. We refer to this as the labelled subset. The light curves and mean spectra of these sources are also part of the training dataset. However, the proposed models are not trained using their class labels. Instead, the variability class labels are solely used to evaluate how well the learned representations distinguish between different variability classes. 

For this evaluation, we adopt a taxonomy comprising 22 classes, including 
\begin{itemize}
    \item Cepheids divided into 1052 $\delta$ Cepheids (\dcep) and 538 type-II Cepheids (\tiicep). These are radially pulsating variables with periods ranging from approximately 1 day to hundreds of days.
    \item RR Lyrae (\rr) stars divided into 1875 type-ab (\rrab), 1313 type-c (\rrc) and 696 type-d (\rrd) pulsators. These are pulsating variables whose dominant radial modes have periods between 0.2 and 1 day.
    \item Eclipsing binary stars (ECL) divided into 1736 Algol-type (ECLa), 2672 $\beta$ Lyrae-type (ECLb) and 1890 W Ursae Majoris-type (ECLw).
    \item Long Period Variables (LPV) including Mira variables and various semiregular variables with 10,751 sources in total.
    \item Other pulsating variables including 111 $\beta$ Cephei (\bcep) stars, 4372 $\delta$ Scuti (\dsct) stars, 381 $\gamma$ Doradus (\gdor) stars, 93 slowly pulsating B (SPB) stars and 149 pulsating white dwarfs (WD).
    \item Rotational variables including 946 RS Canum Venaticorum stars (RS), 248 ellipsoidal variables (ELL), and 5164 generic solar-like variables (SOLAR). Following \citet{rimoldini2023gaia}, we also consider a merged class (ACV|CP) with 979 sources, that combines $\alpha^2$ Canum Venaticorum, magnetic chemically peculiars and generic chemically peculiar stars.
    \item Eruptive variables including 1699 Young Stellar Objects (YSO). Following \citet{rimoldini2023gaia}, we also consider a merged class (BE|GCAS) that combines B-type emission line variables and $\gamma$ Cassiopeiae stars with 752 sources in total.
    \item Cataclysmic variables (CV) of various types with 256 sources in total.
    \item Active Galactic Nuclei (AGN) of different types, with 1067 sources in total. 
\end{itemize} 
We refer the reader to \citet{gavras2023gaia} for a more detailed description of each of the classes. 

We also consider the classifications published in the \gaia DR3 \texttt{vari\_classifier\_result} table to evaluate the learned representations. These labels are the result of the general variability classification carried out by \citet{rimoldini2023gaia}. A total of 3,738,644 sources in our training dataset have this information available. \gaia DR3 also published the results of several Specific Object Study (SOS) pipelines. In this case, we retrieve information from the \texttt{vari\_rrlyrae}, \texttt{vari\_eclipsing\_binary}, \texttt{vari\_long\_period\_variable} and  \texttt{vari\_agn} tables, which contain features extracted by \citet{clementini2023rrlyr}, \citet{Mowlavi2023gaia}, \citet{Lebzelter2023gaia_lpv} and \citet{carnero2023agn}, respectively. The training set includes 48,753 sources from the \rr star SOS table, 486,871 from the eclipsing binary SOS table, 513,208 from the LPV SOS table, and 51,714 from the AGN SOS table.

\section{ANN architecture}
\label{sec:models}

\begin{figure*}[t]
     \centering     
     \includegraphics[width=\textwidth]{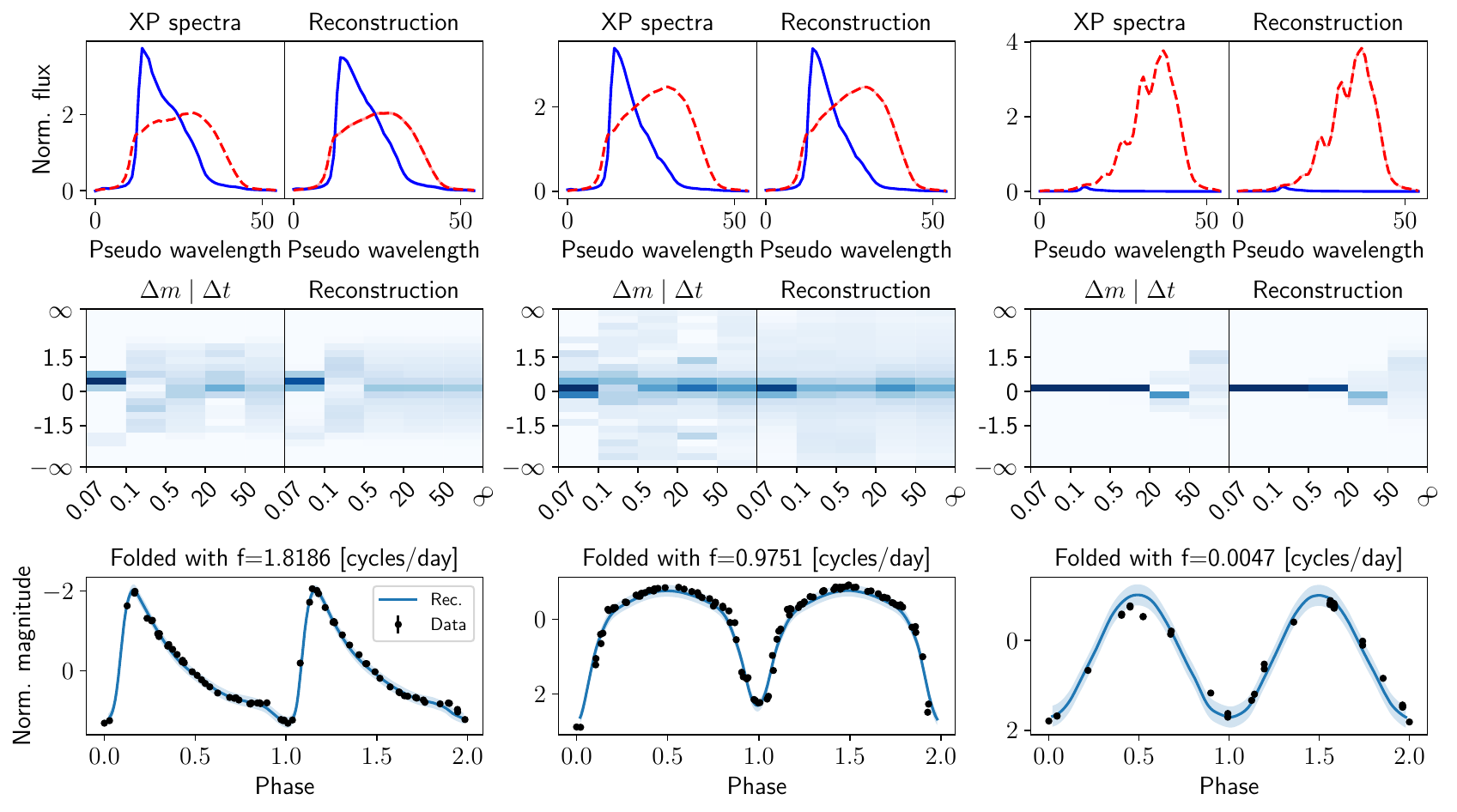}
     \caption{Input data and VAE reconstructions for three variable \gaia DR3 sources. The sources shown in each column are associated with the \rrab star, ECLb, and LPV classes, respectively. Each row corresponds to a specific data modality and shows the preprocessed input data (prior to compression) and its reconstruction (after decompression) using the VAEs described in Sections~\ref{sec:models-xp}, \ref{sec:models-dmdt}, and \ref{sec:models-folded}. The first row corresponds to the sampled representation of the \gaia BP (solid blue) and RP (dashed red) low-resolution spectra. The second row depicts the conditional distribution of magnitude differences given time differences of the \gaia \g band light curves. The third row shows the \gaia \g band light curve folded with its dominant period.
     } 
     \label{fig:example_modalities}
\end{figure*}

Undercomplete AEs are a type of ANN designed to compress complex data into a more compact form \citep{goodfellow2016deep}. They consist of two main components: an encoder, which extracts a compact set of latent variables from the input data, and a decoder, which reconstructs the input data from these latent variables. Compression is achieved by limiting the number of latent variables, that is, the dimensionality of the latent space, to be much smaller than that of the input data while minimising the difference between the input and output of the AE
\begin{align}
\theta^*, \phi^* &= \min_{\theta, \phi} \sum_{i=1}^N L(\bx_i, \theta, \phi) \\
&= \min_{\theta, \phi} \sum_{i=1}^N \left \| \bx_i - \text{Dec}_\theta\left(\text{Enc}_\phi(\bx_i) \right) \right \|^2, \label{eq:mse}
\end{align}
where $\{\bx_i\}_{i=1,\ldots,N}$ with $\bx_i \in \mathbb{R}^D$ is the training dataset and $L(\cdot, \cdot, \cdot)$ is the cost function that measures the error between the input and its reconstruction, the most widely used being the mean square error, in Eq.~\eqref{eq:mse}. The encoder and the decoder are non-linear maps $\text{Enc}_\phi:  \mathbb{R}^D \to \mathbb{R}^K$ and $\text{Dec}_\theta: \mathbb{R}^K \to \mathbb{R}^D$, with trainable parameters by $\phi$ and $\theta$, respectively, and $K \ll D$ is the latent space dimensionality. Once the optimal parameters $\phi^*, \theta^*$ are obtained, the encoder can be used to obtain the latent variables for a given input as $\bz_i = \text{Enc}_{\phi^*}(\bx_i)$.

The VAE \citep{kingma2014} differs from conventional AEs by giving a probabilistic interpretation to the decoder and encoder output. Specifically, the decoder is modelled as a generative process described by the conditional distribution $p_\theta(x|z)$, parametrised by $\theta$, while the encoder is represented by the conditional distribution $q_\phi(z|x)$, parametrised by $\phi$. The encoder serves as a variational approximation of the intractable true posterior $p(z|x)$ and is assumed to follow a Gaussian distribution with diagonal covariance (mean field assumption). In practical terms, this means that the encoder maps the input data to two outputs $\bmu_i~\text{and}~\bsigma_i^2 = \text{Enc}_\phi(\bx_i)$, corresponding to the mean and variance of the latent variables.

The VAE is trained by maximising the Evidence Lower BOund (ELBO), defined as
\begin{equation} \label{eq:elbo}
L(\boldsymbol{x}_i, \theta, \phi) = \mathbb{E}_{\boldsymbol{z}_i\sim q_\phi(z|x_i)} \left[\log p_\theta(x|z_i) \right] - \beta \cdot D_{\rm KL} \left[q_\phi(z|x_i) || p(z) \right],
\end{equation}
where 
\begin{equation} \label{eq:KL}
D_{\rm KL} \left[q_\phi(z|x_i) || p(z) \right] = \frac{1}{2} \sum_{k=1}^K \mu_{ik}^2 + \sigma_{ik}^2  - 1 - \log \sigma_{ik}^2,
\end{equation}
is the Kullback-Leibler (KL) divergence between the encoder distribution and a standard normal prior $p(z) = \mathcal{N}(0, I)$. The first term in Eq.~\eqref{eq:elbo} is the expected value of the log-likelihood of the decoder, and to compute it, one first needs to sample the latent variables using
\begin{equation} \label{eq:reparam_trick}
\boldsymbol{z}_i = \bmu_i + \bsigma_i \cdot \epsilon, \quad \epsilon \sim \mathcal{N}(0, I).
\end{equation}

Maximising the likelihood increases the fidelity of the decoder reconstruction with respect to the original input and can be seen as equivalent to minimising the reconstruction error. On the other hand, minimising the KL divergence provides regularisation to the latent variables. The hyperparameter $\beta$ was introduced in \citet{burgess2018betaVAE} to control the trade-off between reconstruction fidelity and latent space regularisation. It prevents over-regularised solutions where the first term of Eq.~\eqref{eq:elbo} is neglected. 
Since most optimisers are designed for minimisation, the negative ELBO is typically used in practice.

The distribution of $p_\theta(x|z)$, on the other hand, is selected based on the characteristics of the data. In the following sections, we will detail the preprocessing, ANN design and the choice of log-likelihood for each of the modalities considered. This includes a VAE for the XP mean spectra and two VAEs for different representations of the \g band light curve.

\subsection{VAE for XP mean spectra}
\label{sec:models-xp}

Before entering the model, the 55 BP and 55 RP mean spectrum fluxes obtained using GaiaXPy, along with their associated uncertainties, are concatenated into two 110-dimensional vectors, $\bx$ and $\be$. Both vectors are normalised by dividing each element by the standard deviation of $\bx$. This normalisation preserves the relative differences between the BP and RP spectra while eliminating the influence of the source's brightness. The first row in Fig.~\ref{fig:example_modalities} shows the preprocessed XP mean spectra and corresponding VAE reconstructions of three variable sources. The third column in the plot, which corresponds to an LPV, exhibits a considerably stronger RP with respect to the other two sources. Examples for other variability classes can be found in Figs.~\ref{fig:example_modalities_a1}--\ref{fig:example_modalities_a4}.

\begin{figure*}[t]
\begin{center}
\includegraphics[width=0.9\textwidth,keepaspectratio]{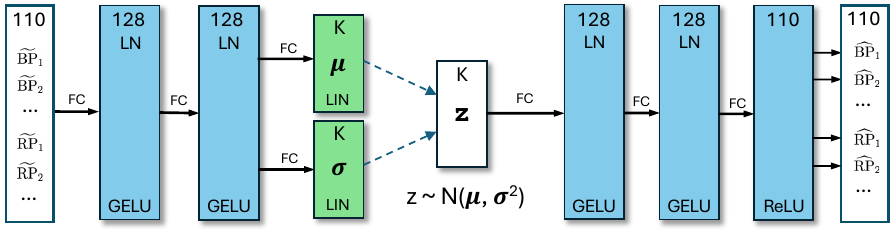}
\caption{\label{fig:ann_xp_architecture}The VAE used to create a $K=5$ dimensional latent space given the 2 $\times$ 55 continuous BP and RP spectra points. Each coloured rectangle represents a layer of the ANN, with the size of the output vector given at the top. The encoder starts with the (standardised) observations denoted in the leftmost white rectangle, and ends with the $K=5$-dimensional latent space denoted in green, representing the position $\bmu$ and its (spherical) uncertainty $\bsigma^2$ of the observed data in the latent space. The decoder starts by sampling a K=5-dimensional value $\mathbf{z}$ from a Gaussian distribution with mean $\bmu$ and the diagonal covariance matrix $\bsigma^2$, and ends with the reconstructed $2 \times 55$ XP spectra values. Some layers use \textit{layer normalisation}, which is denoted by LN. The activation function of the neurons in a particular layer is noted at the bottom: GELU, Linear (LIN), or ReLU.
}
\end{center}
\end{figure*}

After preprocessing, the data is then compressed and reconstructed using a VAE with an architecture based on fully connected (FC) layers that is showcased in Fig.~\ref{fig:ann_xp_architecture}. Three layers are used for the encoder and decoder networks, respectively. Intermediate layers have 128 neurons with Gaussian Error Linear Unit (GELU) activation functions \citep{hendrycks2016gaussian} and layer normalisation  \citep[LN;][]{lei2016layer} to enhance stability during training. The output layer of the encoder, i.e. the mean and standard deviation of the latent variables, do not incorporate LN and use a linear activation (LIN) to avoid constraining their range. The output layer of the decoder, i.e. the reconstructed mean spectra, uses the Rectified Linear Unit (ReLU) $\max(X, 0)$ activation \citep{nair2010rectified, glorot2011deep} to ensure non-negativity. A separate two-layer decoder network is used to predict $\log \sigma_{\rm XP}^2$, an uncertainty term representing the logarithm of the variance of the reconstructed mean spectra. The particular choice for the number of neurons and activation functions of this network is detailed in Appendix~\ref{sec:architectures}. 

This VAE is optimised using Eq.~\eqref{eq:elbo} and the following Gaussian log-likelihood for a given source
\begin{equation}
\log p_\theta(x|z) = - \frac{1}{2} \left[ \log\left(\sum_{j=1}^{110} (e_j^2 + \sigma_{\rm XP}^2) \right) + \sum_{j=1}^{110} \frac{(x_j - \hat x_j)^2}{e_j^2 + \sigma_{\rm XP}^2} \right],
\end{equation}
where $\hat x = \text{Dec}_\theta(z)$ is the reconstructed XP spectra, and terms that do not depend on the model parameters have been omitted.

\subsection{VAE for $\Delta m|\Delta t$  of \g band light curves}
\label{sec:models-dmdt}

Variability across different time scales is a key characteristic for distinguishing between various classes of variable stars \citep{eyer2008variable}. To capture this aspect, the second VAE uses a representation of the \g band light curves inspired by the approaches of \citet{mahabal2017deep} and \citet{soraisam2020classification} as input. Specifically, we employ an estimator of the conditional distribution of magnitude changes ($\Delta m$) given time intervals ($\Delta t$), referred to here as $\Delta m|\Delta t$.

This distribution is estimated non-parametrically from the normalised \g band light curve. A light curve, defined here as a sequence of observations $(t_j, m_j, e_j)_i$ with $j=1,\ldots, L_i$, where $L_i$ is the number of observations of the i-th source, is preprocessed as follows. First, observations with $e_j > \bar e + 3 \sigma_e$, where $\bar e$ and $\sigma_e$ are the mean and standard deviation of the magnitude uncertainties of the source, are removed. Then the magnitudes are centred around zero by subtracting their mean. Finally, the magnitudes and uncertainties are rescaled by dividing by the standard deviation of the magnitudes. 

To compute the $\Delta m | \Delta t$ representation, we first calculate all pairwise differences in magnitudes and observation times for a given light curve. Specifically, this involves computing the tuples $(\Delta m_{j_1 j_2}=m_{j_2} - m_{j_1}, \Delta t_{j_1 j_2}=t_{j_2} - t_{j_1})$, where $1 \leq j_1 < j_2 \leq L_i$. These tuples are then binned into a 2D histogram using the following bin definitions:  
\begin{itemize}
\item For $\Delta t$, we use five bins $b^{(\Delta t)}$: four bins defined by the following edges: $0.07, 0.1, 0.5, 20, 50$ plus an extra bin for when $\Delta t > 50$. This set of bins was designed to capture astrophysically relevant timescales while ensuring that the bins are populated for all sources in the training set, thereby eliminating issues related to missing data. As a result of this design, the bins align with the typical cadences found in \gaia's scanning law \citep[cf.][]{gaia2016mission}, focusing on observations separated by hours or months. For reference, the distribution of $\Delta t$ in the 4M dataset is shown in Fig.~\ref{fig:dt_histogram}.
\item For $\Delta m$ we define 23 bins $b^{(\Delta m)}$: 21 bins with linearly spaced edges between $[-3, 3]$ plus two extra bins for $\Delta m < -3$ and $\Delta m > 3$, respectively. This set of bins was chosen to cover the range of the normalised magnitudes, with outliers being captured in the first and last bins. The normalisation procedure ensures that $\Delta m$ remains centred around zero. For reference, the distribution of $\Delta m$ in the 4M dataset is shown in Fig.~\ref{fig:dm_histogram}.
\end{itemize}

This process produces a $23 \times 5$ sized array, representing the joint distribution of $\Delta m$ and $\Delta t$. To remove the effects of time sampling and obtain the final $\Delta m | \Delta t$ representation, the array is divided by the marginal distribution of $\Delta t$. Concretely, the value at the $k_m$-th row and $k_t$-th column of the $\Delta m | \Delta t$ array is given by
\begin{equation} \label{eq:dmdt}
    p_{k_m  k_t} = \frac{\sum_{j_2=2}^{L_i} \sum_{j_1=1}^{j_2 -1} W\left(\Delta m_{j_1 j_2}, b_{k_m}^{(\Delta m)} \right) \cdot W\left(\Delta t_{j_1j_2}, b_{k_t}^{(\Delta t)} \right)}{
    \sum_{j_2=2}^{L_i} \sum_{j_1=1}^{j_2 -1} W\left(\Delta t_{j_1 j_2}, b_{k_t}^{(\Delta t)} \right)
    } ,
\end{equation}
where
\begin{equation}
W(\Delta x, b) = \begin{cases}
1 & \text{if}~ \Delta x \in b, \\
0 & \text{otherwise}.
\end{cases}
\end{equation}
The numerator in Eq.~\eqref{eq:dmdt} counts the number of magnitude–time pairs for which the magnitude difference $\Delta m$ falls into the $k_m$-th magnitude bin and the time difference $\Delta t$ falls into the $k_t$-th time bin. The denominator counts how many pairs fall into the $k_t$-th time bin, regardless of their magnitude difference. 

The middle row in Fig.~\ref{fig:example_modalities} shows the resulting $\Delta m | \Delta t$ representations and corresponding VAE reconstructions for three variable sources. For example, the third column, which corresponds to an LPV, exhibits near-zero difference in magnitude for time differences below 20 days. The other two sources show variability at much shorter time scales. 

The $\Delta m|\Delta t$ representation is then flattened to a 115-dimensional vector, compressed and reconstructed using a VAE with an FC architecture shown in Fig.~\ref{fig:ann_xp_architecture} and detailed in Appendix~\ref{sec:architectures}. This architecture is similar to the one used for the XP mean spectra but with one additional layer in the encoder and decoder networks, respectively. Additionally, a Softmax activation function is applied to the decoder output to ensure that all values are non-negative and add up to one column-wise.
\begin{figure*}[t]
\begin{center}
\includegraphics[width=0.95\textwidth,keepaspectratio]{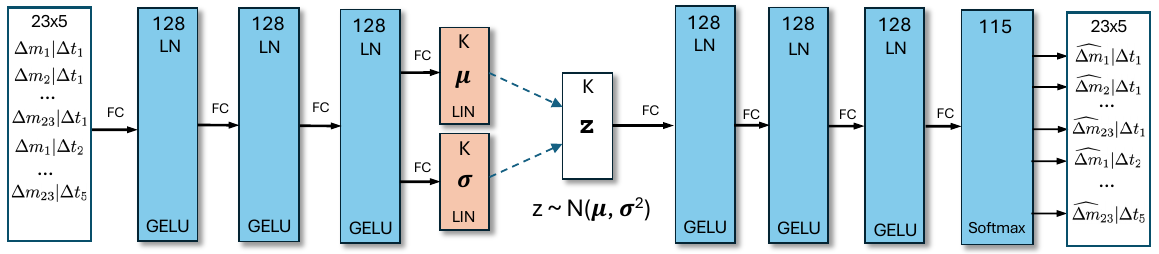}
\caption{\label{fig:ann_dmdt_architecture}The VAE used to create a K=5 dimensional latent space given the 23 $\times$ 5 values of the observed $\Delta m |\Delta t$ histogram.  The explanation of the network is very similar to the one in Fig.~\ref{fig:ann_xp_architecture}.}
\end{center}
\end{figure*}
This VAE is optimised using Eq.~\eqref{eq:elbo} and the binary-cross entropy loss for a given source
\begin{equation}
\log p_\theta(x|z) = \sum_{k_m=1}^{23} \sum_{k_t=1}^{5} p_{k_m k_t} \log (\hat p_{k_m k_t}) + (1-p_{k_m k_t}) \log (1-\hat p_{k_m k_t}),
\end{equation}
where $\hat p = \text{Dec}_\theta(z)$ is the reconstructed $\Delta m | \Delta t$ and terms that do not depend on the model parameters have been omitted.

\citet{mahabal2017deep} treated this representation as an image and processed it using convolutional neural networks (CNNs). However, preliminary experiments showed that FC architectures outperformed CNNs. The low resolution of $\Delta m|\Delta t$ constrains the number of convolutional layers that can be stacked, limiting the representational capacity of CNNs.

\subsection{VAE for folded \g band light curves}
\label{sec:models-folded}

Several of the variability classes found in \gaia DR3 exhibit periodic behaviour \citep{eyer2023vari}. To capitalise on this, the third VAE uses epoch-folded light curves as input, similarly to \citet{naul2018recurrent} and \citet{jamal2020neural}. Folding the light curves yields a more homogeneous and less sparse representation that is considerably less challenging to learn and reconstruct. In addition, preliminary experiments using unfolded light curves resulted in the model producing latent-space overdensities linked to the scanning law and sky position of the sources, further supporting our decision to adopt folded light curves.

After normalising the \g band light curve as described in Section~\ref{sec:models-dmdt}, its fundamental frequency is estimated using the NUFFT implementation of the Lomb-Scargle (LS) periodogram \citep{lomb1976least,scargle1982studies,garrison2024nifty}. A frequency search is carried out in a linearly-spaced grid between $f_{\min}=0.001$ d$^{-1}$ and $f_{\max}=25$ d$^{-1}$ with a step of $\Delta_f = 0.0001$ d$^{-1}$, the latter being roughly $0.1/T$ with $T$ the total time span of the time series. The ten highest local maxima of the periodogram are refined by searching for a higher periodogram value in the range $[f^* - \Delta_f, f^* + \Delta_f]$ with a step of $0.1 \Delta_f$, where $f^*$ is a given local maximum. Finally, the frequency associated with the highest periodogram value among these candidates ($f_1$) is saved. \citet{hey2024confronting} have shown that the vast majority of non-radial pulsators found by \citet{deridder2023gaia} have their dominant frequency confirmed from independent high-cadence TESS photometric light curves, for amplitudes as low as a few mmag.

The preprocessed \g band time series are epoch-folded using
\begin{equation}
    \phi_j = \text{modulo}\left(t_j, 2/ f_1 \right) \cdot f_1 \in [0, 2],
\end{equation}
where $\phi_j$ represents the phase, which replaces $t_j$ in the sequence. Importantly, the light curve is folded using half the dominant frequency (i.e., twice the dominant period) rather than the $f_1$ itself. This choice is deliberate, as it accounts for certain variability classes, such as eclipsing binaries, that are more likely to be detected at half their actual period due to their characteristic non-sinusoidal shapes. Finally, the folded time series is shifted so that the zero phase is aligned to the maximum magnitude (faintest observation), as seen in the bottom row of Fig.~\ref{fig:example_modalities}. 

To train the model efficiently, time series must have a uniform length. In this work, we set the maximum sequence length to $L=100$. Shorter time series are padded with zeros, while longer time series are randomly subsampled during training\footnote{Only 1.64\% of the light curves in the dataset have more than 100 observations.}. Subsampling occurs after preprocessing, folding and phase-aligning. A binary mask $B\in [0,1]^{L}$ is created to indicate the location of non-padded values in each sequence, ensuring that the model parameters are optimised only on valid observations.

The architecture of the VAE for folded light curves is illustrated in Fig.~\ref{fig:ann_lc_architecture} and detailed in Appendix~\ref{sec:architectures}. The encoder receives the phases, the magnitudes and the binary masks of the folded time series and returns the parameters of the latent variables. The decoder, on the other hand, receives the phases and the sampled latent variables and returns the reconstructed magnitudes. Before entering the encoder and decoder, the phases are expanded using trigonometric functions as $\left[\cos(2\pi h\phi_j), \sin(2\pi h \phi_j) \right]_{h=1}^6 \in \mathbb{R}^{12}$. This allows the encoder to treat the limits of the phase range as neighbours and also forces the decoder to produce periodic reconstructions. 

Contrary to the previously presented models, the architecture of the encoder combines recurrent and attention-based neural layers. Recurrent layers process sequences one observation at a time, updating their internal state. This memory-like architecture allows them to model dynamic changes over time effectively. In this case, we select Gated Recurrent Units \citep[GRU;][]{chung2014gru} as building blocks for our architecture. The collection of hidden states obtained by the last GRU layer is aggregated in time using attention as proposed by \citet{Bahdanau2014attention}. 
Preliminary experiments demonstrated that this approach outperformed both simple averaging of all states and relying solely on the final state, in terms of reconstruction and classification \citep{hu2018rnnpooling}. FC layers are used to compress the result from the attention average into two K-dimensional vectors representing the parameters of latent variables using FC layers. The decoder uses FC layers in a similar way to the previously presented VAEs. An FC decoder was chosen over a recurrent one due to preliminary experiments indicating a higher tendency for the latter to overfit. 

The model is trained using Eq.~\eqref{eq:elbo} and the following Gaussian log-likelihood for a given source
\begin{equation}
\log p_\theta(x|z) = - \frac{1}{2L_i} \left[ \log\left(\sum_{j=1}^{L} B_j(e_j^2 + \sigma_{F}^2) \right) + \sum_{j=1}^{L} \frac{B_j(m_j - \hat m_j)^2}{e_j^2 + \sigma_{F}^2} \right],
\end{equation}
where $\hat m$ is the reconstructed magnitude, $B$ is the binary mask indicating the valid elements, and $\log \sigma_{F}^2$ is an error term that represents the log of the variance of the reconstruction. This term is predicted from the latent variables using a decoder with architecture equivalent to the one used to predict $\log \sigma_{\rm XP}^2$. 

\begin{figure*}[t]
\begin{center}
\includegraphics[width=0.95\textwidth,keepaspectratio]{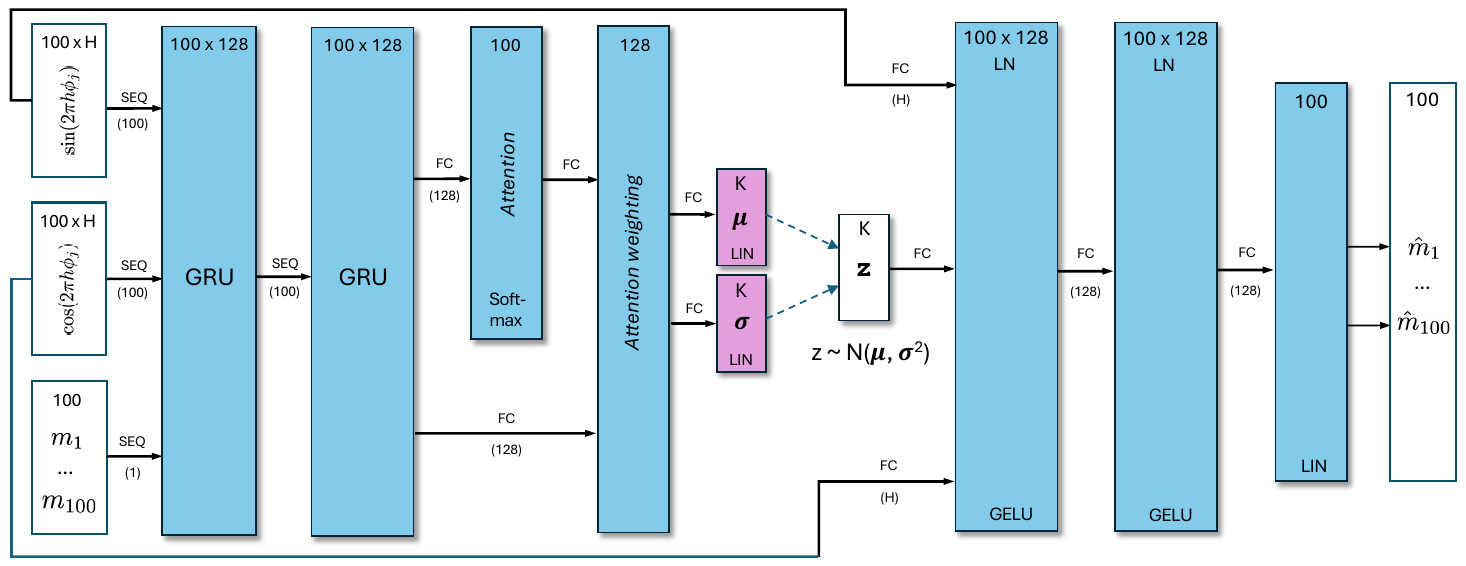}
\caption{\label{fig:ann_lc_architecture}The VAE used to create a $K=5$ dimensional latent space given the 100 points of the folded light curve (phase $\phi_j \in [0,1]$ and the standardised G-magnitude $m_j$). The network does not receive the 100 phase values $\phi_j$ directly, but rather the $100\times H$ values $\sin(2\pi h\phi_j)$ with $h = 1,\ldots, H$, and an equal number of values $\cos(2\pi h\phi_j)$. The first neural layer is a Gated Recurrent Unit (GRU) to which the observations are fed sequentially (SEQ), i.e.~100 consecutive sets of size $1+2H$ corresponding to each time point of the light curve. The result is a matrix of size $100\times 128$. Each of the 100 rows is then fed sequentially to the next GRU layer, resulting in a matrix of the same shape.  
The next layer takes each row of this matrix (of size 128), and computes an attention score corresponding to each point of the light curve. These scores are then used in a subsequent layer to compute a weighted average over the matrix rows, producing a feature set of size 128 where the time points that were deemed more important given the attention score receive a higher weight. In the final layers of the encoder, this feature set is then further reduced to the $K=5$-dimensional latent coordinates (purple rectangles).
The decoder works similarly as explained in Fig.~\ref{fig:ann_xp_architecture}, but now also receives the phase information $(\sin(2\pi h\phi_j), \cos(2\pi h\phi_j))$. The first decoding neural layer results in a matrix of size $100\times 128$. Each row (of size 128) of this matrix is then fed to a neural layer, resulting again in a $100\times 128$ matrix, which is then further reduced to a vector of size 100 by the final decoding layer, resulting in the 100 reconstructed (standardised) magnitudes of the folded light curve. 
}
\end{center}
\end{figure*}
%

\subsection{VAE to visualise the combined latent variables}

For astrophysical interpretation, it is more convenient to use a 2-dimensional latent space, which can be directly visualised. We therefore introduce a final VAE to obtain two latent variables, $\lambda_1$ and $\lambda_2$, which combine the latent variables from the previous three VAEs along with additional features. These features are the standard deviation of the \g band light curve $\sigma_G$, the logarithm of the estimated dominant frequency $\log_{10}  f_1$ of the \g band light curve and the colour estimate \mbox{$\overline{G_{\rm BP}}-\overline{G_{\rm RP}}$}, i.e. the difference between the average magnitude in the \bp band and the \rp band light curves. 

The architecture of this model is illustrated in Fig.~\ref{fig:ann_2D_latent_architecture} and further detailed in Appendix~\ref{sec:architectures}. The model is trained using Eq.~\eqref{eq:elbo} with a Gaussian log-likelihood. No additional variance term is learned for this decoder. The results from this model are used to carry out the astrophysical analysis presented in Section~\ref{sec:astrophysical}. Contrary to other non-linear visualisation methods, the VAE produces an explicit function that maps the two spaces, a capability that is essential for efficiently scaling this method to the future larger \gaia DR4 dataset.

\begin{figure*}[t]
\begin{center}
\includegraphics[width=0.95\textwidth,keepaspectratio]{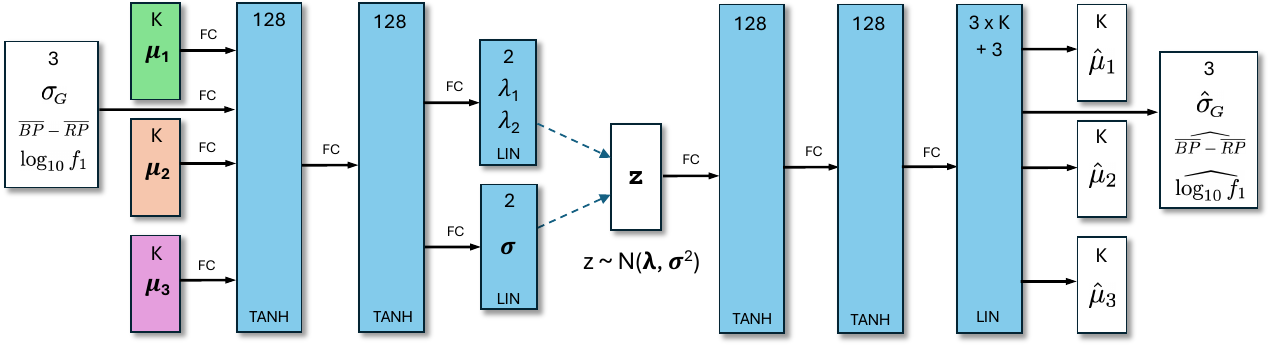}
\caption{\label{fig:ann_2D_latent_architecture}The VAE used to further reduce the latent spaces produced by the networks illustrated in Figs.~\ref{fig:ann_xp_architecture}-\ref{fig:ann_lc_architecture} (the green, orange, and purple rectangles) to a latent space of dimension 2, providing the coordinates $(\lambda_1, \lambda_2)$ that are used for the interpretation in Section~\ref{sec:astrophysical}. Apart from the location in the $3K=15$-dimensional latent space, the VAE also receives the standard deviation of the \g light curve, the colour estimate \mbox{$\overline{G_{\rm BP}}-\overline{G_{\rm RP}}$} and the (logarithm of) the main frequency $f_1$ in the Lomb-Scargle periodogram. The explanation of the ANN is similar as the one in Fig.~\ref{fig:ann_xp_architecture}. 
}
\end{center}
\end{figure*}
%

\subsection{Optimisation and implementation details}
\label{sec:optimization}

The dataset described in Section~\ref{sec:train_data} is randomly divided into training, validation, and test subsets, with proportions of 70\%, 15\%, and 15\%, respectively. 
Models are trained iteratively for a maximum of $1000$ epochs, with parameters initialised randomly using the \citet{he2015initialization} scheme at the start of training.
At the start of each epoch, the sources in the training subset are grouped into batches of size $N_b = 256$. 
Gradients of the cost function are computed for each batch, averaged, and used to update the model parameters using the Adam optimiser \citep{kingmaBa15}, an adaptive variant of Stochastic Gradient Descent. The initial learning rate is set to $0.0001$.

The validation subset is used to monitor the cost function during training. We apply early stopping with a patience of $15$ epochs to avoid overfitting. 
The validation subset is also used to calibrate the most critical hyperparameters of the models, namely the dimensionality of the latent variables $K$ and the trade-off between reconstruction and regularisation $\beta$. The selection of $K$ is discussed in detail in the following section. The value of $\beta$ is calibrated independently for each VAE through a grid search over $[0.001, 0.01, 0.1, 1, 10]$. The optimal values of $\beta$ for the XP, $\Delta m| \Delta t$, folded light curve and 2-dimensional VAEs were found to be $1$, $0.1$, $0.01$, and $1$, respectively. Other hyperparameters are set to their previously defined values and left unmodified.

After the models have converged, the latent variables are evaluated through outlier detection, classification and clustering experiments performed using the test subset. To ensure robustness, each model is trained three times with different random initialisation seeds, and performance is reported as the average of metrics across these runs. All models are implemented using \texttt{PyTorch} \citep{paszke2019}, version 2.4.1, and \texttt{PyTorch Lightning} \citep{falcon2019}, version 2.4. The model implementation, training scripts and guidelines for obtaining the dataset and latent variables are available at \url{https://github.com/IvS-KULeuven/gaia_dr3_autoencoder}. The models are trained using an NVIDIA A4000 GPU. Table \ref{tab:time} lists the computational time needed to obtain the latent variables for each of the described models. Using a single GPU, the latent variables of 100M sources can be obtained in a few hours, ensuring scalability to DR4. 

\subsection{Latent variables assessment}
\label{sec:metrics}

Once the models are trained, the quality of the latent variables is evaluated through classification and clustering tasks. In the supervised classification tests, this evaluation is performed using Logistic Regression (LR) and a \textit{k}-nearest neighbours (\textit{k}-NN) classifier. The LR is a linear model that predicts class probabilities based on a weighted combination of the input features, making it simple yet effective for assessing linear separability. The \textit{k}-NN classifier, on the other hand, is a non-parametric method that assigns labels based on the majority class of the $k$ nearest points in the latent space, providing insight into the local structure of the learned representation. We deliberately restrict this evaluation to simple classifiers because our primary goal is to assess the structure and informativeness of the latent variables. More complex classifiers, such as Random Forest or XGBoost, would likely achieve better classification performance, but they might also compensate for limitations in the learned representation. 

The LR and \textit{k}-NN classifiers are implemented using the \texttt{scikit-learn} library \citep{pedregosa2011sklearn}. The training and validation partitions that were used to train the AEs are also used to fit and select the optimal hyperparameters. Specifically, the level of regularisation for LR and the number of neighbours for \textit{k}-NN are tuned during this process. The input to the classifiers include the mean of the latent variables, the uncertainty of the XP and \g band folded light curve decoders, $\sigma_{\rm XP}$ and $\sigma_{F}$, the standard deviation of the \g band light curve $\sigma_G$\footnote{This quantity is removed during preprocessing and is therefore unavailable to the models.} and the estimated fundamental frequency $f_1$. The classifiers' performance is evaluated on the test partition, which consists of the sources that were not used during the training of the VAEs and the classifiers.

By defining $n_{ij}$ as the number of sources labelled as class $i$ and predicted as class $j$, the performance of the classifier is visually assessed using row-normalised confusion matrices, where the cell in the i-th row and j-th column corresponds to $n_{ij}/\sum_j n_{ij}$. Performance for a given class $c$ is summarised using $F_1\text{-score}_c$, which is the harmonic mean between $\text{precision}_c$, i.e. the proportion of correctly classified sources out of all sources predicted as belonging to $c$ and $\text{recall}_c$, i.e. the proportion of correctly classified sources out of all actual sources of class $c$. 

As the distribution of classes in the labelled subset is highly unbalanced, performance summaries for the entire test partition are presented using both macro and weighted averages. The macro average computes the unweighted mean of the $F_1$-scores across all classes, allowing minority classes to have a significant influence on the overall score. In contrast, the weighted average takes into account each class’s $F_1$-score according to its proportion in the dataset, meaning the overall score is more heavily influenced by the performance on the majority classes. The $F_1$ macro is used to select the best model in cross-validation.

\section{Selecting the dimensionality of the latent space}
\label{sec:latent_variables}

\begin{figure*}[t]
\centering
\includegraphics[width=0.95\textwidth]{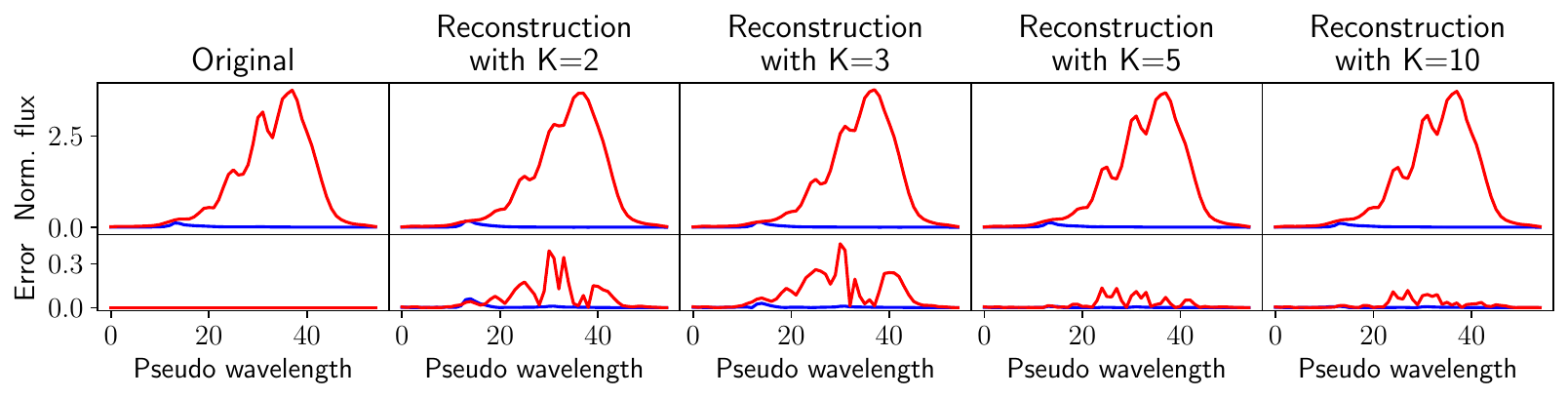}
\caption{XP mean spectra of \gaia DR3 source \texttt{4037765863920151040}, corresponding to a long period variable (first column). This data is compressed using the VAE and then reconstructed. The following columns show the reconstruction and the squared difference between the original and its reconstruction as the number of latent variables $K$ increases.}
\label{fig:rec_lv_xp}
\end{figure*}

\begin{figure*}[t]
\centering
\includegraphics[width=0.95\textwidth]{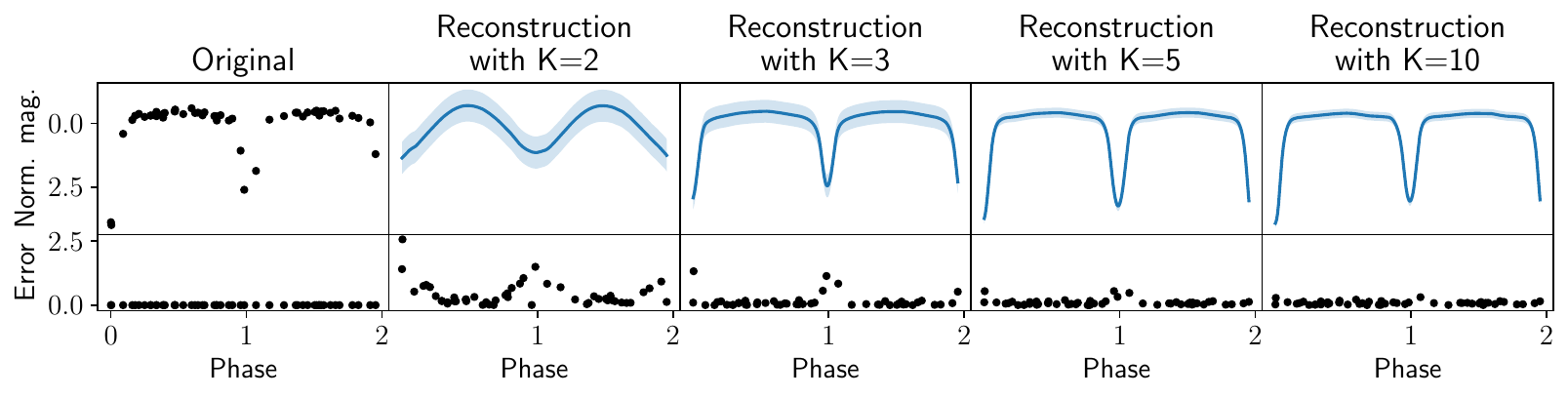}
\caption{Folded \g band light curve of \gaia source DR3 \texttt{5410617628478565248} corresponding to an eclipsing binary (first column). This data is compressed using the VAE and then reconstructed. The following columns show the reconstruction and the squared difference between the original as $K$ increases.}
\label{fig:rec_lv_fold}
\end{figure*}

\begin{figure}[t]
\centering
\includegraphics[width=0.5\textwidth]{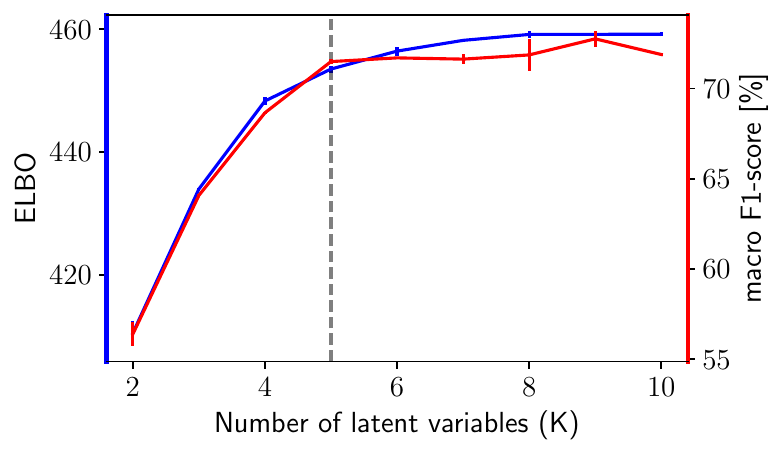}
\caption{Average Evidence Lower BOund (ELBO, in blue) and macro $F_1$-score (in red) in the validation set as a function of the number of latent variables (K). Higher is better for both metrics. A black dashed line marks $K=5$.}
\label{fig:latent_variables}
\end{figure}

Among the hyperparameters of the VAE, the dimensionality of the latent variable vector, $K$, has the most significant influence on the results and we analyse its impact in this section. $K$ controls the amount of compression that an AE applies to the data. A higher $K$ allows the encoder to obtain more informative representations, leading to better reconstructions. However, choosing $K$ requires a balance: too small a value results in a latent space with limited or no discriminative power, while too large a value risks overfitting to noise in the data, enabling near-perfect reconstructions at the expense of the usefulness of the latent variables for classification, clustering and other downstream tasks. 

Figure~\ref{fig:rec_lv_xp} illustrates how the reconstruction of the XP mean spectra for a given \gaia DR3 source evolves with $K$. Even with just two latent variables, the reconstruction captures the general shape and relative differences between the BP and RP spectra, although key details are smoothed out. In contrast, the reconstructions with $K=5$ and $K=10$ are much closer to the original spectra, with only marginal differences between them, as reflected in the errors. Figure~\ref{fig:rec_lv_fold} shows an example of the compression of a folded \g band light curve of an eclipsing binary star. With $K=2$, the reconstruction fails to capture the correct shape, defaulting to a sinusoidal curve, probably the most common pattern found in the dataset for this modality. In contrast, $K=5$ and $K=10$ accurately capture the shape of the light curve, including the distinctive features of the eclipses. Similar to what was observed for the XP mean spectra, the difference between the $K=5$ and $K=10$ reconstructions is marginal.
 
For the evaluation of $K$, we consider two metrics measured in the validation partition of our dataset. The first is the average between the ELBOs (Eq.~\ref{eq:elbo}) from the VAEs of the XP mean spectra, $\Delta m|\Delta t$ and \g band folded light curve. 
The ELBO takes into account the likelihood of the reconstructed data, and hence, we expect it to increase with $K$. Although not optimal, we adopt the same number of latent variables for each VAE to facilitate the analysis. The second metric is the macro $F_1$-score measured through a LR classifier using the combined latent variables as described in Section~\ref{sec:metrics}. We use this downstream task to measure the discriminative power of the latent variables, which, contrary to the likelihood, is not expected to always increase with $K$. Figure~\ref{fig:latent_variables} illustrates these two metrics as $K$ is increased from two to ten. The validation ELBO shows increments above the error bars up to $K=8$. Not growing beyond this value is due to the influence of the regularisation term in Eq.~\eqref{eq:elbo}. The first five latent variables contribute most to this metric. Similarly, the $F_1$-score rises sharply up to $K=5$, after which it levels off. This suggests diminishing returns in both reconstruction quality and classification performance for higher values of $K$. Using fewer latent variables also facilitates the interpretation. Based on these results, we select $K=5$ for the remainder of our analysis.

\section{Identification of outliers using the latent variables}
\label{sec:outliers}

\begin{figure}[t]
\centering
\includegraphics[width=0.48\textwidth]{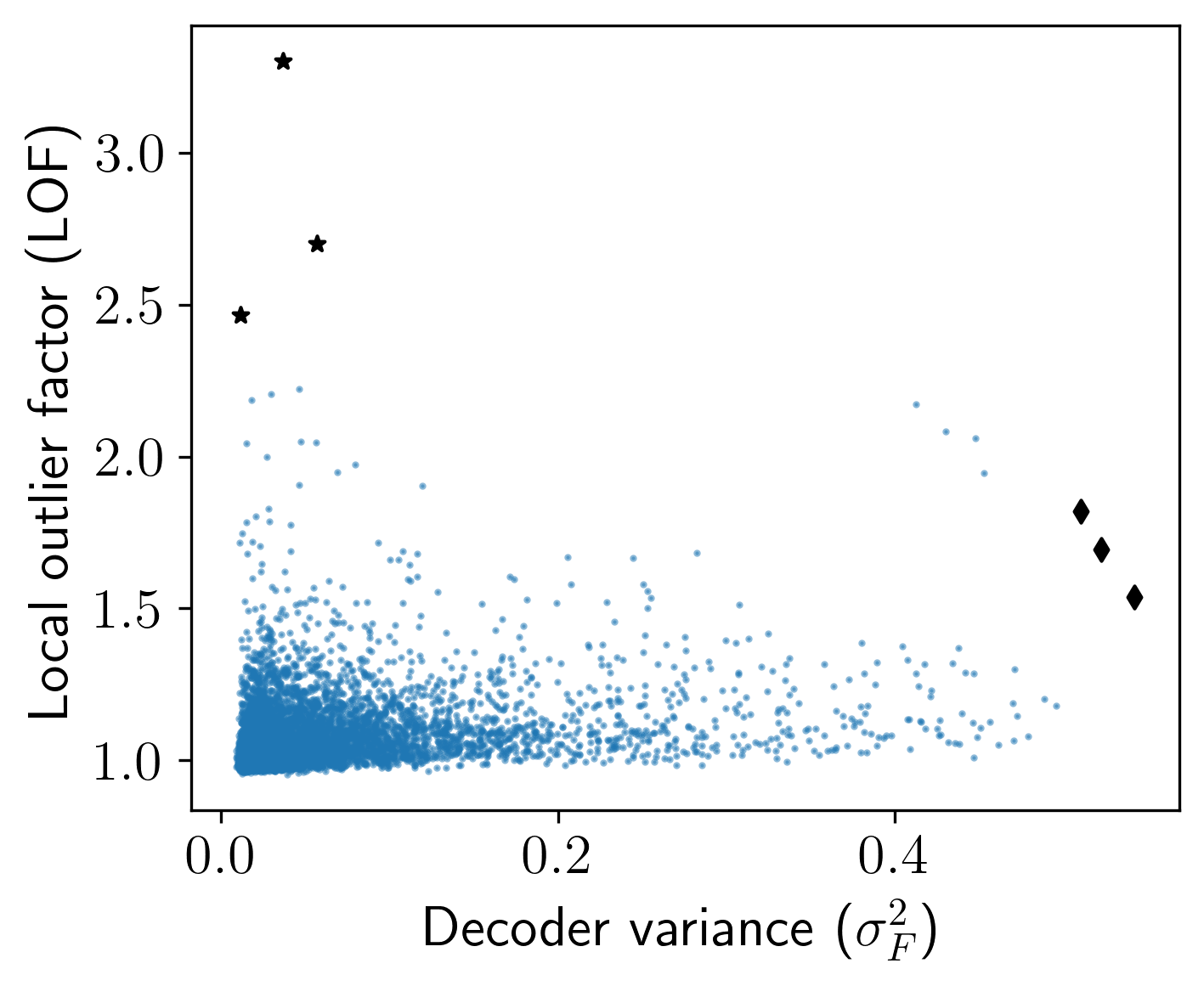}
\caption{Local outlier factor (LOF) vs the variance of the \g band folded light curve decoder ($\sigma_F^2$), for the 7431 sources from the \rr SOS table within the test partition. A large LOF indicates that the source is isolated in the latent space. A large $\sigma_F^2$ is associated with a noisy reconstruction of the folded light curve. Black stars show the three sources with the largest LOF, while black diamonds show the three sources with the largest $\sigma_F^2$.} 
\label{fig:outlier_score}
\end{figure}

\begin{figure}[t]
\centering
\includegraphics[width=0.48\textwidth]{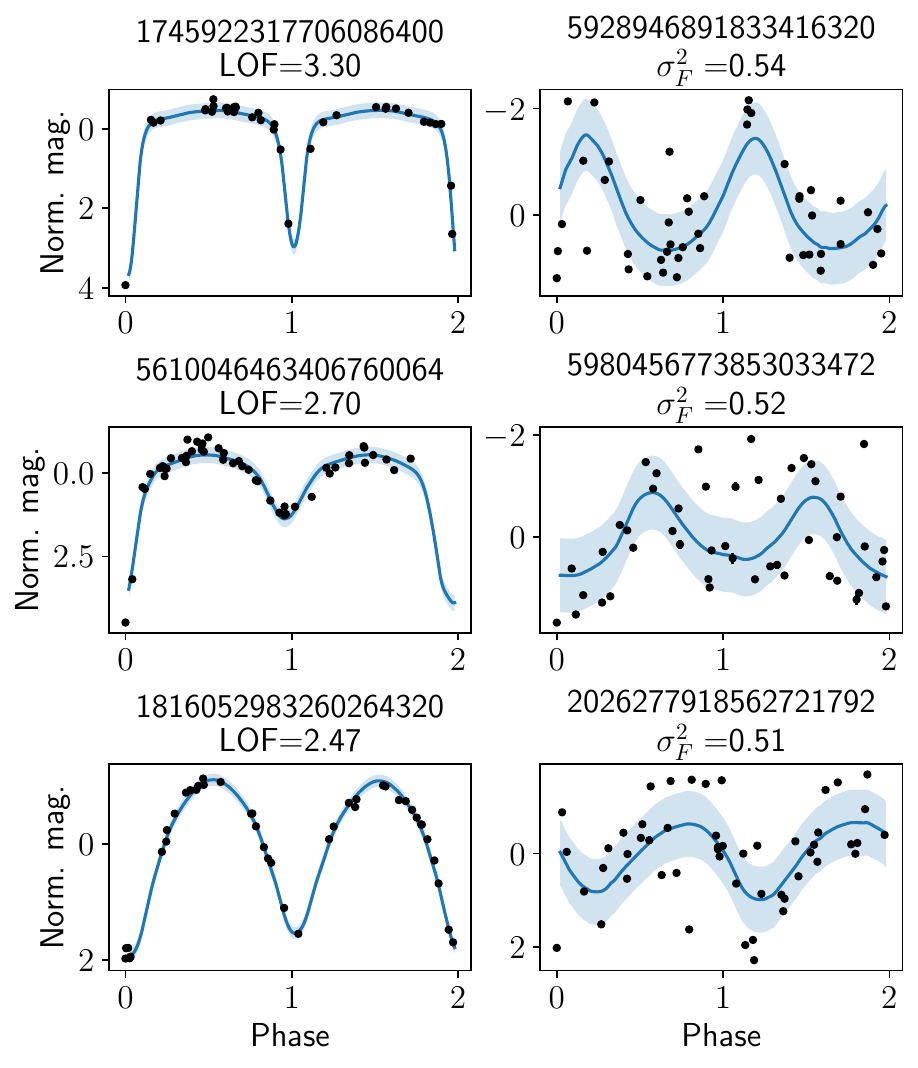}
\caption{Folded light curves (black dots) and their reconstructions (blue), for the three sources in the \rr SOS table with highest LOF (left panels) and highest $\sigma_F^2$ (right panels), respectively. \gaia DR3 source ids are shown in the title.}
\label{fig:outlier_examples}
\end{figure}

In this section, we show that the latent space produced by our ANN can be used to identify outliers or anomalous sources. One important application of this capability is quality control of the variability classifications performed by DPAC-CU7, in which several of the current authors are involved. Given the immense size of the datasets, these classifications are often carried out using automated algorithms. Consequently, there is significant interest in exploring whether part of the verification process can also be automated.

As an example, we consider the 7431 sources from the \texttt{vari\_rrlyrae} table that are included in the test partition of our dataset. This table, compiled through a rigorous analysis by \citet{clementini2023rrlyr}, contains a clean catalogue of \rr stars. The aim is to identify the most atypical \rr sources using the latent variables as input for an outlier detection algorithm. Specifically, we use the Local Outlier Factor (LOF) algorithm \citep{breunig2000lof}, which assigns a score indicating how isolated a source is relative to its local neighbourhood. The larger the LOF, the more isolated the source is; inliers tend to have values close to one. The \texttt{scikit-learn} implementation of LOF with its default hyperparameters is used. As input, we consider the five latent variables derived from the folded \g band time series, i.e. the outliers are to be interpreted as the most atypical in the latent space of the folded time series. However, the same procedure outlined here can be applied to the other latent variables to identify the most atypical in the XP or $\Delta m| \Delta t$ representations.

A widely used heuristic for anomaly detection that is specific to AEs consists of using the quality of the reconstructed data to identify outlier examples. This assumes that anomalous sources are harder to reconstruct because the model minimises the overall reconstruction error, thereby prioritising normal instances \citep{pang2020anomaly, sanchez2021agns}. In this case, this corresponds to identifying sources for which the decoder variance $\sigma_F^2$ is large. In what follows, we use both the LOF and $\sigma_F^2$ to identify outlier candidates and analyse the differences between these approaches.

Figure~\ref{fig:outlier_score} shows the LOF versus $\sigma_F^2$ distribution for the 7431 sources from \citet{clementini2023rrlyr} in the test-set partition of our training set. Most sources are concentrated around LOF values close to one and have low $\sigma_F^2$. In general, the correlation between LOF and $\sigma_F^2$ is weak, i.e. the most isolated sources are not necessarily due to the model being unable to reconstruct them correctly. The three sources with the highest LOF values are marked with stars. Notably, these sources also exhibit low values of $\sigma_F^2$, suggesting that their reconstructions are reliable. The folded light curves of these sources are shown in the left panel of Fig.~\ref{fig:outlier_examples}, ranked by their LOF from top to bottom. The shape of the light curves suggests that these sources are likely eclipsing binaries. This is further confirmed using the amplitude ratio between the \bp and \rp band time series, as indicated by \citet{clementini2023rrlyr}. Moreover, the second and third sources have been classified as binaries using ATLAS data \citep{heinze2018atlas}. The full potential of latent-space neighbour analysis for identifying contaminants in \gaia data will be examined in greater detail in Wyrzykowski et al. (in prep.).

The second column of Fig.~\ref{fig:outlier_examples} shows the three sources with the highest decoder variance, representing those that the model struggles to reconstruct accurately. These sources have very noisy periodograms, which complicates the extraction of the correct frequency and leads to finding a high-frequency alias instead, resulting in a seemingly random pattern in the folded light curve. Analysis using the amplitude ratio between the \bp and \rp band time series suggests that these sources are indeed \rr stars. Moreover, the first and third sources have been classified as \rr stars using ATLAS data. In contrast to the LOF method, the candidates identified using $\sigma_F^2$ are not contaminants from other astrophysical classes. Instead, they primarily result from noisy time series or preprocessing errors. Nonetheless, combining both metrics may offer added value. This will be explored in future work.

\section{Supervised classification in the latent space}
\label{sec:classification}

\begin{table}[t]
\caption{Supervised classification results in the latent space}
\label{tab:f1s}
\centering
\begin{tabular}{r|cccc}
\toprule \toprule
Latent& \multicolumn{4}{c}{$F_1$-score [\%]} \\
variables &  \multicolumn{2}{c}{Logistic regressor} & \multicolumn{2}{c}{\textit{k}-NN classifier}\\
from: & Macro & Weighted & Macro & Weighted \\
\midrule
XP spectra & 39.9(1.1) & 55.1(0.9) & 52.3(0.8) & 68.1(0.5) \\[1mm]
$\Delta m|\Delta t$ LC& 28.4(0.5) & 39.7(0.8) & 27.8(0.7) & 51.6(0.1) \\[1mm]
Folded LC & 36.1(0.1) & 47.9(0.1) & 35.8(1.2) & 55.2(0.1) \\[1mm]
$\Delta m|\Delta t$\&Fold & 43.1(0.7) & 66.6(0.5) & 39.3(0.4) & 64.5(0.3) \\[1mm]
All inputs & 68.2(0.4) & 79.9(0.4) & 66.1(0.3) & 80.5(0.2) \\
\bottomrule
\end{tabular}
\tablefoot{Average $F_1$-scores on the test set for logistic regression and k-nearest neighbours classifiers, trained using latent variables from XP mean spectra, folded light curves (LC), $\Delta m|\Delta t$ from LC and combinations between them. Standard deviations are shown in parentheses. A higher $F_1$-score implies a better classification.}
\end{table}

\begin{figure*}[t]
     \centering
     \begin{subfigure}[t]{0.49\textwidth}
     \centering
         \includegraphics[width=\textwidth]{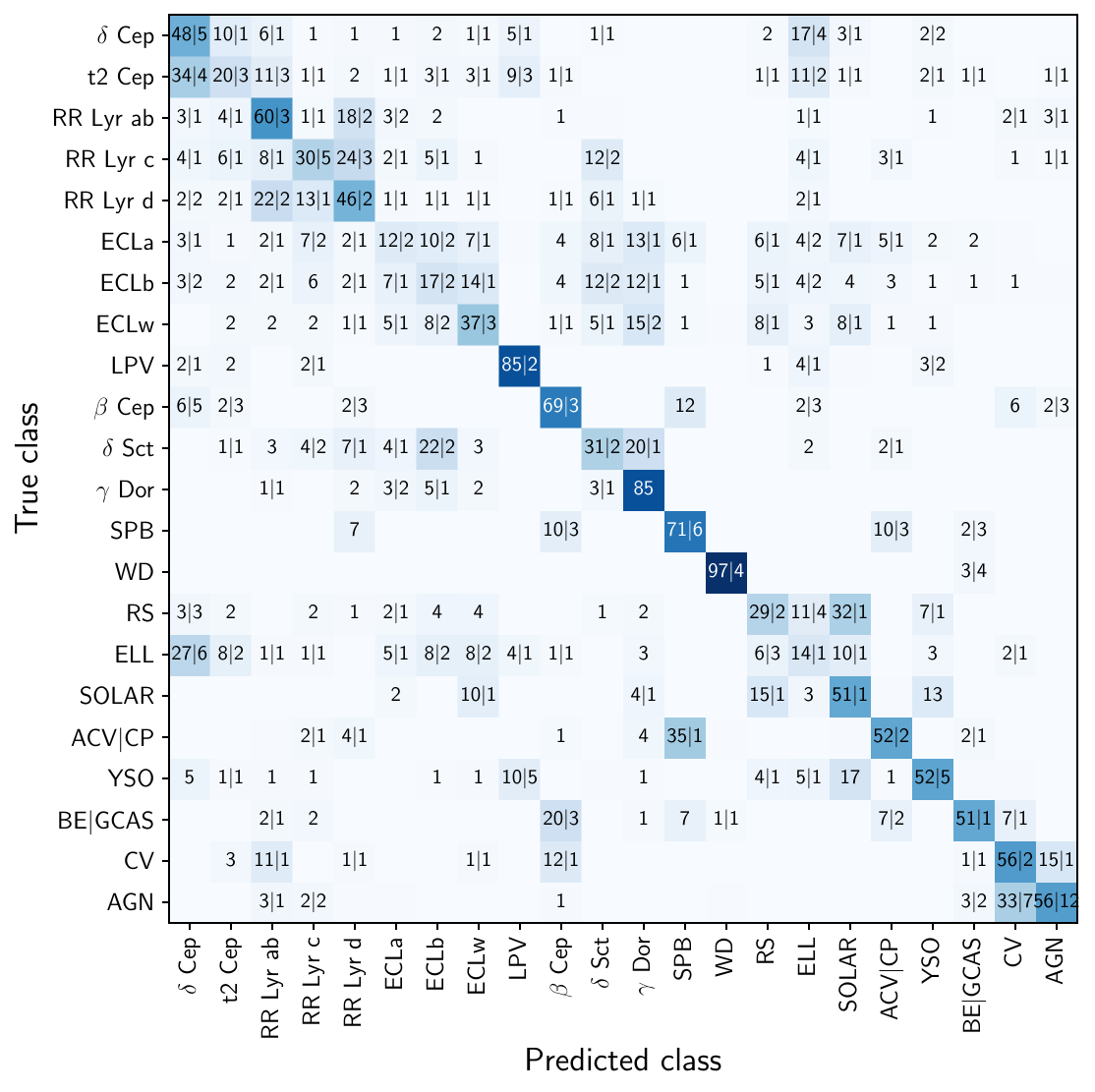}
         \caption{Using only the latent variables from the XP mean spectra.}
         \label{fig:cm_xp}
     \end{subfigure}
     \hfill
     \begin{subfigure}[t]{0.49\textwidth}
     \centering
         \includegraphics[width=\textwidth]{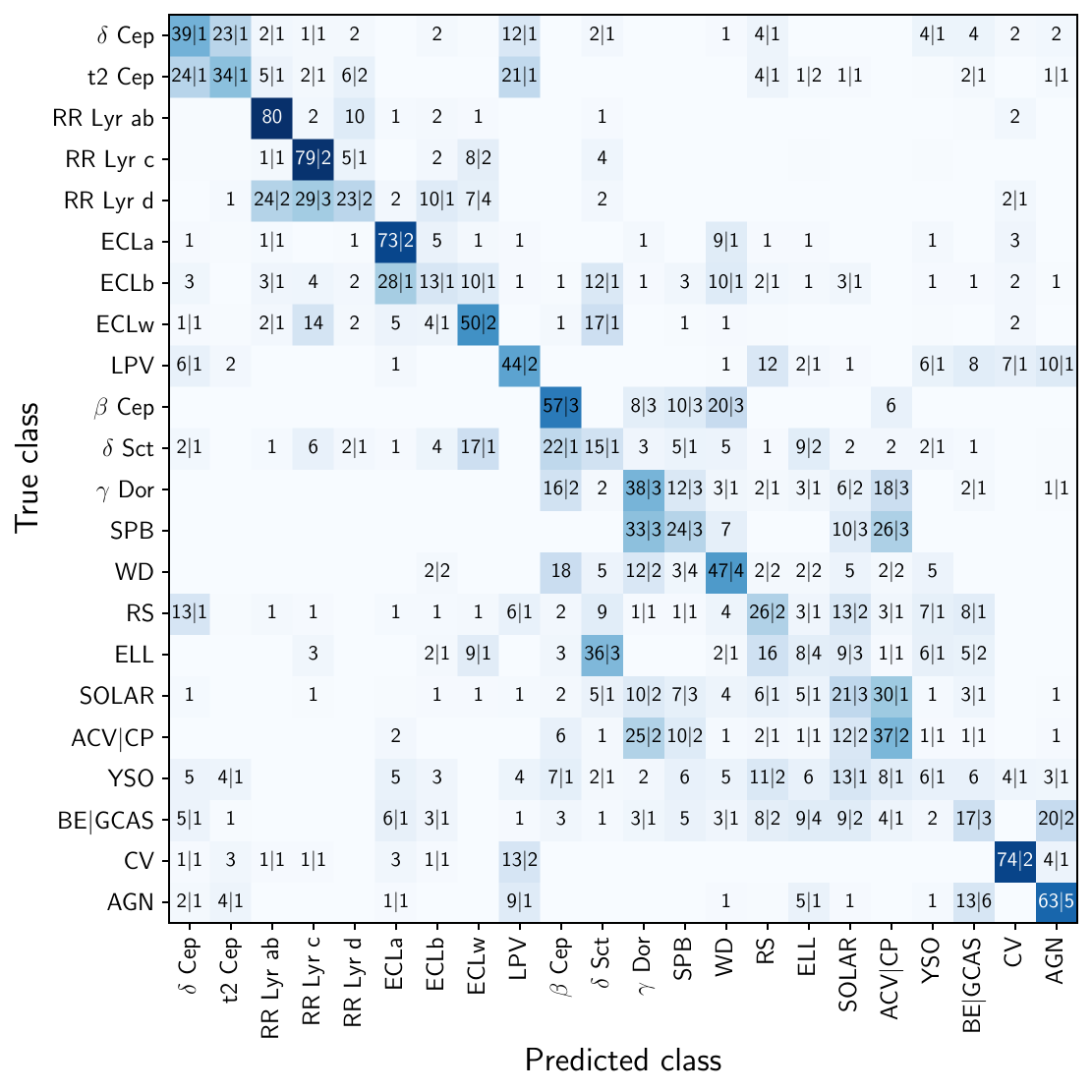}
         \caption{Using only the latent variables obtained from the \g band $\Delta m|\Delta t$.}
         \label{fig:cm_dmdt}
     \end{subfigure}
     \vfill
     \begin{subfigure}[t]{0.49\textwidth}
     \centering
         \includegraphics[width=\textwidth]{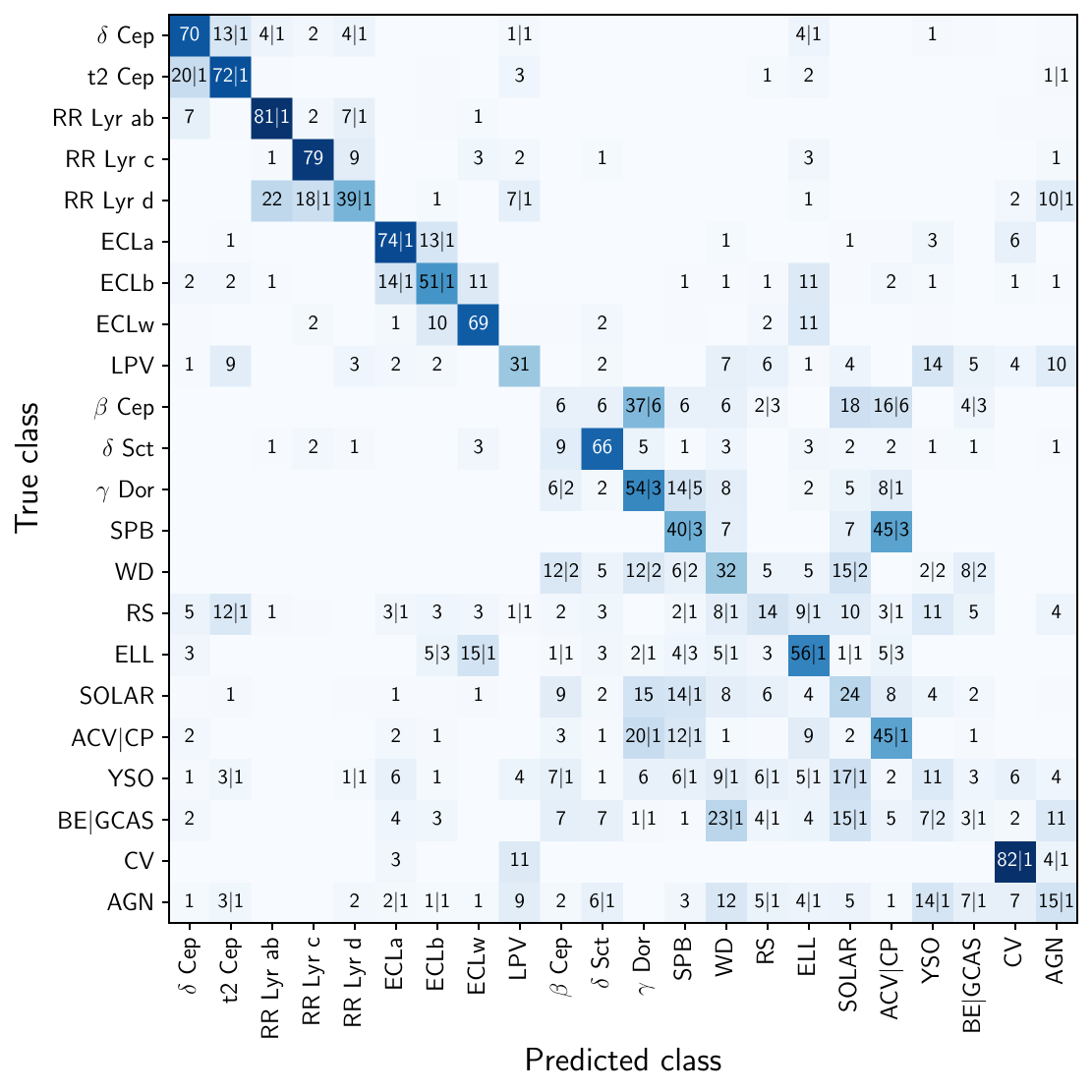}
         \caption{Using only the latent variables obtained from the folded \g band LCs.}
         \label{fig:cm_fold}
     \end{subfigure}
     \begin{subfigure}[t]{0.49\textwidth}
     \centering
         \includegraphics[width=\textwidth]{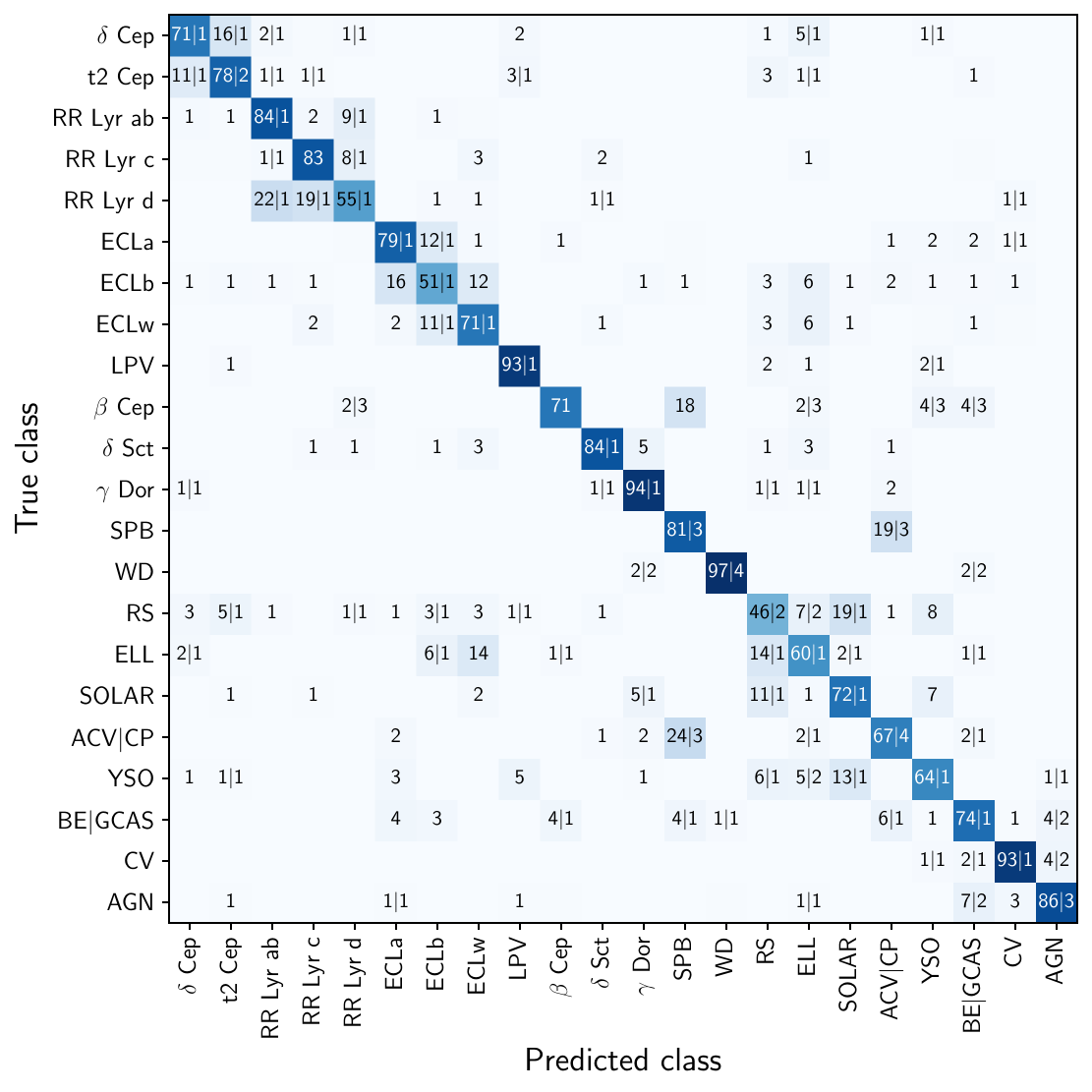}
         \caption{Using the latent variables from all the data products as input.}
         \label{fig:cm_all}
     \end{subfigure}
     \caption{Confusion matrices for the test partition sources of the labelled subset. The rows correspond to the ground-truth class selected from \cite{gavras2023gaia}, and the columns to the class predicted by the classifier using the latent variables as input. The cell in the i-th row and j-th column gives the percentage of sources labelled as class $i$ and predicted as class $j$ with respect to the total number of sources of class $i$. Each matrix represents the averaged results of a linear classifier trained on different sets of latent variables. Standard deviations are shown after the $\mid$ symbol. Percentage rates and corresponding standard deviations below 1\% are not shown for clarity.}
     \label{fig:cm}
\end{figure*}

The latent space was constructed in a self-supervised way without any variability class labels known from the literature. Although it primarily serves to discover unanticipated overdensities, one could ask if the $3K = 15$ coordinates of the latent space can also be used as new features for supervised classification. To verify this, we perform a classification experiment using the 38,740 labelled sources described in Section~\ref{sec:test_data}.
Table \ref{tab:f1s} presents the test-set $F_1$-scores achieved by the LR and the \textit{k}-NN classifiers (cf.~Section~\ref{sec:metrics}) trained on different subsets of the latent variables. 
We present results where the classifiers were applied to 1) the $K$ latent variables coming from the XP spectra,
2) the $K$ latent variables coming from the
$\Delta m|\Delta t$ distribution, 3) the $K$ latent variables coming from the folded light curves, 4) the $2K$ latent variables coming from both the folded light curves and the 
$\Delta m|\Delta t$ distribution combined, and 5) all the $3K$ latent variables combined.

The classifiers show comparable performance for the single-modality $K$ input latent variables, except for the XP mean spectra, where \textit{k}-NN significantly outperforms LR. This indicates that the latent space of the XP spectra contains intricate local structures that cannot easily be linearly separated. For the LR classifier, the best single-modality performance is achieved by the \g band folded light curves.
Across both classifiers, the best performance is observed when all latent variables are combined, highlighting the advantages of a multimodal approach. Even though both classifiers achieve a similar 81\% weighted $F_1$-score, the LR slightly outperforms the \textit{k}-NN in terms of macro $F_1$-score, meaning that the former handles the minority classes better without affecting the overall performance. In what follows, we perform a comparison at the individual class level, focusing on the results of the LR classifier.

Figure~\ref{fig:cm} shows the test-set confusion matrices of the LR classifier trained using different combinations of latent variables. Through these figures, we can see that the performance across classes varies considerably depending on the modality. From the results of the XP VAE, shown in Fig.~\ref{fig:cm_xp}, we can see that:
\begin{itemize}
\item WD are almost perfectly separable despite being indistinguishable from other classes in the light curve associated modalities. 
\item Most LPVs are correctly recovered. The sources classified as LPV are characterised by their strong red spectra.
\item SPB stars are the primary source of misclassification for ACV|CP and \bcep stars. Conversely, the most significant source of confusion for the SPB stars are ACV|CP and \bcep stars. Additionally, \bcep stars are the leading source of misclassification for BE|GCAS.  All these classes are characterised by their strong blue spectra.
\end{itemize}

The results from the $\Delta m|\Delta t$ VAE, shown in Fig.~\ref{fig:cm_dmdt}, reveal interesting patterns despite notable contamination among, for example:
\begin{itemize}
\item Most \rrab and \rrc stars are correctly recovered due to their constrained period range. \rrd stars, which pulsate in double mode are mostly confused with the former. 
\item \dcep and \tiicep stars are confused with one another, as well as with LPVs. The period range of the classes partially overlap. Furthermore, some AGN and CV are also misclassified as LPV. All of these sources are characterized by variability on long timescales.
\item Most ECLa are correctly recovered. This is due to the model capturing the pattern of the top and bottom magnitude bins being filled by the sharp eclipses of these sources. 
\end{itemize}

The results from the folded light curve VAE, shown in Fig.~\ref{fig:cm_fold}, also show interesting properties:
\begin{itemize}
\item ECL yield recalls above 50\%, with most of the contamination confined to their sub-classes. This demonstrates the model's ability to effectively capture the distinctive phase patterns of these sources.
\item The majority of Cepheid and \rr stars are accurately classified. As with the ECL, these classes also have distinct and recognisable phase patterns. 
\item There is significant misclassification between SOLAR, YSO, SPB, ACV$\mid$CP, \bcep and \gdor stars. The folded light curves of these sources exhibit shapes resembling noisy sine waves.
\item AGNs are frequently misclassified as YSOs, WDs, and LPVs. Similarly, LPVs are often misclassified as YSOs and AGNs. For non-periodic and semiregular variables, the estimated periods are, in most cases, related to \gaia scan-angle dependent signals \citep{holl2023gaia}.
\end{itemize}

Figure~\ref{fig:cm_all} shows the results of the LR classifier using the latent variables from all modalities as input. For reference, the precision, recall, and $F_1$-scores per class for this classifier are listed in Table~\ref{tab:pre_rec_f1}. Performance is superior to the individual modalities across all individual classes, which is reflected in the strong diagonal of the confusion matrix. 
The following classes exhibit a relative increase in $F_1$-score of more than 50\% with respect to the best single modality: \bcep stars, \gdor stars, ELL, RS, SPB, \rrd stars, \tiicep, CV and YSO.

Some classes remain challenging to classify correctly, for example:
\begin{itemize}
\item RS, ELL and YSO: These classes are confused between them and with the SOLAR class. RS, ELL and SOLAR are all examples of rotational variables. Confusion between these classes was also reported by \citet{rimoldini2023gaia}  (cf. Table 3).
\item SPB: Approximately 20\% of sources from this minority class are missclassified as ACV$\mid$CP. Conversely, a notable fraction of ACV$\mid$CP, and \bcep stars are erroneously predicted as SPB. All these classes share similar XP properties and periodicities. Confusion between SPB and ACV$\mid$CP was also reported by \citet{rimoldini2023gaia}.
\item \rrd stars, which are multimode pulsators, are difficult to distinguish from other members of their parent class. A similar situation can be observed for ECLb. The folded light curve morphology of ECLb is a transition between ECLa and ECLw and this is captured by the latent variables. 
\end{itemize}

Next, we compare the latent variables against the classifications published in \gaia DR3 \citep{rimoldini2023gaia}. These were obtained using an ML classifier trained with labels similar to those employed in this study \citep{gavras2023gaia}, but with notably different inputs: the \gaia DR3 classifier did not utilise information from the XP mean spectra (but it did use astrometric features). Moreover, the DR3 predictions were refined through post-processing based on attribute filters to enhance the reliability of the classifier's results.

\begin{table}[t]
\caption{Comparison between VAE and DR3 classifications}
\label{tab:f1s_dr3}
\centering
\begin{tabular}{r|ccc}
\toprule \toprule
Class& VAE all inputs & DR3  & \# Sources\\
\midrule
\dcep|\tiicep & 81.0(1.3) & 98.1 & 106 \\
\rr & 92.9(0.4) & 96.7 & 259 \\
ECL & 92.3(0.2) & 94.8 & 651 \\
LPV & 97.7(0.2) & 98.9 & 1329 \\
\dsct$\mid$\gdor & 89.7(0.4) & 93.5 & 383 \\
RS & 0.0(0.0) & 0.0 & 183 \\
SOLAR & 88.4(0.4) & 75.1 & 262 \\
ACV$\mid$CP & 87.3(0.7) & 55.0 & 12 \\
YSO & 52.3(1.4) & 90.4 & 37 \\
AGN & 97.2(1.1) & 100.0 & 74 \\
\bottomrule
macro avg.& 81.5(1.1)  & 80.3 & 3300  \\
weighted avg.& 88.4(0.3) & 89.5 & 3300 \\
\bottomrule
\end{tabular}
\tablefoot{$F_1$-scores per class for the LR trained using all the latent variables (with merged subclasses) and the DR3 predictions from \citet{rimoldini2023gaia} after removing the sources belonging to the training set of the latter.}
\end{table}

This comparison is carried out using a subset of 3300 sources obtained by removing from our test set all the sources that \citet{rimoldini2023gaia} used for training. The results presented in Table~\ref{tab:f1s_dr3} show overall agreement between the two, though significant differences arise in some of the classes, such as ACV, CEP, SOLAR and YSO. However, given the limited sample size, these differences may not be fully representative of the broader dataset. Notably, both models are not able to recover the RS sources in this sample.

\section{Analysis of clusters in the latent space}
\label{sec:clustering}

\begin{figure}[t]
\centering
     \begin{subfigure}[t]{0.49\textwidth}
    \includegraphics[width=\textwidth]{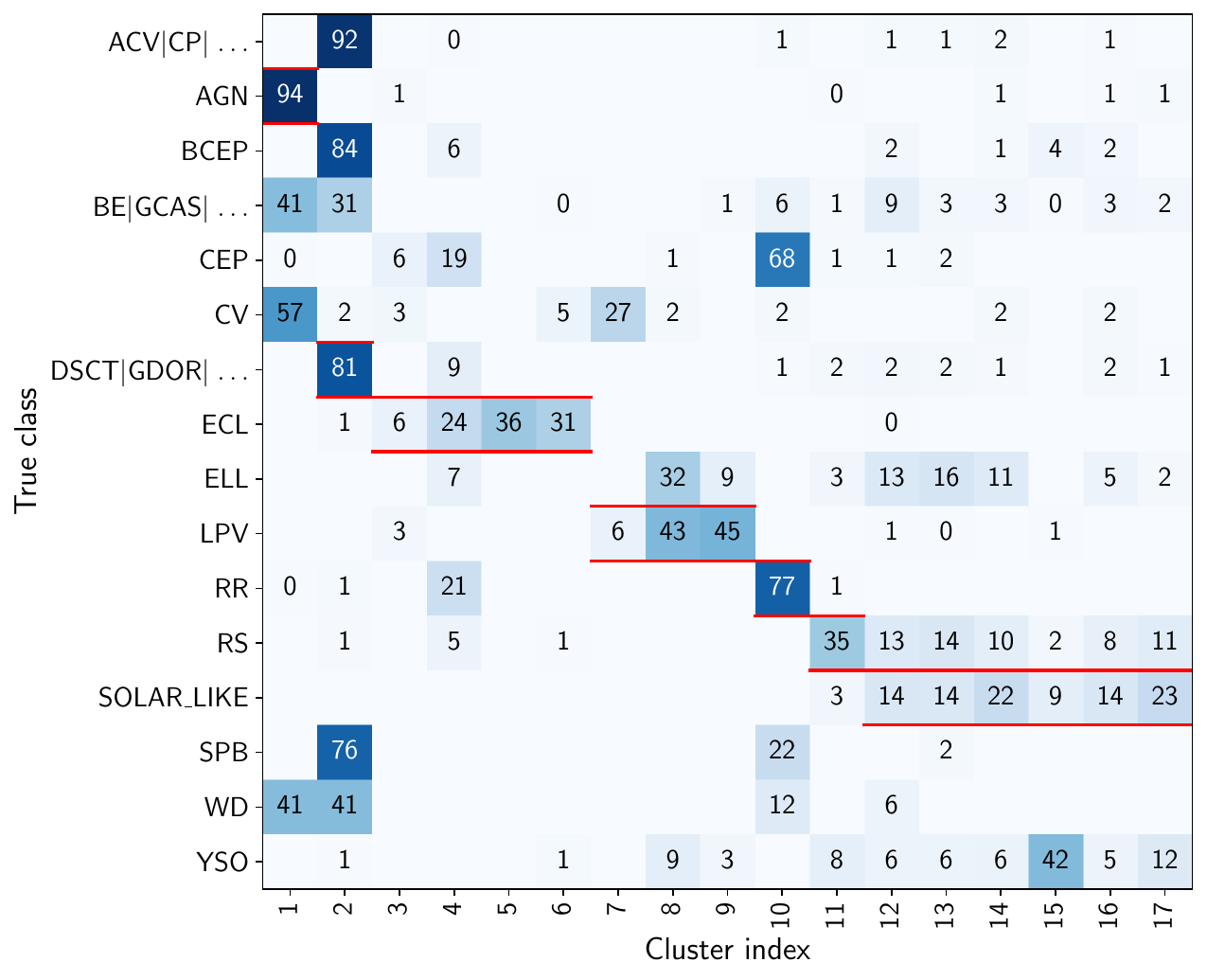}
    \end{subfigure}
    \begin{subfigure}[t]{0.49\textwidth}
    \includegraphics[width=\textwidth]{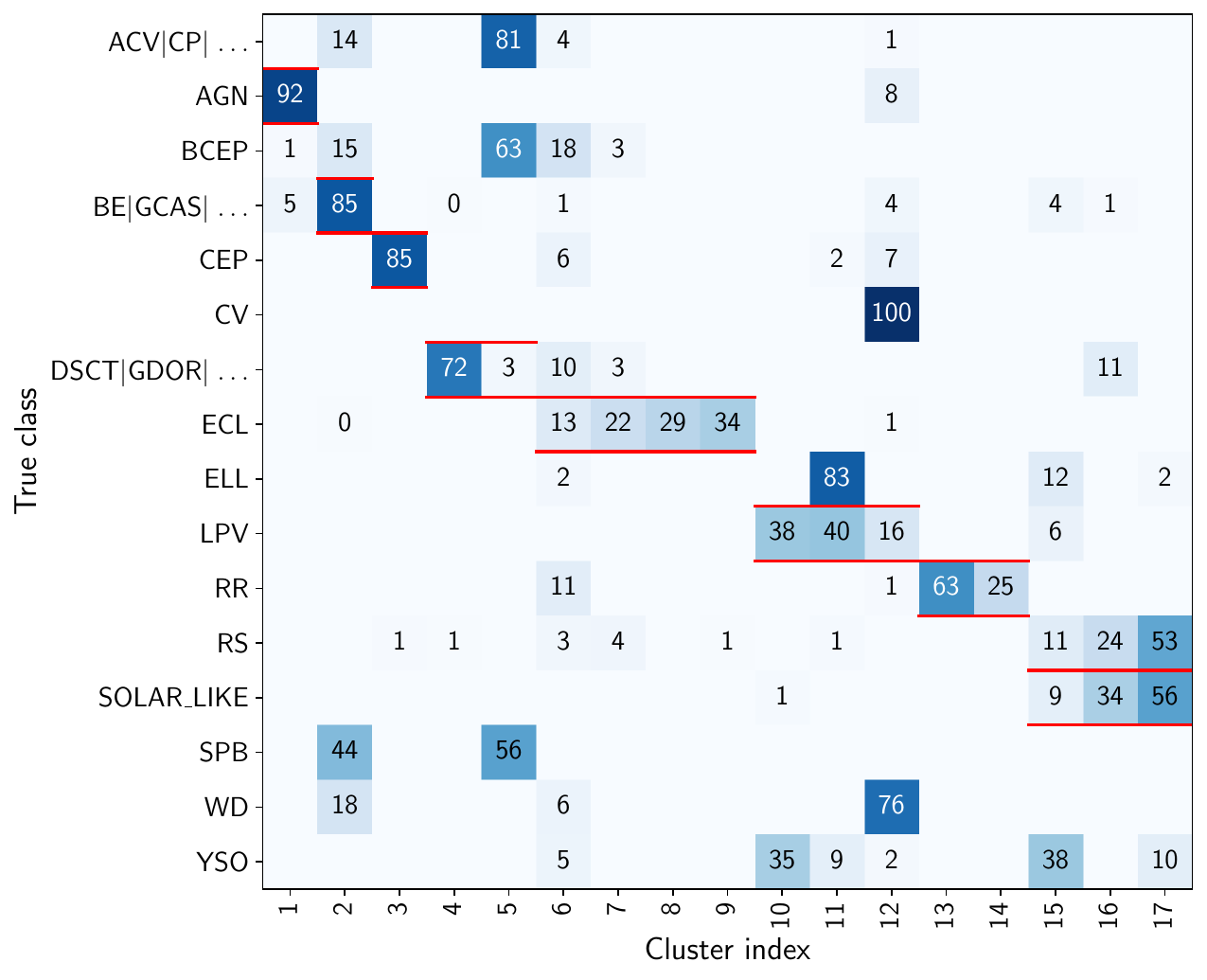}
    \end{subfigure}
    
\caption{Clustering results of the latent variables into 17 groups using $k$-means (top panel) and Gaussian mixture model with full covariance (bottom panel). Rows show the distribution of the classes across the clusters (rows add up to 100). Red lines highlight the class that constitutes the majority within each cluster.}
\label{fig:clustering}
\end{figure}

We now aim to identify and extract in an automated way the different clusters present in the $3K=15$-dimensional latent space, and explore their content by using the 301,778 test-set sources with labels from \citet{rimoldini2023gaia}.

We begin by evaluating the $k$-means algorithm for clustering the latent variables. $k$-means was selected due to its computational efficiency, which is essential for application to the significantly larger dataset expected for \gaia DR4. In this case, we set the number of clusters to 17, as it yields the minimum Davies-Bouldin index \citep{Davies1979} as shown in Fig.~\ref{fig:dbi}. 

The top panel of Fig.~\ref{fig:clustering} presents the resulting confusion matrix between the reference labels and the $k$-means clusters. Each row displays the distribution of a given variability class across the clusters (recall). To aid visualisation, clusters are sorted by the dominant class within them, marked with red lines. For example, the first cluster is dominated by AGNs, with contamination from CVs, BE|GCAS and WD. The confusion matrix reveals that more than half of the clusters are dominated by three classes. This is consistent with the imbalanced distribution of classes in the subset, where SOLAR\_LIKE, LPV, and ECL make up 40\%, 20\%, and 13\% of the dataset, respectively. 

As a second clustering method, we consider a Gaussian mixture model (GMM) with full covariance matrices. The marginal likelihood for this model is defined as:
\begin{equation}
L(\vec \pi, \vec \mu, \vec \Sigma) = \sum_{i} \sum_{c=1}^C \pi_c \mathcal{N}(x_i|\mu_c, \Sigma_c),
\end{equation}
where $\vec \pi$ are the mixture coefficients and $C$ is the number of clusters. $k$-means can be considered as a special case of GMM, where $\pi_c=\frac{1}{C}$, i.e. equal density, and $\Sigma_c = \sigma^2 I$, i.e. a shared spherical covariance. While GMMs are computationally more complex than $k$-means and are unlikely to scale efficiently for the larger \gaia DR4, their increased flexibility offers the potential to better handle class imbalance.

The bottom panel of Fig.~\ref{fig:clustering} shows the resulting confusion matrix between the reference labels and the GMM clusters. Compared to $k$-means, the GMM reduces the dominance of the majority class (SOLAR\_LIKE), enabling better representation of medium-sized classes such as CEP (0.2\%) and RR (2\%). The RS class (10\%), despite being adequately represented in terms of examples, cannot be distinguished from the SOLAR\_LIKE class. In general, classes such as RS, SPB, and ELL, which were incorrectly classified in the supervised experiments, exhibit similar behaviour in the clustering scenario.
While medium-sized classes show improvement over the $k$-means results, very small classes such as \bcep, CV and WD\footnote{Together, these four classes make less than 0.1\% of the subset.} remain intermingled as contaminants within larger clusters. Increasing the number of clusters to 100 allows for the identification of pure clusters associated with these classes, but at the cost of making the clustering results impractical to use and interpret. In what follows, we describe the clusters found by GMM:
\begin{enumerate}
\item Cluster 1 contains the majority of AGN with slight contamination from BE|GCAS.
\item Cluster 2 is dominated by BE|GCAS with contamination from ACV|CP, \bcep stars, SPB and WD. All of these sources are characterised by their strong blue spectra, as inferred from the reconstruction of the cluster mean (Fig.~\ref{fig:cluster2}).
\item Cluster 3 contains the majority of Cepheids. 
\item Clusters 4 and 5 are dominated by \dsct|\gdor stars, where cluster 4 is much purer. Interestingly, cluster 5 showcases the same contaminants as cluster 2. The reconstruction of the mean of cluster 5 (Fig.~\ref{fig:cluster5}) suggests that these sources vary more rapidly than those in cluster 2.
\item Clusters 6 to 9 are ECL clusters. Cluster 6 is contaminated with several other classes. Other ECL clusters exhibit high purity.
\item Clusters 10 to 12 are LPV clusters with different degrees of contamination, mainly from YSO, ELL, CV and Cepheids.
\item Clusters 13 and 14 are pure \rr star clusters. The reconstructions from the means of these clusters (Figs.~\ref{fig:cluster13}--\ref{fig:cluster14}) suggest that they are associated with \rrab and \rrc stars, respectively.
\item The last three clusters are dominated by SOLAR\_LIKE with high contamination from RS and YSO. This is similar to what was observed in the supervised analysis.
\end{enumerate}

The observations extracted from Fig.~\ref{fig:clustering} are contrasted in the following section, where a manual analysis of the latent variables through a two-dimensional projection is carried out.

\section{Astrophysical interpretation of the latent space}
\label{sec:astrophysical}

\begin{figure}
\begin{center}
\includegraphics[width=0.5\textwidth,keepaspectratio]{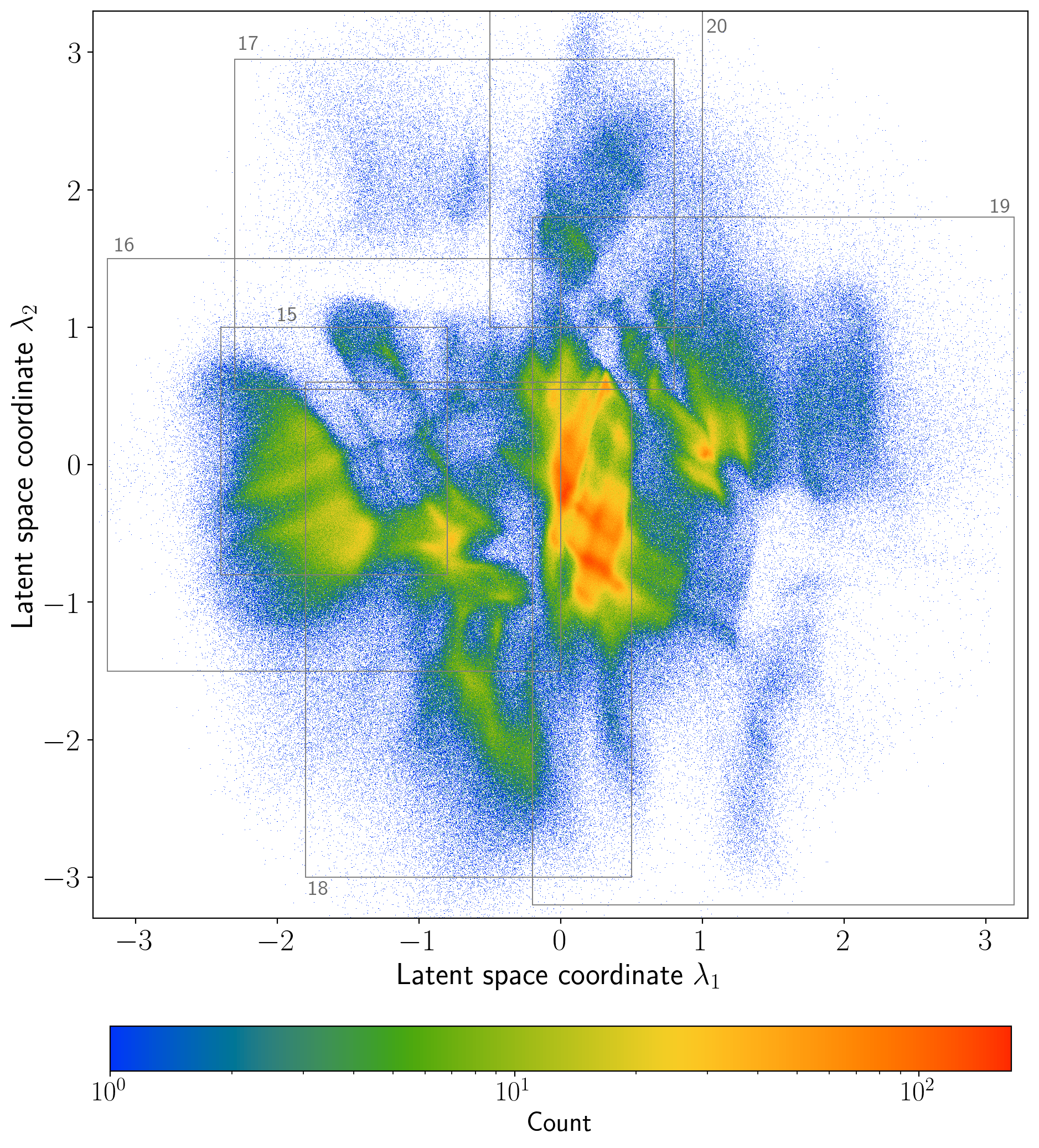}
\caption{\label{fig:density_heatmap} Distribution of the 4,136,544 sources in the two-dimensional latent output space of our ANN designed to perform an unsupervised clustering of variable sources (see Section~\ref{sec:models} and Fig.~\ref{fig:ann_2D_latent_architecture} for more details). Different overdensities of points correspond to 
different (sub)types of variability. A first identification of the overdensities can be found in Fig.~\ref{fig:purity_class_heatmap}. For future reference we also depict the parts of this latent space shown in Figs.~\ref{fig:rrlyr_heatmap}-\ref{fig:agn_f1_heatmap} with rectangles and provide the figure number is in one of the corners.}
\end{center}
\end{figure}

In this section, we analyse in more depth the most important regions and overdensities in the latent space from an astrophysical point of view. We work in the 2-dimensional latent space shown in Fig.~\ref{fig:density_heatmap} produced by the ANN described in Fig.~\ref{fig:ann_2D_latent_architecture}, rather than the $3K=15$-dimensional latent space, as the former is easier to visualise.
Our analysis uses the 38,740 sources of \cite{gavras2023gaia} for which a variability class label is available (cf.~Section~\ref{sec:datasets}), as well as the set of sources for which the SOS catalogues in \gaia DR3 published a class label. 

\begin{figure}
\begin{center}
\includegraphics[width=0.5\textwidth,keepaspectratio]{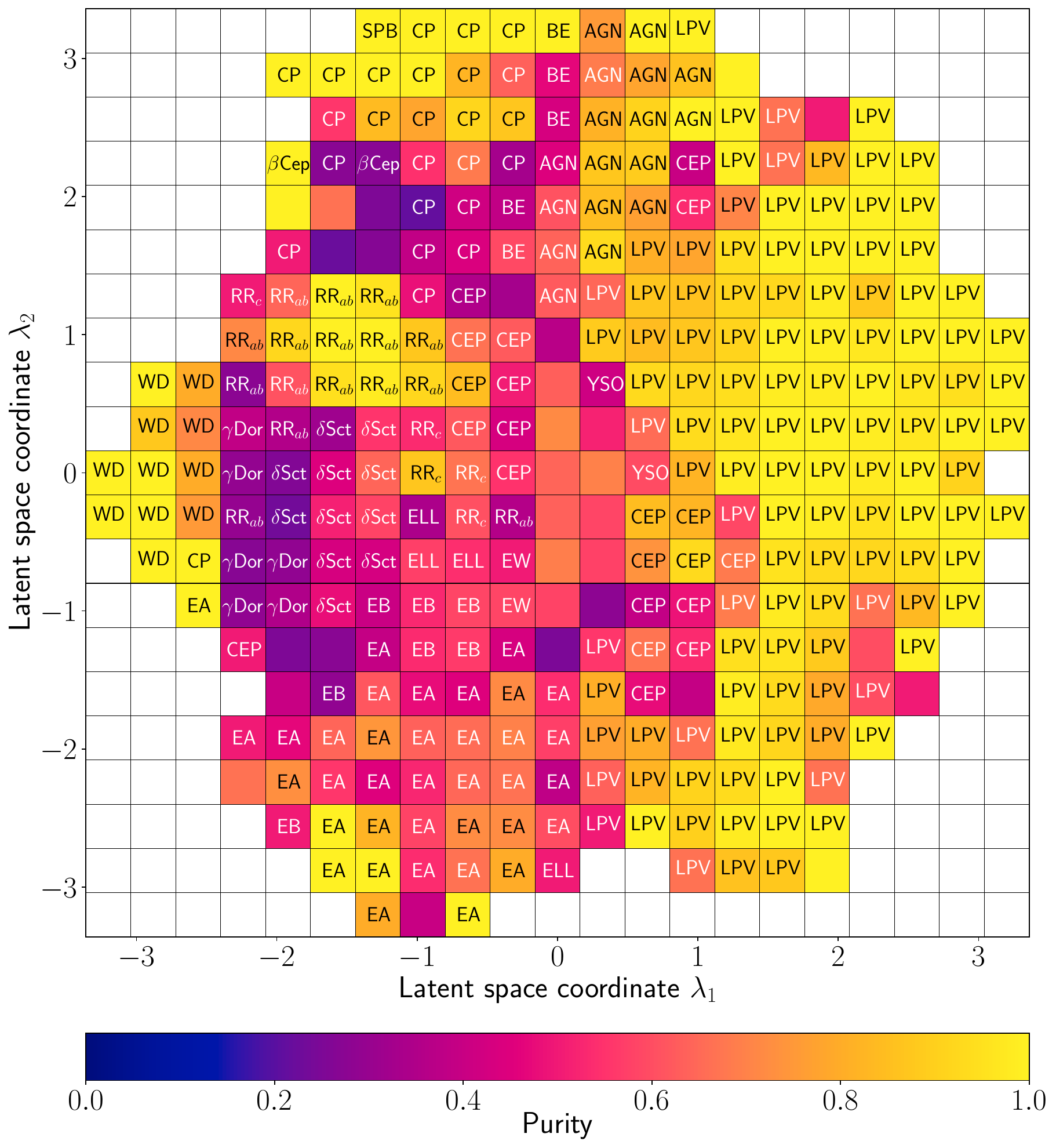}
\caption{\label{fig:purity_class_heatmap} The most frequently occurring variability class for each bin of the latent space shown in Fig.~\ref{fig:density_heatmap}, for the set of variable stars that have a variability class label available (cf.~Section~\ref{sec:datasets}). The bins are colour coded with the fraction of stars in the bin that have the shown label. Only the variability classes SPB, \gdor, \bcep, \dsct, chemically peculiar (CP) stars, active galactic nuclei (AGN), long-period variables (LPV), \rr (ab, c, and d), Cepheids, white dwarfs (WD), eclipsing binaries (EA, EB, EW), ellipsoidal (ELL) variables, Be (BE) variables, and Young Stellar Object variables (YSO) were taken into account to generate this figure.}
\end{center}
\end{figure}

For a first identification of the different overdensities in Fig.~\ref{fig:density_heatmap}, we divided the latent space into bins, and provide the most frequently occurring variability class for each bin in
Fig.~\ref{fig:purity_class_heatmap}.
The colour code indicates the fraction of stars in the bin that have the shown label. Despite the fact that the ANN does not have access to the published class label, many of the known variability classes are well separated in the latent space. Moreover, not only are the majority of bins with the same class label grouped together, variability classes that are expected to be similar or overlapping (e.g. eclipsing EA, EB, and EW stars) have bins close to each other, while sources that a human classifier would set apart, e.g. white dwarfs (WD) and long-period red giants (LPV) are positioned at opposite sides of the latent space. 

Some Gaia sources have time series that show calibration-induced spurious variability, introduced by an incomplete modelling of the on-sky source structure (e.g. multiple unresolved sources or extended objects like galaxies), resulting in a scan angle dependent signal \citep{holl2023gaia}. Figure~\ref{fig:spurious_heatmap} plots the known such sources and shows that the centre of the latent space is particularly affected, which is the reason that the centre bins of Fig.~\ref{fig:purity_class_heatmap} do not show a class label.

In the subsections below, we investigate some of the overdensities in the latent space by overplotting them with sources for which we have a variability class label as outlined in Section~\ref{sec:datasets}.

\subsection{\rr stars}
\begin{figure}
\begin{center}
\includegraphics[width=0.5\textwidth,keepaspectratio]{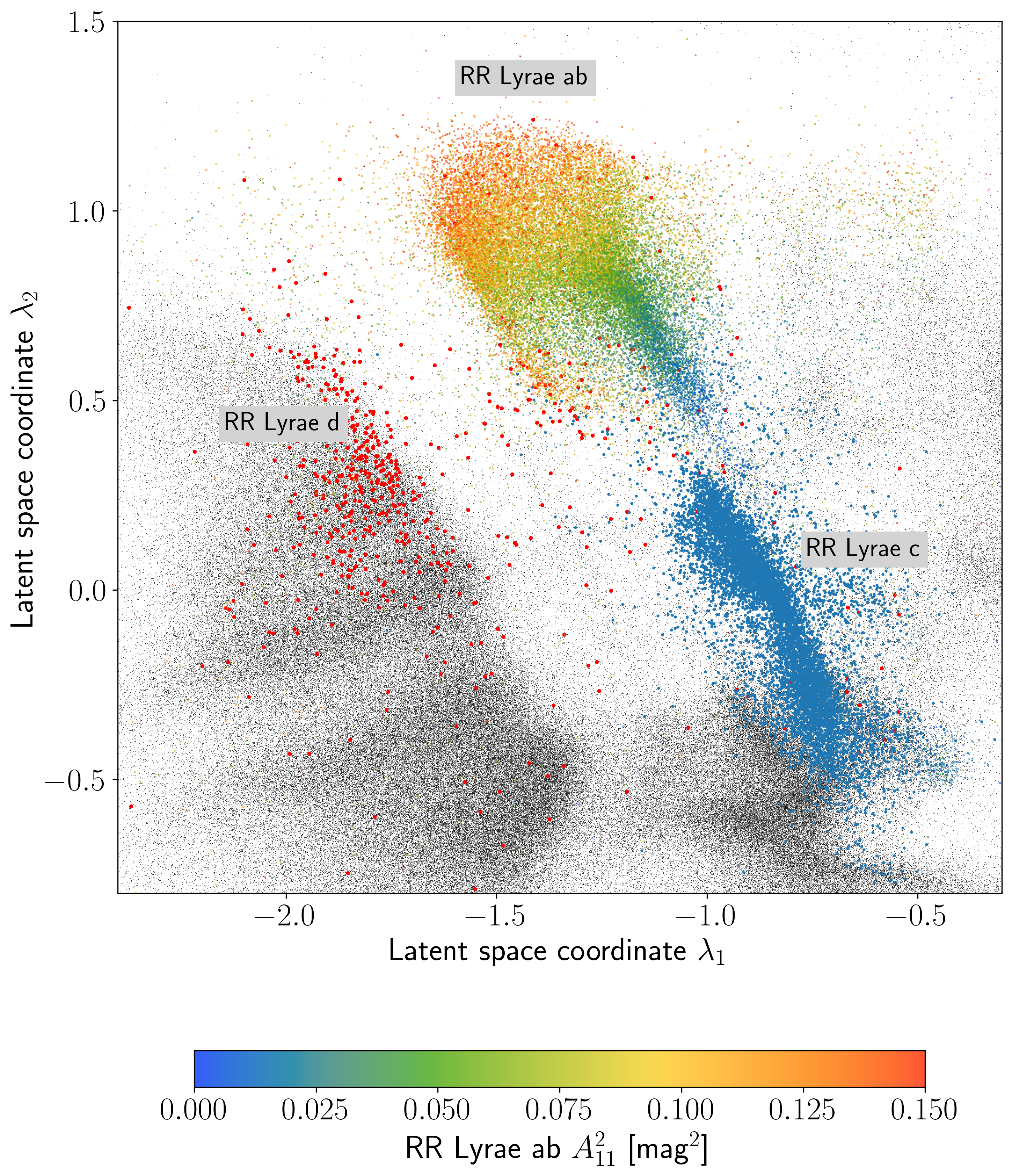}
\caption{\label{fig:rrlyr_heatmap}The region of the latent space where the \rr stars are located. The \rr variables for which a ``RRab'' classification label was available in the \gaia DR3 catalogue are overplotted in colour, where we used the pulsation amplitude $A_{11}$ for the colour scale. The \rrc and \rrd variables are plotted in blue resp.~red.}
\end{center}
\end{figure}

The overdensities in the latent space around coordinates $(\lambda_1, \lambda_2)=(-1.3, 0.9)$ relate to \rr variables. A close-up of this region is shown in Fig.~\ref{fig:rrlyr_heatmap}, where we overplotted the sources with an \rr variability class label from the literature.  
The ANN separates the three different subgroups of \rr (ab, c, and d). The \rrd variables occupy the same region as the \dsct and \gdor variables. For the sake of conciseness, we used two different quantities for the colour scales to overplot the \rrab and \rrc samples. The colour of the \rrab variables relates to the amplitude $A_{11}$ of the first harmonic of the main frequency as published in the \gaia DR3 SOS variability catalogue \citep[see][for more details]{clementini2023rrlyr}. A significant gradient of this amplitude can be seen, explaining the extent of the \rrab cluster in the latent space.  The \rrab cluster contains two substructures best seen in Fig.\ref{fig:dsct_gdor_wd_heatmap} around $(\lambda_1, \lambda_2)= (-1.4, 0.9)$. We discuss this in more depth in Appendix~\ref{app:astrophysical_interpretation}, in which we provide support for the hypothesis that the well-known Oosterhoff dichotomy is the underlying reason.
Finally, we mention that the \rrc variables extend from close to the \rrab variables down to the region in the latent space where the eclipsing binaries are located (see the subsection on eclipsing binaries below).

\subsection{\dsct stars, \gdor stars, and white dwarfs}

\begin{figure}
\begin{center}
\includegraphics[width=0.5\textwidth,keepaspectratio]{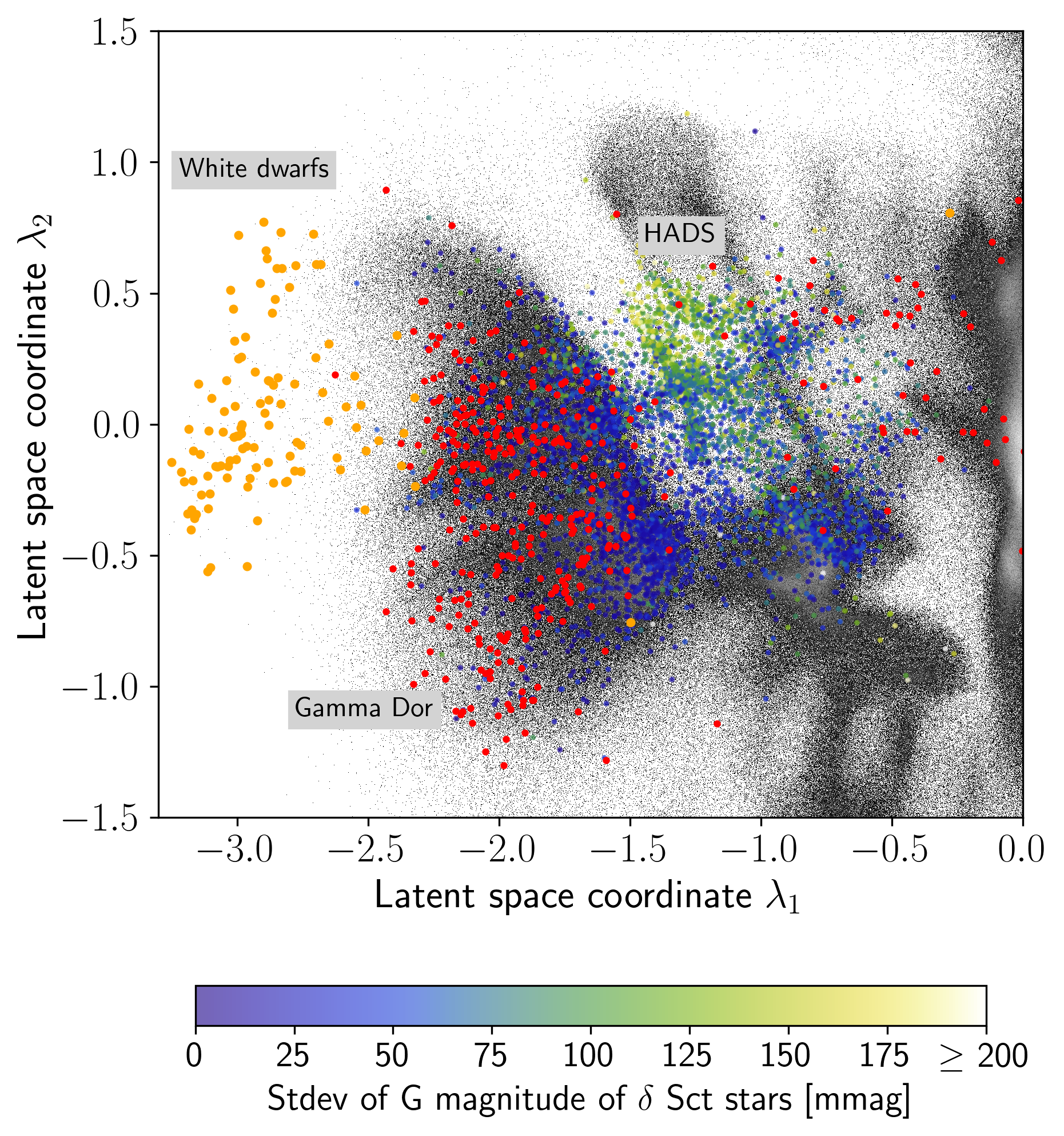}
\caption{\label{fig:dsct_gdor_wd_heatmap}The sources for which we have a \gdor or a white dwarf classification label from the literature are shown in red and orange, respectively. For the known \dsct variables, we used a colour scheme that shows the standard deviation in the \gaia \g passband as a proxy for the oscillation amplitude. We indicate the position of the high-amplitude \dsct stars (HADS), with amplitude above 100 mmag.}
\end{center}
\end{figure}

The location of the \dsct stars in the latent space is close to that of the \rr stars and the \gdor stars. Figure~\ref{fig:dsct_gdor_wd_heatmap} offers a zoom of the relevant region in the latent space, in which we highlighted those sources for which we have a \dsct variability label from the literature (cf.~Section~\ref{sec:datasets}). 
The colour scheme represents the standard deviation in the \gaia \g light curve which serves as a proxy for the oscillation amplitude. We indicate the location of the high-amplitude \dsct stars (HADS) in the plot,  close to the \rr stars. The low-amplitude \dsct stars (in dark blue), well separated from the HADS, are close to the \gdor stars (in red), and partly overlap with them. This is expected from an astrophysical point of view, given that they have similar folded light curves and are only slightly redder. The main way to distinguish them is the oscillation frequency (long period $g$-modes vs short-period $p$-modes), but in practice overlap is expected. Indeed, hybrid pulsators which show both $p$- and $g$-mode oscillations 
with overlapping frequencies due to fast rotation are common 
\citep{Aerts2021,Aerts2024}. In addition, the \gaia time sampling is such that a \gdor frequency can be aliased into the \dsct frequency domain and vice versa \citep{deridder2023gaia,hey2024confronting}.

The white dwarf oscillators occur as a separate group in Fig.~\ref{fig:dsct_gdor_wd_heatmap}. The \gaia time sampling is too coarse to detect the high-frequency oscillations ($P\approx 10$ min), so this clustering is based on their distinct XP spectra. This is supported by the difference in classification performance between the individual modalities (Section~\ref{sec:classification}).

\subsection{Hot variable stars}

\begin{figure}[b]
\begin{center}
\includegraphics[width=0.5\textwidth,keepaspectratio]{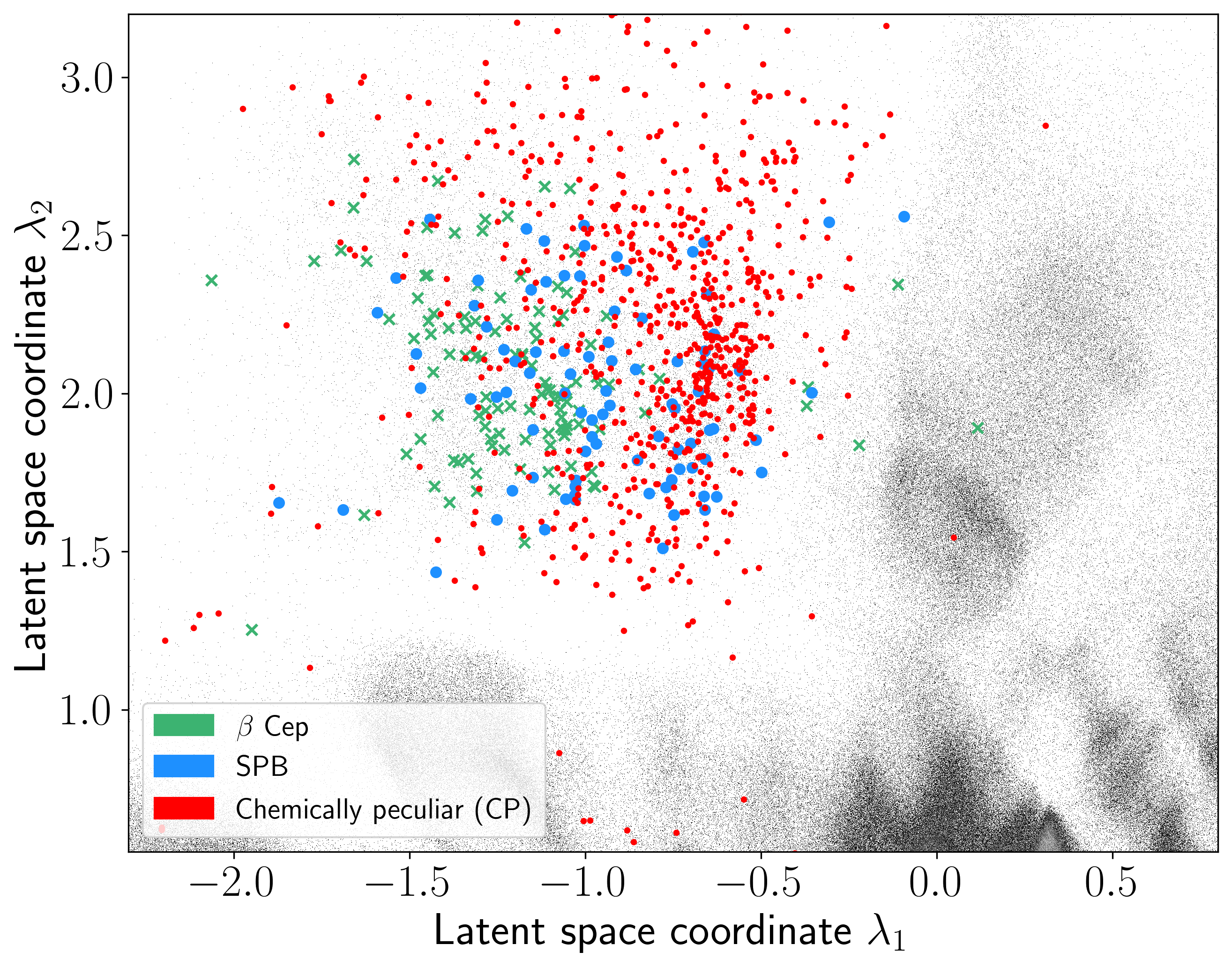}
\caption{\label{fig:ob_stars_heatmap}The region of the latent space where the variable B stars are projected.
The \bcep, SPB and CP variables for which we have a classification from the DPAC-CU7 training set mentioned in Section~\ref{sec:datasets} are plotted in colour.}
\end{center}
\end{figure}

The variable B stars occupy a separate region in the latent space around $(\lambda_1, \lambda_2)=(-1.0, 2.3)$.
In Fig.~\ref{fig:ob_stars_heatmap} we show a zoom of this region, and overplot the \bcep stars, 
SPB, and CP stars from the \gaia DR3 training set as outlined in Section~\ref{sec:datasets}.

The \bcep stars and SPB variables largely overlap in the latent space, although the former occur predominantly at lower $\lambda_1$ coordinates in this region, which is mainly correlated with the main frequency $f_1$ as illustrated in Fig.~\ref{fig:main_freq_heatmap}. This overlap is not an artefact of projecting the data in a 2D latent space. Section \ref{sec:classification} reveals that also in the 15D latent space it is hard to separate the SPB variables with the current dataset. This is in line with the results from \citet{Mombarg2024} and \citet{Fritzewski2025}, who found large populations of hot $g-$mode and hybrid $p-/g-$mode pulsators to occupy an uninterrupted area along the main sequence, stretching all the way from the \gdor stars up to the \bcep instability strip. 

The CP stars occupy the same region in the latent space but are more concentrated towards
higher $\lambda_1$ and $\lambda_2$ values. The reason that the ANN is able to discriminate them is that
they more frequently have a late B or early A spectral type, and are therefore redder than many pulsating B stars, 
which the ANN picks up through the XP spectra. 
In addition, CP stars can show variability at longer time scales in the $\Delta m|\Delta t$ histogram, 
related to their rotation period, discriminating them from the short-period \bcep stars.  
It is well known that CP stars and SPB pulsators occupy the same region in the Hertzsprung-Russell diagram \citep{Briquet2007}.

\subsection{Eclipsing binary stars}

\begin{figure}[t]
\begin{center}
\includegraphics[width=0.5\textwidth,keepaspectratio]{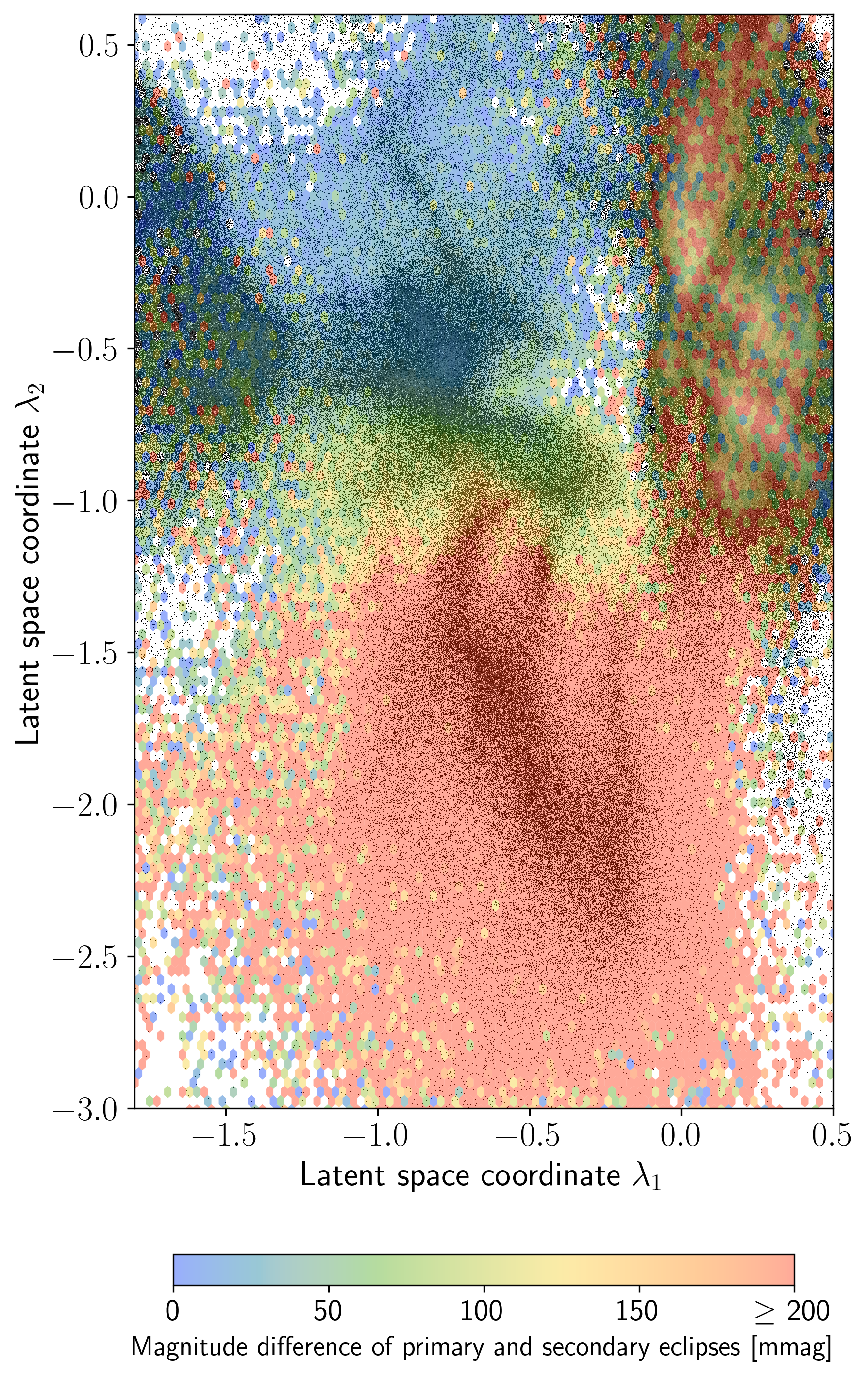}
\caption{\label{fig:eclipsing_depth_diff_heatmap}The part of the latent space where the ANN projects
eclipsing and ellipsoidal binaries. The grey background shows the same density as in Fig.~\ref{fig:density_heatmap}.
Sources which were published in the \gaia DR3 Variability catalogue \texttt{gaiadr3.vari\_eclipsing\_binary} are
overplotted in a 2D histogram of small hexagons where the colour of each hexagon corresponds to the average magnitude difference between primary and secondary eclipses of the sources inside that bin.}
\end{center}
\end{figure}

The overdensities in the 2D latent space around $(\lambda_1, \lambda_2)=(-0.5, -1.4)$ are dominantly occupied
by binary stars as shown in Fig.~\ref{fig:purity_class_heatmap}. The substructure in that region is largely
determined by the morphology of the (folded) light curve. To demonstrate this, we follow
\citet{Mowlavi2023gaia} who used a simple but effective model to characterise the variability of \gaia DR3 binary candidates. 
In their work, the eclipses were modelled using two Gaussian functions, while an 
ellipsoidal variation was modelled with a cosine function. In Figs.~\ref{fig:eclipsing_depth_diff_heatmap}, \ref{fig:eclipsing_duration_heatmap} and \ref{fig:ellipsoidal_amplitude_heatmap}, we plot for the binary candidates in their catalogue, respectively, the 
magnitude difference between two consecutive eclipses, the duration of the primary eclipse as a fraction of the 
orbital period, and the amplitude of the cosine component (if any), which serves as a proxy for the amplitude of 
the ellipsoidal variation.  

These figures show that the overdensity in the bottom relates to eclipsing binaries of type EA with clearly 
detached eclipses and no ellipsoidal variation. The magnitude differences between the eclipse depths range from 0 to more
than 200 mmag, but there is no clear gradient present. 
Visual inspection of the folded light curves of 
the sources revealed that this region also includes sources for which outliers were incorrectly interpreted 
as eclipses and for which the folded light curve does not show the expected morphology of a type EA 
eclipsing binary. This had an impact on the performance of the ANN in substructuring this type of binaries. Points flagged by DPAC-CU7 as unreliable and points with large uncertainties were removed during preprocessing. No other outlier-removing procedure was applied to avoid that eclipses would be erroneously removed. 

In the centre of the binary region, the ANN places the EB binaries with detached eclipses,
but for which Fig.~\ref{fig:ellipsoidal_amplitude_heatmap} shows a detectable variation outside the eclipses.
The ANN shows a gradient in the amplitude of the ellipsoidal variation, increasing from left to right. The top part of the region is occupied by the close over-contact EW binaries for which Fig.~\ref{fig:eclipsing_duration_heatmap} shows that the eclipses take a 
large fraction of the orbital period and with a continuous variation outside of the eclipses. The colour map in Fig.~\ref{fig:eclipsing_depth_diff_heatmap}
shows that most systems have either equal or similar minima.

\subsection{Long-period variables}

\begin{figure}[t]
\begin{center}
\includegraphics[width=0.5\textwidth,keepaspectratio]{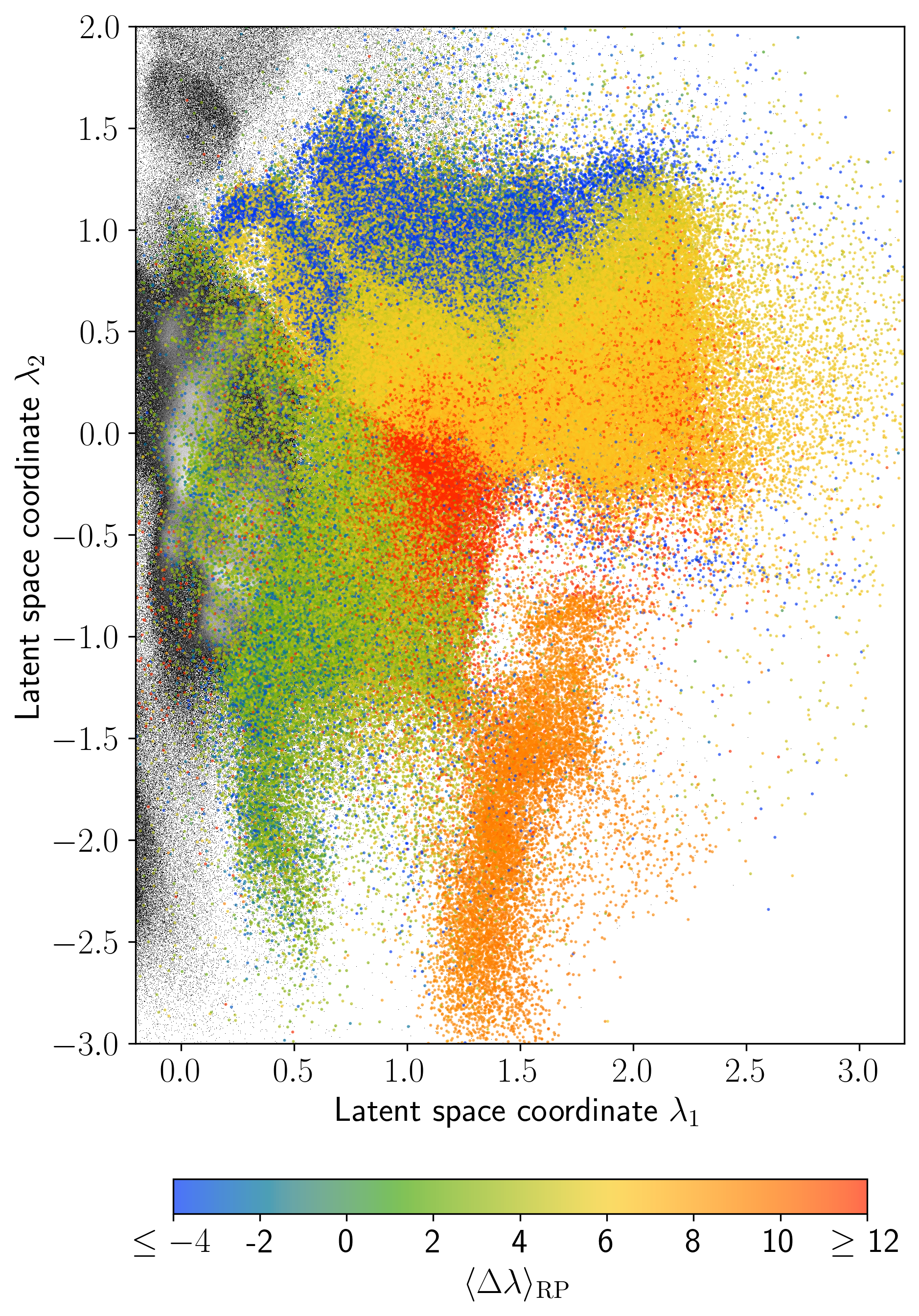}
\caption{\label{fig:lpv_xp_heatmap}The part of the latent space where the ANN projects
the long period variables. The grey background shows the same density as in Fig.~\ref{fig:density_heatmap}.
In colour, we overplotted the sources which were published in the \gaia DR3 Variability catalogue as a candidate 
O-rich or C-type LPVs. The quantity $\langle\Delta\lambda\rangle_{\rm RP}$ was taken from the same catalogue
(column name \texttt{median\_delta\_wl\_rp}) and probes particular molecular bands in the RP spectrum (cf.~text).
}
\end{center}
\end{figure}

Figure~\ref{fig:purity_class_heatmap} shows that most of the area defined by $\lambda_1 > 0.3, \lambda_2 < 2 $ in the latent space is occupied by LPVs, such as Mira and red semi-regular variables. Figure~\ref{fig:density_heatmap} shows substructure in the density in that region.
In Fig.~\ref{fig:lpv_xp_heatmap} we show a zoom of this region, and overplot the LPV candidates published in \gaia DR3 (table \texttt{gaiadr3.vari\_long\_period\_variable}, selecting $G \le 19$) with a classification of \texttt{LPV/C} or \texttt{LPV/O rich}. The colour coding was done using the quantity $\langle\Delta\lambda\rangle_{\rm RP}$ 
which is published as \texttt{median\_delta\_wl\_rp} in the same table.
This quantity is the difference in pseudo-wavelength between the two highest peaks in the \gaia RP spectrum,
and probes the existence of particular molecular bands. 
It allows to detect carbon stars (C stars) which show strong imprints of carbon-based molecules such as CN or C$_2$ in their spectrum, in contrast with oxygen-rich LPVs which show molecular bands of TiO and VO.
\citet{Lebzelter2023gaia_lpv} identify the interval 
$\langle\Delta\lambda\rangle_{\rm RP} \in [7, 11]$ and $G\le 19$ as reliable criteria to select C stars. This corresponds to the
group of orange points in the latent space. The ANN was never given the value of 
$\langle\Delta\lambda\rangle_{\rm RP}$, so the fact that the C stars form an isolated cluster in the latent space 
implies that the ANN was able to recognise the relevant feature in the \gaia RP spectrum. 

Sources with $\langle\Delta\lambda\rangle_{\rm RP}\le 7$ are identified by \citet{Lebzelter2023gaia_lpv} as  
oxygen-rich LPVs. The ANN separates well the sources with $\langle\Delta\lambda\rangle_{\rm RP}\approx 6$ (yellow in Fig.~\ref{fig:lpv_xp_heatmap}) from the sources with $\langle\Delta\lambda\rangle_{\rm RP}\in [2,5]$ (greenish) in the latent space. 
Figure~5 in \citet{Lebzelter2023gaia_lpv} shows that values of $\langle\Delta\lambda\rangle_{\rm RP}$
between 4 and 5 can be caused by ZrO bands in addition to TiO bands, occurring for S stars.
Sources at $\langle\Delta\lambda\rangle_{\rm RP}\ge 12$ were identified by these authors as originating from a molecular band at shorter wavelengths identified as the second largest 
peak in the RP spectrum. Although they occur in a large part of the LPV region in the latent space, 
the ANN does produce an overdensity with a strong concentration of these stars. 

Figure~\ref{fig:lpv_f1_heatmap} shows the same LPV region in the latent space, but this time colour coded with the main period $P_1$ found in the Lomb-Scargle periodogram. The ANN used this information to insert further structure in the overdensities. In particular, the Mira variables with long periods (red points) appear at larger values of $\lambda_1$. The blue dots are sources for which a main frequency $f_1 \approx 12 \ d^{-1}$ originating from \gaia's scanning law was found, so it is doubtful whether the overdensity at the left of the region has an astrophysical origin.

\subsection{Active galactic nuclei}

\begin{figure}[t]
\begin{center}
\includegraphics[width=0.5\textwidth,keepaspectratio]{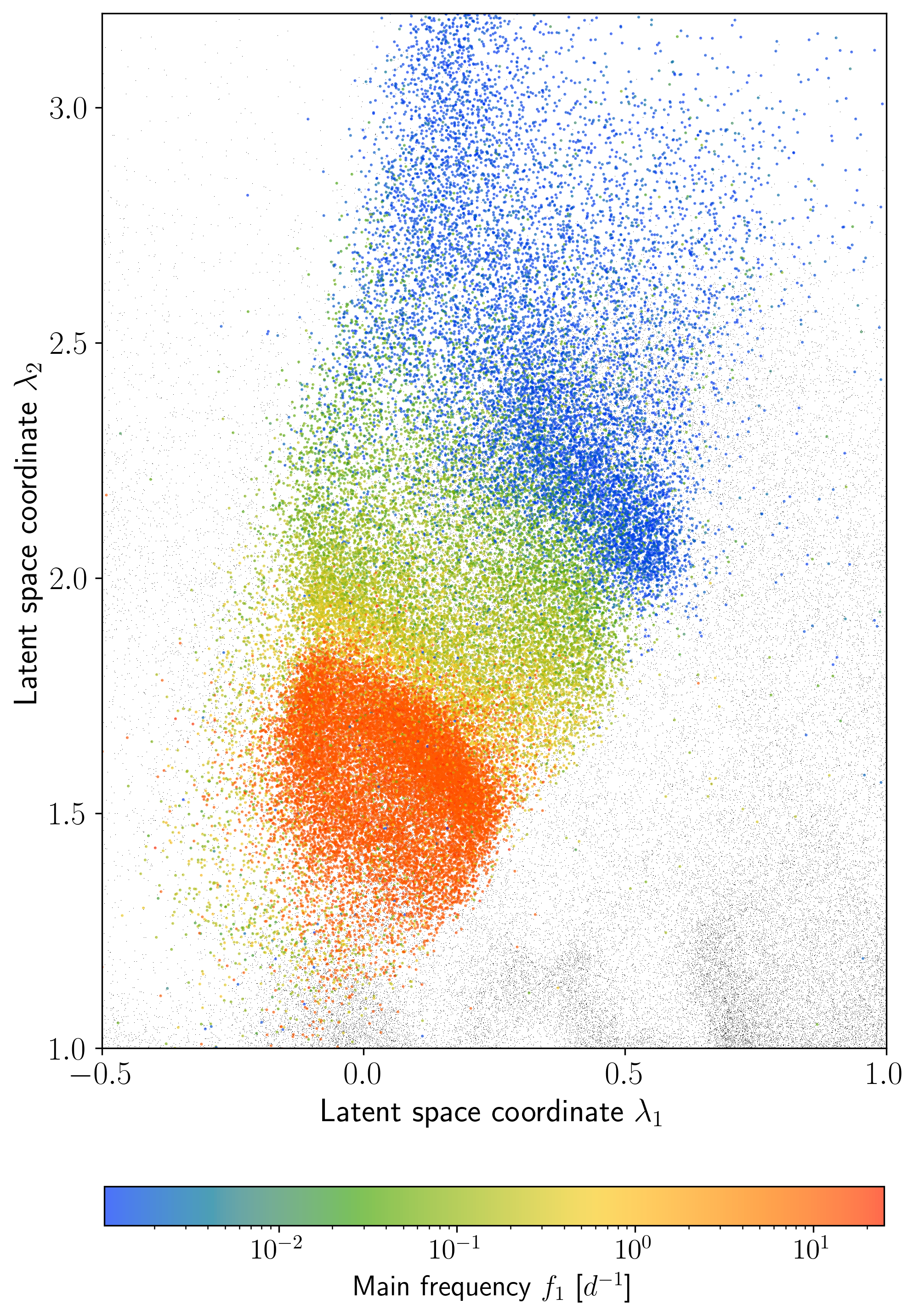}
\caption{\label{fig:agn_f1_heatmap}The region of the latent space occupied by active galactic nuclei. For the AGN candidates published in the \gaia DR3 release, we colour coded the frequency with the highest amplitude found in
the Lomb-Scargle frequencygram computed from the light curves in the \gaia \g passband.
}
\end{center}
\end{figure}

As already shown by the substructure in Fig.~\ref{fig:density_heatmap}, the variable active galactic nuclei (AGNs) form a fairly isolated cluster around latent coordinates $(\lambda_1, \lambda_2)=(0.25, 2.0)$. However, as discussed at the beginning of this Section, the extendedness of the AGN is a known origin
of spurious scan-angle dependent signals. Much of the same substructures can therefore also be found in Fig.~\ref{fig:spurious_heatmap}, containing known sources with a spurious signal, in particular the overdensity around $(\lambda_1, \lambda_2)=(0.1, 1.6)$. Although AGNs are usually not periodic, we computed the main frequency $f_1$ for each AGN candidate published in \gaia DR3 (table \texttt{gaiadr3.vari\_agn}, see \citet{carnero2023agn} for details), and plotted the result in Fig.~\ref{fig:agn_f1_heatmap}. This figure shows that the main overdensity originates from sources having $f_1 = 12\ d^{-1}$, originating from \gaia's scanning law. 

The group of points coloured green in Fig.~\ref{fig:agn_f1_heatmap} contain sources with 
a main period listed in \citet{holl2023gaia} relating to \gaia's scanning law, for example 
$P_1 = f_1^{-1} = 13.4\ {\rm d}, 16.3\ {\rm d}, 25.0\ {\rm d}$, and $31.5\ {\rm d}$. 
The blue sources in Fig.~\ref{fig:agn_f1_heatmap}, on the other hand, show a main frequency that is consistent
with the low-frequency red noise that is typical for AGNs. 
The ANN projected the vast majority of AGN candidates published in \gaia DR3 
in the same region in the latent space, independently confirming their candidate status, even the ones where the
highest peak in the Fourier spectrum is instrumental. However, care needs to be
taken when analysing overdensities. If the observed data still contains instrumental features, the ANN will likely 
pick it up, resulting in spurious overdensities in the latent space.

\section{Conclusions}
\label{sec:conclusions}

In this work, we presented a self-supervised ML methodology based on VAEs to derive latent representations of variable sources. Our evaluation of the latent variables using \gaia DR3 data demonstrated that the VAEs effectively encode the characteristic variability patterns, enabling downstream applications such as classification and clustering.
These models extract information from two modalities, namely epoch photometry and low-resolution XP mean spectra. Classification experiments show that there is a significant beneficial synergy when these \gaia DR3 products are used together. Notable examples include the improvement in $F_1$-score for classes such as \bcep and \gdor stars, which were highly-contaminated in single-modality setups. The latent variables also prove effective in detecting atypical sources. This was exemplified with \rr stars, where the latent representations successfully highlighted atypical sources, which were found to correspond to misclassifications or issues arising from a wrong estimation of the period. Our preliminary results will be expanded in the future as we expect this capability to be particularly useful for \gaia DR4. Clustering results also show that partitioning the latent space allows for the recovery of the known major variability classes, but handling dataset imbalance remains a challenge for a fully unsupervised variability classification. Overall, the classification performance achieved with the latent variables is competitive with that of DR3, though differences arise for specific classes. This suggests that incorporating the self-supervised latent variables as additional features could enhance supervised classification efforts for DR4. Conversely, it also indicates that the latent space could gain greater discriminative power by integrating the features and data modalities used in \citet{rimoldini2023gaia}.

Our analysis of the 2D projection of the latent space shown in Fig.~\ref{fig:density_heatmap} showed that several overdensities and stratifications have a strong correlation with astrophysical properties. A notable example is the LPVs in which the ANN was able to identify the O-rich and C-type LPVs from the light curve and the RP continuous spectrum it was given. These results are encouraging for \gaia DR4 where a much larger and unbiased sample will be available, which has more potential for novelty discovery.

Although our methodology already contains several novel aspects, it can still be improved, for example, by including additional modalities, such as astrometric information, \bp and \rp time series and RVS spectra. In particular, we will focus our attention on new modalities to better distinguish challenging classes such as ELL, RS, SPB and YSO.
In this work, each modality was compressed independently, with their latent variables subsequently concatenated. The three modalities can also be trained jointly to learn a single shared latent space, which might enable the model to better exploit correlations between them. Another compelling direction for future research is to replace the standard normal prior in the VAE with a Gaussian Mixture Model (GMM), as proposed by \citet{jiang2017variational}, which would allow for simultaneous dimensionality reduction and clustering. Additionally, semi-supervised approaches could be explored, such as training a classifier alongside the AEs or fine-tuning the self-supervised representation using a small set of labelled examples, as demonstrated in \citet{rizhko2024self}.

Since our analysis was based exclusively on DR3 sources with public epoch photometry and XP mean spectra, it was inherently limited to sources already classified as variables. In contrast, DR4 will publish data for all sources, adding large amounts of non-variable sources and potentially increasing the imbalance between variability classes. Furthermore, DR4 will include new data products, such as low-resolution XP, RVS spectrum, and radial velocity time series. More sophisticated training strategies and architectures will be required to address these challenges and exploit the full range of data offered by \gaia DR4.

\begin{acknowledgements} 

PH, JDR and CA acknowledge support from the BELgian federal Science Policy Office (BELSPO) through various PROgramme de Développement d’Expériences scientifiques (PRODEX) grants. JDR and CA acknowledge funding from the Research Foundation Flanders (FWO) under grant agreement G089422N.
CA, PH, and MV acknowledge financial support from the European Research Council (ERC) under the Horizon Europe programme (Synergy Grant agreement N◦101071505: 4D-STAR).
While partially funded by the European Union, views and opinions expressed are however those of the author(s) only and do not necessarily reflect those of the European Union or the European Research Council. Neither the European Union nor the granting authority can be held responsible for them.
PH also acknowledges support from ANID – Millennium Science Initiative Program – ICN12\_009 awarded to the Millennium Institute of Astrophysics MAS. 
CA, AK, DJF and HW acknowledge financial support from the Flemish Government under the long-term structural Methusalem funding program by means of the project SOUL: Stellar evOlution in fUll gLory, grant METH/24/012 at KU Leuven.
VV gratefully acknowledges support from the Research Foundation Flanders (FWO) under grant agreement N°1156923N (PhD Fellowship).
This work has made use of data from the European Space Agency (ESA) mission \gaia (\url{https://www.cosmos.esa.int/gaia}), processed by the \gaia Data Processing and Analysis Consortium (DPAC,
\url{https://www.cosmos.esa.int/web/gaia/dpac/consortium}). Funding for the DPAC has been provided by national institutions, in particular the institutions participating in the \gaia Multilateral Agreement. This work has made use of the Python package GaiaXPy, developed and maintained by members of the \gaia Data Processing and Analysis Consortium (DPAC) and in particular, Coordination Unit 5 (CU5), and the Data Processing Centre located at the Institute of Astronomy, Cambridge, UK (DPCI).
This research was supported by the Patag\'{o}n supercomputer of Universidad Austral de Chile (FONDEQUIP EQM180042).

\end{acknowledgements}

\bibliographystyle{aa_url} 
\bibliography{references}

\begin{thebibliography}{87}
\expandafter\ifx\csname natexlab\endcsname\relax\def\natexlab#1{#1}\fi

\bibitem[{{Abdurro'uf} {et~al.}(2022){Abdurro'uf}, {Accetta}, {Aerts}, {Silva
  Aguirre}, {Ahumada}, {Ajgaonkar}, {Filiz Ak}, {Alam}, {Allende Prieto},
  {Almeida}, {Anders}, {Anderson}, {Andrews}, {Anguiano}, {Aquino-Ort{\'\i}z},
  {Arag{\'o}n-Salamanca}, {Argudo-Fern{\'a}ndez}, {Ata}, {Aubert},
  {Avila-Reese}, {Badenes}, {Barb{\'a}}, {Barger}, {Barrera-Ballesteros},
  {Beaton}, {Beers}, {Belfiore}, {Bender}, {Bernardi}, {Bershady}, {Beutler},
  {Bidin}, {Bird}, {Bizyaev}, {Blanc}, {Blanton}, {Boardman}, {Bolton},
  {Boquien}, {Borissova}, {Bovy}, {Brandt}, {Brown}, {Brownstein}, {Brusa},
  {Buchner}, {Bundy}, {Burchett}, {Bureau}, {Burgasser}, {Cabang}, {Campbell},
  {Cappellari}, {Carlberg}, {Wanderley}, {Carrera}, {Cash}, {Chen}, {Chen},
  {Cherinka}, {Chiappini}, {Choi}, {Chojnowski}, {Chung}, {Clerc}, {Cohen},
  {Comerford}, {Comparat}, {da Costa}, {Covey}, {Crane}, {Cruz-Gonzalez},
  {Culhane}, {Cunha}, {Dai}, {Damke}, {Darling}, {Davidson}, {Davies},
  {Dawson}, {De Lee}, {Diamond-Stanic}, {Cano-D{\'\i}az}, {S{\'a}nchez},
  {Donor}, {Duckworth}, {Dwelly}, {Eisenstein}, {Elsworth}, {Emsellem},
  {Eracleous}, {Escoffier}, {Fan}, {Farr}, {Feng}, {Fern{\'a}ndez-Trincado},
  {Feuillet}, {Filipp}, {Fillingham}, {Frinchaboy}, {Fromenteau}, {Galbany},
  {Garc{\'\i}a}, {Garc{\'\i}a-Hern{\'a}ndez}, {Ge}, {Geisler}, {Gelfand},
  {G{\'e}ron}, {Gibson}, {Goddy}, {Godoy-Rivera}, {Grabowski}, {Green},
  {Greener}, {Grier}, {Griffith}, {Guo}, {Guy}, {Hadjara}, {Harding},
  {Hasselquist}, {Hayes}, {Hearty}, {Hern{\'a}ndez}, {Hill}, {Hogg},
  {Holtzman}, {Horta}, {Hsieh}, {Hsu}, {Hsu}, {Huber}, {Huertas-Company},
  {Hutchinson}, {Hwang}, {Ibarra-Medel}, {Chitham}, {Ilha}, {Imig}, {Jaekle},
  {Jayasinghe}, {Ji}, {Johnson}, {Jones}, {J{\"o}nsson}, {Katkov}, {Khalatyan},
  {Kinemuchi}, {Kisku}, {Knapen}, {Kneib}, {Kollmeier}, {Kong}, {Kounkel},
  {Kreckel}, {Krishnarao}, {Lacerna}, {Lane}, {Langgin}, {Lavender}, {Law},
  {Lazarz}, {Leung}, {Leung}, {Lewis}, {Li}, {Li}, {Lian}, {Liang}, {Lin},
  {Lin}, {Lin}, {Lintott}, {Long}, {Longa-Pe{\~n}a}, {L{\'o}pez-Cob{\'a}},
  {Lu}, {Lundgren}, {Luo}, {Mackereth}, {de la Macorra}, {Mahadevan},
  {Majewski}, {Manchado}, {Mandeville}, {Maraston}, {Margalef-Bentabol},
  {Masseron}, {Masters}, {Mathur}, {McDermid}, {Mckay}, {Merloni},
  {Merrifield}, {Meszaros}, {Miglio}, {Di Mille}, {Minniti}, {Minsley}, \&
  {Monachesi}}]{abdurrouf2022sdssdr17}
{Abdurro'uf}, {Accetta}, K., {Aerts}, C., {et~al.} 2022,
  \href{http://dx.doi.org/10.3847/1538-4365/ac4414}{\color{magenta}\apjs},
  \href{https://ui.adsabs.harvard.edu/abs/2022ApJS..259...35A}{259, 35}

\bibitem[{{Aerts}(2021)}]{Aerts2021}
{Aerts}, C. 2021,
  \href{http://dx.doi.org/10.1103/RevModPhys.93.015001}{\color{magenta}Reviews
  of Modern Physics},
  \href{https://ui.adsabs.harvard.edu/abs/2021RvMP...93a5001A}{93, 015001}

\bibitem[{{Aerts} \& {Tkachenko}(2024)}]{Aerts2024}
{Aerts}, C. \& {Tkachenko}, A. 2024,
  \href{http://dx.doi.org/10.1051/0004-6361/202348575}{\color{magenta}\aap},
  \href{https://ui.adsabs.harvard.edu/abs/2024A&A...692R...1A}{692, R1}

\bibitem[{{Bahdanau} {et~al.}(2014){Bahdanau}, {Cho}, \&
  {Bengio}}]{Bahdanau2014attention}
{Bahdanau}, D., {Cho}, K., \& {Bengio}, Y. 2014, in International Conference on
  Learning Representations 2015,
  \href{https://ui.adsabs.harvard.edu/abs/2014arXiv1409.0473B}{arXiv:1409.0473}

\bibitem[{{Bellm} {et~al.}(2019){Bellm}, {Kulkarni}, {Graham}, {Dekany},
  {Smith}, {Riddle}, {Masci}, {Helou}, {Prince}, {Adams}, {Barbarino},
  {Barlow}, {Bauer}, {Beck}, {Belicki}, {Biswas}, {Blagorodnova}, {Bodewits},
  {Bolin}, {Brinnel}, {Brooke}, {Bue}, {Bulla}, {Burruss}, {Cenko}, {Chang},
  {Connolly}, {Coughlin}, {Cromer}, {Cunningham}, {De}, {Delacroix}, {Desai},
  {Duev}, {Eadie}, {Farnham}, {Feeney}, {Feindt}, {Flynn}, {Franckowiak},
  {Frederick}, {Fremling}, {Gal-Yam}, {Gezari}, {Giomi}, {Goldstein},
  {Golkhou}, {Goobar}, {Groom}, {Hacopians}, {Hale}, {Henning}, {Ho}, {Hover},
  {Howell}, {Hung}, {Huppenkothen}, {Imel}, {Ip}, {Ivezi{\'c}}, {Jackson},
  {Jones}, {Juric}, {Kasliwal}, {Kaspi}, {Kaye}, {Kelley}, {Kowalski},
  {Kramer}, {Kupfer}, {Landry}, {Laher}, {Lee}, {Lin}, {Lin}, {Lunnan},
  {Giomi}, {Mahabal}, {Mao}, {Miller}, {Monkewitz}, {Murphy}, {Ngeow},
  {Nordin}, {Nugent}, {Ofek}, {Patterson}, {Penprase}, {Porter}, {Rauch},
  {Rebbapragada}, {Reiley}, {Rigault}, {Rodriguez}, {van Roestel}, {Rusholme},
  {van Santen}, {Schulze}, {Shupe}, {Singer}, {Soumagnac}, {Stein}, {Surace},
  {Sollerman}, {Szkody}, {Taddia}, {Terek}, {Van Sistine}, {van Velzen},
  {Vestrand}, {Walters}, {Ward}, {Ye}, {Yu}, {Yan}, \&
  {Zolkower}}]{bellm2019ztf}
{Bellm}, E.~C., {Kulkarni}, S.~R., {Graham}, M.~J., {et~al.} 2019,
  \href{http://dx.doi.org/10.1088/1538-3873/aaecbe}{\color{magenta}\pasp},
  \href{https://ui.adsabs.harvard.edu/abs/2019PASP..131a8002B}{131, 018002}

\bibitem[{Bengio {et~al.}(2013)Bengio, Courville, \&
  Vincent}]{bengio2013representation}
Bengio, Y., Courville, A., \& Vincent, P. 2013,
  \href{http://dx.doi.org/10.48550/arXiv.1206.5538}{\color{magenta}IEEE
  transactions on pattern analysis and machine intelligence},
  \href{https://ui.adsabs.harvard.edu/abs/2012arXiv1206.5538B}{35, 1798}

\bibitem[{{Blagorodnova} {et~al.}(2018){Blagorodnova}, {Neill}, {Walters},
  {Kulkarni}, {Fremling}, {Ben-Ami}, {Dekany}, {Fucik}, {Konidaris}, {Nash},
  {Ngeow}, {Ofek}, {O' Sullivan}, {Quimby}, {Ritter}, \&
  {Vyhmeister}}]{blagorodnova2018sedm}
{Blagorodnova}, N., {Neill}, J.~D., {Walters}, R., {et~al.} 2018,
  \href{http://dx.doi.org/10.1088/1538-3873/aaa53f}{\color{magenta}\pasp},
  \href{https://ui.adsabs.harvard.edu/abs/2018PASP..130c5003B}{130, 035003}

\bibitem[{{Bla{\v{z}}ko}(1907)}]{Blazko1907}
{Bla{\v{z}}ko}, S. 1907,
  \href{http://dx.doi.org/10.1002/asna.19071752002}{\color{magenta}Astronomische
  Nachrichten},
  \href{https://ui.adsabs.harvard.edu/abs/1907AN....175..325B}{175, 325}

\bibitem[{{Breiman}(2001)}]{breiman2001random}
{Breiman}, L. 2001,
  \href{http://dx.doi.org/10.1023/A:1010933404324}{\color{magenta}Machine
  Learning}, \href{https://ui.adsabs.harvard.edu/abs/2001MachL..45....5B}{45,
  5}

\bibitem[{Breunig {et~al.}(2000)Breunig, Kriegel, Ng, \&
  Sander}]{breunig2000lof}
Breunig, M.~M., Kriegel, H.-P., Ng, R.~T., \& Sander, J. 2000,
  \href{http://dx.doi.org/10.1145/335191.335388}{\color{magenta}SIGMOD Rec.},
  29, 93–104

\bibitem[{{Briquet} {et~al.}(2007){Briquet}, {Hubrig}, {De Cat}, {Aerts},
  {North}, \& {Sch{\"o}ller}}]{Briquet2007}
{Briquet}, M., {Hubrig}, S., {De Cat}, P., {et~al.} 2007,
  \href{http://dx.doi.org/10.1051/0004-6361:20066940}{\color{magenta}\aap},
  \href{https://ui.adsabs.harvard.edu/abs/2007A&A...466..269B}{466, 269}

\bibitem[{{Buck} \& {Schwarz}(2024)}]{buck2024deep}
{Buck}, T. \& {Schwarz}, C. 2024, in NIPS 2024 Workshop on Machine Learning and
  the Physical Sciences,
  \href{https://ui.adsabs.harvard.edu/abs/2024arXiv241016081B}{arXiv:2410.16081}

\bibitem[{{Burgess} {et~al.}(2018){Burgess}, {Higgins}, {Pal}, {Matthey},
  {Watters}, {Desjardins}, \& {Lerchner}}]{burgess2018betaVAE}
{Burgess}, C.~P., {Higgins}, I., {Pal}, A., {et~al.} 2018, in NIPS 2017
  Workshop on Learning Disentangled Representations,
  \href{https://ui.adsabs.harvard.edu/abs/2018arXiv180403599B}{arXiv:1804.03599}

\bibitem[{{Carnerero} {et~al.}(2023){Carnerero}, {Raiteri}, {Rimoldini},
  {Busonero}, {Licata}, {Mowlavi}, {Lecoeur-Ta{\"\i}bi}, {Audard}, {Holl},
  {Gavras}, {Nienartowicz}, {Jevardat de Fombelle}, {Carballo}, {Clementini},
  {Delchambre}, {Klioner}, {Lattanzi}, \& {Eyer}}]{carnero2023agn}
{Carnerero}, M.~I., {Raiteri}, C.~M., {Rimoldini}, L., {et~al.} 2023,
  \href{http://dx.doi.org/10.1051/0004-6361/202244035}{\color{magenta}\aap},
  \href{https://ui.adsabs.harvard.edu/abs/2023A&A...674A..24C}{674, A24}

\bibitem[{{Carrasco} {et~al.}(2021){Carrasco}, {Weiler}, {Jordi}, {Fabricius},
  {De Angeli}, {Evans}, {van Leeuwen}, {Riello}, \&
  {Montegriffo}}]{carrasco2021xpcalibration}
{Carrasco}, J.~M., {Weiler}, M., {Jordi}, C., {et~al.} 2021,
  \href{http://dx.doi.org/10.1051/0004-6361/202141249}{\color{magenta}\aap},
  \href{https://ui.adsabs.harvard.edu/abs/2021A&A...652A..86C}{652, A86}

\bibitem[{{Catelan}(2009)}]{Catelan09}
{Catelan}, M. 2009,
  \href{http://dx.doi.org/10.1007/s10509-009-9987-8}{\color{magenta}\apss},
  \href{https://ui.adsabs.harvard.edu/abs/2009Ap&SS.320..261C}{320, 261}

\bibitem[{{Catelan} \& {Smith}(2015)}]{Catelan15}
{Catelan}, M. \& {Smith}, H.~A. 2015, {Pulsating Stars} (John Wiley \& Sons)

\bibitem[{Chen \& Guestrin(2016)}]{chen2016xgboost}
Chen, T. \& Guestrin, C. 2016, in Proceedings of the 22nd {ACM} {SIGKDD}
  international conference on knowledge discovery and data mining, KDD '16,
  \href{https://ui.adsabs.harvard.edu/abs/2016arXiv160302754C}{785--794}

\bibitem[{{Chung} {et~al.}(2014){Chung}, {Gulcehre}, {Cho}, \&
  {Bengio}}]{chung2014gru}
{Chung}, J., {Gulcehre}, C., {Cho}, K., \& {Bengio}, Y. 2014, in NIPS 2014 Deep
  Learning and Representation Learning Workshop,
  \href{https://ui.adsabs.harvard.edu/abs/2014arXiv1412.3555C}{arXiv:1412.3555}

\bibitem[{{Clementini} {et~al.}(2023){Clementini}, {Ripepi}, {Garofalo},
  {Molinaro}, {Muraveva}, {Leccia}, {Rimoldini}, {Holl}, {Jevardat de
  Fombelle}, {Sartoretti}, {Marchal}, {Audard}, {Nienartowicz}, {Andrae},
  {Marconi}, {Szabados}, {Evans}, {Lecoeur-Taibi}, {Mowlavi}, {Musella}, \&
  {Eyer}}]{clementini2023rrlyr}
{Clementini}, G., {Ripepi}, V., {Garofalo}, A., {et~al.} 2023,
  \href{http://dx.doi.org/10.1051/0004-6361/202243964}{\color{magenta}\aap},
  \href{https://ui.adsabs.harvard.edu/abs/2023A&A...674A..18C}{674, A18}

\bibitem[{{Cui} {et~al.}(2012){Cui}, {Zhao}, {Chu}, {Li}, {Li}, {Zhang}, {Su},
  {Yao}, {Wang}, {Xing}, {Li}, {Zhu}, {Wang}, {Gu}, {Luo}, {Xu}, {Zhang},
  {Liu}, {Zhang}, {Yang}, {Cao}, {Chen}, {Chen}, {Chen}, {Chen}, {Chu}, {Feng},
  {Gong}, {Hou}, {Hu}, {Hu}, {Hu}, {Jia}, {Jiang}, {Jiang}, {Jiang}, {Jin},
  {Li}, {Li}, {Li}, {Liu}, {Liu}, {Lu}, {Mao}, {Men}, {Qi}, {Qi}, {Shi},
  {Tang}, {Tao}, {Wang}, {Wang}, {Wang}, {Wang}, {Wang}, {Wang}, {Wang},
  {Wang}, {Wang}, {Wang}, {Wang}, {Wang}, {Xu}, {Xu}, {Yang}, {Yu}, {Yuan},
  {Yuan}, {Zhai}, {Zhang}, {Zhang}, {Zhang}, {Zhao}, {Zhou}, {Zhou}, {Zhu}, \&
  {Zou}}]{cui2012lamost}
{Cui}, X.-Q., {Zhao}, Y.-H., {Chu}, Y.-Q., {et~al.} 2012,
  \href{http://dx.doi.org/10.1088/1674-4527/12/9/003}{\color{magenta}Research
  in Astronomy and Astrophysics},
  \href{https://ui.adsabs.harvard.edu/abs/2012RAA....12.1197C}{12, 1197}

\bibitem[{Davies \& Bouldin(1979)}]{Davies1979}
Davies, D.~L. \& Bouldin, D.~W. 1979,
  \href{http://dx.doi.org/10.1109/tpami.1979.4766909}{\color{magenta}IEEE
  Transactions on Pattern Analysis and Machine Intelligence}, PAMI-1, 224–227

\bibitem[{{De Angeli} {et~al.}(2023){De Angeli}, {Weiler}, {Montegriffo},
  {Evans}, {Riello}, {Andrae}, {Carrasco}, {Busso}, {Burgess}, {Cacciari},
  {Davidson}, {Harrison}, {Hodgkin}, {Jordi}, {Osborne}, {Pancino},
  {Altavilla}, {Barstow}, {Bailer-Jones}, {Bellazzini}, {Brown}, {Castellani},
  {Cowell}, {Delchambre}, {De Luise}, {Diener}, {Fabricius}, {Fouesneau},
  {Fr{\'e}mat}, {Gilmore}, {Giuffrida}, {Hambly}, {Hidalgo}, {Holland},
  {Kostrzewa-Rutkowska}, {van Leeuwen}, {Lobel}, {Marinoni}, {Miller},
  {Pagani}, {Palaversa}, {Piersimoni}, {Pulone}, {Ragaini}, {Rainer},
  {Richards}, {Rixon}, {Ruz-Mieres}, {Sanna}, {Sarro}, {Rowell}, {Sordo},
  {Walton}, \& {Yoldas}}]{deangeli2023xp}
{De Angeli}, F., {Weiler}, M., {Montegriffo}, P., {et~al.} 2023,
  \href{http://dx.doi.org/10.1051/0004-6361/202243680}{\color{magenta}\aap},
  \href{https://ui.adsabs.harvard.edu/abs/2023A&A...674A...2D}{674, A2}

\bibitem[{{DESI Collaboration} {et~al.}(2024){DESI Collaboration}, {Adame},
  {Aguilar}, {Ahlen}, {Alam}, {Aldering}, {Alexander}, {Alfarsy}, {Allende
  Prieto}, {Alvarez}, {Alves}, {Anand}, {Andrade-Oliveira}, {Armengaud},
  {Asorey}, {Avila}, {Aviles}, {Bailey}, {Balaguera-Antol{\'\i}nez},
  {Ballester}, {Baltay}, {Bault}, {Bautista}, {Behera}, {Beltran}, {BenZvi},
  {Beraldo e Silva}, {Bermejo-Climent}, {Berti}, {Besuner}, {Beutler},
  {Bianchi}, {Blake}, {Blum}, {Bolton}, {Brieden}, {Brodzeller}, {Brooks},
  {Brown}, {Buckley-Geer}, {Burtin}, {Cabayol-Garcia}, {Cai}, {Canning},
  {Cardiel-Sas}, {Carnero Rosell}, {Castander}, {Cervantes-Cota}, {Chabanier},
  {Chaussidon}, {Chaves-Montero}, {Chen}, {Chen}, {Chuang}, {Claybaugh},
  {Cole}, {Cooper}, {Cuceu}, {Davis}, {Dawson}, {de Belsunce}, {de la Cruz},
  {de la Macorra}, {Della Costa}, {de Mattia}, {Demina}, {Demirbozan},
  {DeRose}, {Dey}, {Dey}, {Dhungana}, {Ding}, {Ding}, {Doel}, {Doshi},
  {Douglass}, {Edge}, {Eftekharzadeh}, {Eisenstein}, {Elliott}, {Ereza},
  {Escoffier}, {Fagrelius}, {Fan}, {Fanning}, {Fawcett}, {Ferraro}, {Flaugher},
  {Font-Ribera}, {Forero-Romero}, {Forero-S{\'a}nchez}, {Frenk},
  {G{\"a}nsicke}, {Garc{\'\i}a}, {Garc{\'\i}a-Bellido}, {Garcia-Quintero},
  {Garrison}, {Gil-Mar{\'\i}n}, {Golden-Marx}, {Gontcho A Gontcho},
  {Gonzalez-Morales}, {Gonzalez-Perez}, {Gordon}, {Graur}, {Green}, {Gruen},
  {Guy}, {Hadzhiyska}, {Hahn}, {Han}, {Hanif}, {Herrera-Alcantar}, {Honscheid},
  {Hou}, {Howlett}, {Huterer}, {Ir{\v{s}}i{\v{c}}}, {Ishak}, {Jacques}, {Jana},
  {Jiang}, {Jimenez}, {Jing}, {Joudaki}, {Joyce}, {Jullo}, {Juneau},
  {Kara{\c{c}}ayl{\i}}, {Karim}, {Kehoe}, {Kent}, {Khederlarian}, {Kim},
  {Kirkby}, {Kisner}, {Kitaura}, {Kizhuprakkat}, {Kneib}, {Koposov},
  {Kov{\'a}cs}, {Kremin}, {Krolewski}, {L'Huillier}, {Lahav}, {Lambert},
  {Lamman}, {Lan}, {Landriau}, {Lang}, {Lange}, {Lasker}, {Leauthaud}, {Le
  Guillou}, {Levi}, {Li}, {Linder}, {Lyons}, {Magneville}, {Manera}, {Manser},
  {Margala}, {Martini}, {McDonald}, {Medina}, {Medina-Varela}, {Meisner},
  {Mena-Fern{\'a}ndez}, {Meneses-Rizo}, {Mezcua}, {Miquel}, {Montero-Camacho},
  {Moon}, {Moore}, {Moustakas}, {Mueller}, {Mundet}, {Mu{\~n}oz-Guti{\'e}rrez},
  {Myers}, {Nadathur}, {Napolitano}, {Neveux}, {Newman}, {Nie}, {Nikutta},
  {Niz}, {Norberg}, {Noriega}, {Paillas}, {Palanque-Delabrouille}, {Palmese},
  {Pan}, {Parkinson}, {Penmetsa}, {Percival}, {P{\'e}rez-Fern{\'a}ndez},
  {P{\'e}rez-R{\`a}fols}, {Pieri}, {Poppett}, {Porredon}, \&
  {Pothier}}]{desi2024desi}
{DESI Collaboration}, {Adame}, A.~G., {Aguilar}, J., {et~al.} 2024,
  \href{http://dx.doi.org/10.3847/1538-3881/ad3217}{\color{magenta}\aj},
  \href{https://ui.adsabs.harvard.edu/abs/2024AJ....168...58D}{168, 58}

\bibitem[{{Donoso-Oliva} {et~al.}(2023){Donoso-Oliva}, {Becker}, {Protopapas},
  {Cabrera-Vives}, {Vishnu}, \& {Vardhan}}]{donoso2023astromer}
{Donoso-Oliva}, C., {Becker}, I., {Protopapas}, P., {et~al.} 2023,
  \href{http://dx.doi.org/10.1051/0004-6361/202243928}{\color{magenta}\aap},
  \href{https://ui.adsabs.harvard.edu/abs/2023A&A...670A..54D}{670, A54}

\bibitem[{{Evans} {et~al.}(2023){Evans}, {Eyer}, {Busso}, {Riello}, {De
  Angeli}, {Burgess}, {Audard}, {Clementini}, {Garofalo}, {Holl}, {Jevardat de
  Fombelle}, {Lanzafame}, {Lecoeur-Taibi}, {Mowlavi}, {Nienartowicz},
  {Palaversa}, \& {Rimoldini}}]{evans2023gaps}
{Evans}, D.~W., {Eyer}, L., {Busso}, G., {et~al.} 2023,
  \href{http://dx.doi.org/10.1051/0004-6361/202244204}{\color{magenta}\aap},
  \href{https://ui.adsabs.harvard.edu/abs/2023A&A...674A...4E}{674, A4}

\bibitem[{{Eyer} {et~al.}(2023){Eyer}, {Audard}, {Holl}, {Rimoldini},
  {Carnerero}, {Clementini}, {De Ridder}, {Distefano}, {Evans}, {Gavras},
  {Gomel}, {Lebzelter}, {Marton}, {Mowlavi}, {Panahi}, {Ripepi}, {Wyrzykowski},
  {Nienartowicz}, {Jevardat de Fombelle}, {Lecoeur-Taibi}, {Rohrbasser},
  {Riello}, {Garc{\'\i}a-Lario}, {Lanzafame}, {Mazeh}, {Raiteri}, {Zucker},
  {{\'A}brah{\'a}m}, {Aerts}, {Aguado}, {Anderson}, {Bashi}, {Binnenfeld},
  {Faigler}, {Garofalo}, {Karbevska}, {K{\'o}sp{\'a}l}, {Kruszy{\'n}ska},
  {Kun}, {Lanza}, {Leccia}, {Marconi}, {Messina}, {Molinaro}, {Moln{\'a}r},
  {Muraveva}, {Musella}, {Nagy}, {Pagano}, {Palaversa}, {Plachy}, {Pr{\v{s}}a},
  {Rybicki}, {Shahaf}, {Szabados}, {Szegedi-Elek}, {Trabucchi}, {Barblan},
  {Grenon}, {Roelens}, \& {S{\"u}veges}}]{eyer2023vari}
{Eyer}, L., {Audard}, M., {Holl}, B., {et~al.} 2023,
  \href{http://dx.doi.org/10.1051/0004-6361/202244242}{\color{magenta}\aap},
  \href{https://ui.adsabs.harvard.edu/abs/2023A&A...674A..13E}{674, A13}

\bibitem[{{Eyer} \& {Mowlavi}(2008)}]{eyer2008variable}
{Eyer}, L. \& {Mowlavi}, N. 2008, in Journal of Physics Conference Series, Vol.
  118, Journal of Physics Conference Series (IOP),
  \href{https://ui.adsabs.harvard.edu/abs/2008JPhCS.118a2010E}{012010}

\bibitem[{Falcon {et~al.}(2020)Falcon, Borovec, W\"{a}lchli, Eggert, Schock,
  Jordan, Skafte, {Ir1dXD}, Bereznyuk, Harris, {Tullie Murrell}, Yu, Præsius,
  Addair, Zhong, Lipin, Uchida, {Shreyas Bapat}, Schr\"{o}ter, Dayma,
  Karnachev, {Akshay Kulkarni}, {Shunta Komatsu}, {Martin.B}, {Jean-Baptiste
  SCHIRATTI}, Mary, Byrne, {Cristobal Eyzaguirre}, {Cinjon}, \&
  Bakhtin}]{falcon2019}
Falcon, W., Borovec, J., W\"{a}lchli, A., {et~al.} 2020,
  PyTorchLightning/pytorch-lightning: 0.7.6 release

\bibitem[{{Fritzewski} {et~al.}(2025){Fritzewski}, {Vanrespaille}, {Aerts},
  {Guo}, {Hey}, \& {De Ridder}}]{Fritzewski2025}
{Fritzewski}, D.~J., {Vanrespaille}, M., {Aerts}, C., {et~al.} 2025,
  \href{http://dx.doi.org/10.1051/0004-6361/202451721}{\color{magenta}\aap},
  \href{https://ui.adsabs.harvard.edu/abs/2025A&A...698A.253F}{698, A253}

\bibitem[{{Gaia Collaboration} {et~al.}(2023{\natexlab{a}}){Gaia
  Collaboration}, {De Ridder}, {Ripepi}, {Aerts}, {Palaversa}, {Eyer}, {Holl},
  {Audard}, {Rimoldini}, {Brown}, \& et~al.}]{deridder2023gaia}
{Gaia Collaboration}, {De Ridder}, J., {Ripepi}, V., {et~al.}
  2023{\natexlab{a}},
  \href{http://dx.doi.org/10.1051/0004-6361/202243767}{\color{magenta}\aap},
  \href{https://ui.adsabs.harvard.edu/abs/2023A&A...674A..36G}{674, A36}

\bibitem[{{Gaia Collaboration} {et~al.}(2016){Gaia Collaboration}, {Prusti},
  {de Bruijne}, {Brown}, {Vallenari}, {Babusiaux}, {Bailer-Jones}, {Bastian},
  {Biermann}, {Evans}, {Eyer}, {Jansen}, {Jordi}, {Klioner}, {Lammers},
  {Lindegren}, {Luri}, {Mignard}, {Milligan}, {Panem}, {Poinsignon},
  {Pourbaix}, {Randich}, {Sarri}, {Sartoretti}, {Siddiqui}, {Soubiran},
  {Valette}, {van Leeuwen}, {Walton}, {Aerts}, {Arenou}, {Cropper}, {Drimmel},
  {H{\o}g}, {Katz}, {Lattanzi}, {O'Mullane}, {Grebel}, {Holland}, {Huc},
  {Passot}, {Bramante}, {Cacciari}, {Casta{\~n}eda}, {Chaoul}, {Cheek}, {De
  Angeli}, {Fabricius}, {Guerra}, {Hern{\'a}ndez}, {Jean-Antoine-Piccolo},
  {Masana}, {Messineo}, {Mowlavi}, {Nienartowicz}, {Ord{\'o}{\~n}ez-Blanco},
  {Panuzzo}, {Portell}, {Richards}, {Riello}, {Seabroke}, {Tanga},
  {Th{\'e}venin}, {Torra}, {Els}, {Gracia-Abril}, {Comoretto},
  {Garcia-Reinaldos}, {Lock}, {Mercier}, {Altmann}, {Andrae}, {Astraatmadja},
  {Bellas-Velidis}, {Benson}, {Berthier}, {Blomme}, {Busso}, {Carry},
  {Cellino}, {Clementini}, {Cowell}, {Creevey}, {Cuypers}, {Davidson}, {De
  Ridder}, {de Torres}, {Delchambre}, {Dell'Oro}, {Ducourant}, {Fr{\'e}mat},
  {Garc{\'\i}a-Torres}, {Gosset}, {Halbwachs}, {Hambly}, {Harrison}, {Hauser},
  {Hestroffer}, {Hodgkin}, {Huckle}, {Hutton}, {Jasniewicz}, {Jordan},
  {Kontizas}, {Korn}, {Lanzafame}, {Manteiga}, {Moitinho}, {Muinonen},
  {Osinde}, {Pancino}, {Pauwels}, {Petit}, {Recio-Blanco}, {Robin}, {Sarro},
  {Siopis}, {Smith}, {Smith}, {Sozzetti}, {Thuillot}, {van Reeven}, {Viala},
  {Abbas}, {Abreu Aramburu}, {Accart}, {Aguado}, {Allan}, {Allasia},
  {Altavilla}, {{\'A}lvarez}, {Alves}, {Anderson}, {Andrei}, {Anglada Varela},
  {Antiche}, {Antoja}, {Ant{\'o}n}, {Arcay}, {Atzei}, {Ayache}, {Bach},
  {Baker}, {Balaguer-N{\'u}{\~n}ez}, {Barache}, {Barata}, {Barbier}, {Barblan},
  {Baroni}, {Barrado y Navascu{\'e}s}, {Barros}, {Barstow}, {Becciani},
  {Bellazzini}, {Bellei}, {Bello Garc{\'\i}a}, {Belokurov}, {Bendjoya},
  {Berihuete}, {Bianchi}, {Bienaym{\'e}}, {Billebaud}, {Blagorodnova},
  {Blanco-Cuaresma}, {Boch}, {Bombrun}, {Borrachero}, {Bouquillon}, {Bourda},
  {Bouy}, {Bragaglia}, {Breddels}, {Brouillet}, {Br{\"u}semeister},
  {Bucciarelli}, {Budnik}, {Burgess}, {Burgon}, {Burlacu}, {Busonero}, {Buzzi},
  {Caffau}, {Cambras}, {Campbell}, {Cancelliere}, {Cantat-Gaudin}, {Carlucci},
  {Carrasco}, {Castellani}, {Charlot}, {Charnas}, {Charvet}, {Chassat},
  {Chiavassa}, {Clotet}, {Cocozza}, {Collins}, {Collins}, \&
  {Costigan}}]{gaia2016mission}
{Gaia Collaboration}, {Prusti}, T., {de Bruijne}, J.~H.~J., {et~al.} 2016,
  \href{http://dx.doi.org/10.1051/0004-6361/201629272}{\color{magenta}\aap},
  \href{https://ui.adsabs.harvard.edu/abs/2016A&A...595A...1G}{595, A1}

\bibitem[{{Gaia Collaboration} {et~al.}(2023{\natexlab{b}}){Gaia
  Collaboration}, {Vallenari}, {Brown}, {Prusti}, {de Bruijne}, {Arenou},
  {Babusiaux}, {Biermann}, {Creevey}, {Ducourant}, {Evans}, {Eyer}, {Guerra},
  {Hutton}, {Jordi}, {Klioner}, {Lammers}, {Lindegren}, {Luri}, {Mignard},
  {Panem}, {Pourbaix}, {Randich}, {Sartoretti}, {Soubiran}, {Tanga}, {Walton},
  {Bailer-Jones}, {Bastian}, {Drimmel}, {Jansen}, {Katz}, {Lattanzi}, {van
  Leeuwen}, {Bakker}, {Cacciari}, {Casta{\~n}eda}, {De Angeli}, {Fabricius},
  {Fouesneau}, {Fr{\'e}mat}, {Galluccio}, {Guerrier}, {Heiter}, {Masana},
  {Messineo}, {Mowlavi}, {Nicolas}, {Nienartowicz}, {Pailler}, {Panuzzo},
  {Riclet}, {Roux}, {Seabroke}, {Sordo}, {Th{\'e}venin}, {Gracia-Abril},
  {Portell}, {Teyssier}, {Altmann}, {Andrae}, {Audard}, {Bellas-Velidis},
  {Benson}, {Berthier}, {Blomme}, {Burgess}, {Busonero}, {Busso},
  {C{\'a}novas}, {Carry}, {Cellino}, {Cheek}, {Clementini}, {Damerdji},
  {Davidson}, {de Teodoro}, {Nu{\~n}ez Campos}, {Delchambre}, {Dell'Oro},
  {Esquej}, {Fern{\'a}ndez-Hern{\'a}ndez}, {Fraile}, {Garabato},
  {Garc{\'\i}a-Lario}, {Gosset}, {Haigron}, {Halbwachs}, {Hambly}, {Harrison},
  {Hern{\'a}ndez}, {Hestroffer}, {Hodgkin}, {Holl}, {Jan{\ss}en}, {Jevardat de
  Fombelle}, {Jordan}, {Krone-Martins}, {Lanzafame}, {L{\"o}ffler}, {Marchal},
  {Marrese}, {Moitinho}, {Muinonen}, {Osborne}, {Pancino}, {Pauwels},
  {Recio-Blanco}, {Reyl{\'e}}, {Riello}, {Rimoldini}, {Roegiers}, {Rybizki},
  {Sarro}, {Siopis}, {Smith}, {Sozzetti}, {Utrilla}, {van Leeuwen}, {Abbas},
  {{\'A}brah{\'a}m}, {Abreu Aramburu}, {Aerts}, {Aguado}, {Ajaj},
  {Aldea-Montero}, {Altavilla}, {{\'A}lvarez}, {Alves}, {Anders}, {Anderson},
  {Anglada Varela}, {Antoja}, {Baines}, {Baker}, {Balaguer-N{\'u}{\~n}ez},
  {Balbinot}, {Balog}, {Barache}, {Barbato}, {Barros}, {Barstow},
  {Bartolom{\'e}}, {Bassilana}, {Bauchet}, {Becciani}, {Bellazzini},
  {Berihuete}, {Bernet}, {Bertone}, {Bianchi}, {Binnenfeld}, {Blanco-Cuaresma},
  {Blazere}, {Boch}, {Bombrun}, {Bossini}, {Bouquillon}, {Bragaglia},
  {Bramante}, {Breedt}, {Bressan}, {Brouillet}, {Brugaletta}, {Bucciarelli},
  {Burlacu}, {Butkevich}, {Buzzi}, {Caffau}, {Cancelliere}, {Cantat-Gaudin},
  {Carballo}, {Carlucci}, {Carnerero}, {Carrasco}, {Casamiquela}, {Castellani},
  {Castro-Ginard}, {Chaoul}, {Charlot}, {Chemin}, {Chiaramida}, {Chiavassa},
  {Chornay}, {Comoretto}, {Contursi}, {Cooper}, {Cornez}, {Cowell}, {Crifo},
  {Cropper}, {Crosta}, {Crowley}, {Dafonte}, {Dapergolas}, {David}, {David},
  {de Laverny}, {De Luise}, \& {De March}}]{gaia2023dr3}
{Gaia Collaboration}, {Vallenari}, A., {Brown}, A.~G.~A., {et~al.}
  2023{\natexlab{b}},
  \href{http://dx.doi.org/10.1051/0004-6361/202243940}{\color{magenta}\aap},
  \href{https://ui.adsabs.harvard.edu/abs/2023A&A...674A...1G}{674, A1}

\bibitem[{{Garrison} {et~al.}(2024){Garrison}, {Foreman-Mackey}, {Shih}, \&
  {Barnett}}]{garrison2024nifty}
{Garrison}, L.~H., {Foreman-Mackey}, D., {Shih}, Y.-h., \& {Barnett}, A. 2024,
  \href{http://dx.doi.org/10.3847/2515-5172/ad82cd}{\color{magenta}Research
  Notes of the American Astronomical Society},
  \href{https://ui.adsabs.harvard.edu/abs/2024RNAAS...8..250G}{8, 250}

\bibitem[{{Gavras} {et~al.}(2023){Gavras}, {Rimoldini}, {Nienartowicz}, {de
  Fombelle}, {Holl}, {{\'A}brah{\'a}m}, {Audard}, {Carnerero}, {Clementini},
  {De Ridder}, {Distefano}, {Garcia-Lario}, {Garofalo}, {K{\'o}sp{\'a}l},
  {Kruszy{\'n}ska}, {Kun}, {Lecoeur-Ta{\"\i}bi}, {Marton}, {Mazeh}, {Mowlavi},
  {Raiteri}, {Ripepi}, {Szabados}, {Zucker}, \& {Eyer}}]{gavras2023gaia}
{Gavras}, P., {Rimoldini}, L., {Nienartowicz}, K., {et~al.} 2023,
  \href{http://dx.doi.org/10.1051/0004-6361/202244367}{\color{magenta}\aap},
  \href{https://ui.adsabs.harvard.edu/abs/2023A&A...674A..22G}{674, A22}

\bibitem[{{Ginsburg} {et~al.}(2019){Ginsburg}, {Sip{\H{o}}cz}, {Brasseur},
  {Cowperthwaite}, {Craig}, {Deil}, {Guillochon}, {Guzman}, {Liedtke}, {Lian
  Lim}, {Lockhart}, {Mommert}, {Morris}, {Norman}, {Parikh}, {Persson},
  {Robitaille}, {Segovia}, {Singer}, {Tollerud}, {de Val-Borro}, {Valtchanov},
  {Woillez}, {Astroquery Collaboration}, \& {a subset of astropy
  Collaboration}}]{ginsburg2019astroquery}
{Ginsburg}, A., {Sip{\H{o}}cz}, B.~M., {Brasseur}, C.~E., {et~al.} 2019,
  \href{http://dx.doi.org/10.3847/1538-3881/aafc33}{\color{magenta}\aj},
  \href{https://ui.adsabs.harvard.edu/abs/2019AJ....157...98G}{157, 98}

\bibitem[{Glorot {et~al.}(2011)Glorot, Bordes, \& Bengio}]{glorot2011deep}
Glorot, X., Bordes, A., \& Bengio, Y. 2011, in Proceedings of Machine Learning
  Research, Vol.~15, Proceedings of the Fourteenth International Conference on
  Artificial Intelligence and Statistics (PMLR),
  \href{https://proceedings.mlr.press/v15/glorot11a.html}{315--323}

\bibitem[{Goodfellow {et~al.}(2016)Goodfellow, Bengio, \&
  Courville}]{goodfellow2016deep}
Goodfellow, I., Bengio, Y., \& Courville, A. 2016, Deep Learning (The MIT
  Press)

\bibitem[{Gui {et~al.}(2024)Gui, Chen, Zhang, Cao, Sun, Luo, \&
  Tao}]{gui2024sslsurvey}
Gui, J., Chen, T., Zhang, J., {et~al.} 2024,
  \href{http://dx.doi.org/10.48550/arXiv.2301.05712}{\color{magenta}IEEE
  Transactions on Pattern Analysis and Machine Intelligence},
  \href{https://ui.adsabs.harvard.edu/abs/2023arXiv230105712G}{46, 9052}

\bibitem[{{Guiglion} {et~al.}(2024){Guiglion}, {Nepal}, {Chiappini},
  {Khoperskov}, {Traven}, {Queiroz}, {Steinmetz}, {Valentini}, {Fournier},
  {Vallenari}, {Youakim}, {Bergemann}, {M{\'e}sz{\'a}ros}, {Lucatello},
  {Sordo}, {Fabbro}, {Minchev}, {Tautvai{\v{s}}ien{\.{e}}}, {Mikolaitis}, \&
  {Montalb{\'a}n}}]{guiglion2024beyond}
{Guiglion}, G., {Nepal}, S., {Chiappini}, C., {et~al.} 2024,
  \href{http://dx.doi.org/10.1051/0004-6361/202347122}{\color{magenta}\aap},
  \href{https://ui.adsabs.harvard.edu/abs/2024A&A...682A...9G}{682, A9}

\bibitem[{{Guo} {et~al.}(2019){Guo}, {Wang}, \& {Wang}}]{guo2019deep}
{Guo}, W., {Wang}, J., \& {Wang}, S. 2019,
  \href{http://dx.doi.org/10.1109/ACCESS.2019.2916887}{\color{magenta}IEEE
  Access}, \href{https://ui.adsabs.harvard.edu/abs/2019IEEEA...763373G}{7,
  63373}

\bibitem[{He {et~al.}(2015)He, Zhang, Ren, \& Sun}]{he2015initialization}
He, K., Zhang, X., Ren, S., \& Sun, J. 2015, in Proceedings of the IEEE
  International Conference on Computer Vision (ICCV), IEEE,
  \href{https://ui.adsabs.harvard.edu/abs/2015arXiv150201852H}{1026--1034}

\bibitem[{{Heinze} {et~al.}(2018){Heinze}, {Tonry}, {Denneau}, {Flewelling},
  {Stalder}, {Rest}, {Smith}, {Smartt}, \& {Weiland}}]{heinze2018atlas}
{Heinze}, A.~N., {Tonry}, J.~L., {Denneau}, L., {et~al.} 2018,
  \href{http://dx.doi.org/10.3847/1538-3881/aae47f}{\color{magenta}\aj},
  \href{https://ui.adsabs.harvard.edu/abs/2018AJ....156..241H}{156, 241}

\bibitem[{{Hendrycks} \& {Gimpel}(2016)}]{hendrycks2016gaussian}
{Hendrycks}, D. \& {Gimpel}, K. 2016,
  \href{https://ui.adsabs.harvard.edu/abs/2016arXiv160608415H}{\href{http://dx.doi.org/10.48550/arXiv.1606.08415F}{\color{magenta}arXiv
  e-prints}, arXiv:1606.08415}

\bibitem[{{Hey} \& {Aerts}(2024)}]{hey2024confronting}
{Hey}, D. \& {Aerts}, C. 2024,
  \href{http://dx.doi.org/10.1051/0004-6361/202450489}{\color{magenta}\aap},
  \href{https://ui.adsabs.harvard.edu/abs/2024A&A...688A..93H}{688, A93}

\bibitem[{Hochreiter {et~al.}(1997)Hochreiter, Schmidhuber, \&
  Elvezia}]{hochreiter1997lstm}
Hochreiter, S., Schmidhuber, J., \& Elvezia, C. 1997,
  \href{http://dx.doi.org/10.1162/neco.1997.9.8.1735}{\color{magenta}Neural
  Computation}, 9, 1735

\bibitem[{{Holl} {et~al.}(2023){Holl}, {Fabricius}, {Portell}, {Lindegren},
  {Panuzzo}, {Bernet}, {Casta{\~n}eda}, {Jevardat de Fombelle}, {Audard},
  {Ducourant}, {Harrison}, {Evans}, {Busso}, {Sozzetti}, {Gosset}, {Arenou},
  {De Angeli}, {Riello}, {Eyer}, {Rimoldini}, {Gavras}, {Mowlavi},
  {Nienartowicz}, {Lecoeur-Ta{\"\i}bi}, {Garc{\'\i}a-Lario}, \&
  {Pourbaix}}]{holl2023gaia}
{Holl}, B., {Fabricius}, C., {Portell}, J., {et~al.} 2023,
  \href{http://dx.doi.org/10.1051/0004-6361/202245353}{\color{magenta}\aap},
  \href{https://ui.adsabs.harvard.edu/abs/2023A&A...674A..25H}{674, A25}

\bibitem[{Hu \& Tan(2018)}]{hu2018rnnpooling}
Hu, W. \& Tan, Y. 2018, in AAAI Workshops,
  \href{https://ui.adsabs.harvard.edu/abs/2017arXiv170508131H}{245--251}

\bibitem[{{Huertas-Company} {et~al.}(2023){Huertas-Company}, {Sarmiento}, \&
  {Knapen}}]{huertas2023brief}
{Huertas-Company}, M., {Sarmiento}, R., \& {Knapen}, J.~H. 2023,
  \href{http://dx.doi.org/10.1093/rasti/rzad028}{\color{magenta}RAS Techniques
  and Instruments},
  \href{https://ui.adsabs.harvard.edu/abs/2023RASTI...2..441H}{2, 441}

\bibitem[{{Jamal} \& {Bloom}(2020)}]{jamal2020neural}
{Jamal}, S. \& {Bloom}, J.~S. 2020,
  \href{http://dx.doi.org/10.3847/1538-4365/aba8ff}{\color{magenta}\apjs},
  \href{https://ui.adsabs.harvard.edu/abs/2020ApJS..250...30J}{250, 30}

\bibitem[{{Jayasinghe} {et~al.}(2019){Jayasinghe}, {Stanek}, {Kochanek},
  {Shappee}, {Holoien}, {Thompson}, {Prieto}, {Dong}, {Pawlak}, {Pejcha},
  {Shields}, {Pojmanski}, {Otero}, {Britt}, \& {Will}}]{jayasinghe2019asassnvs}
{Jayasinghe}, T., {Stanek}, K.~Z., {Kochanek}, C.~S., {et~al.} 2019,
  \href{http://dx.doi.org/10.1093/mnras/stz844}{\color{magenta}\mnras},
  \href{https://ui.adsabs.harvard.edu/abs/2019MNRAS.486.1907J}{486, 1907}

\bibitem[{Jiang {et~al.}(2017)Jiang, Zheng, Tan, Tang, \&
  Zhou}]{jiang2017variational}
Jiang, Z., Zheng, Y., Tan, H., Tang, B., \& Zhou, H. 2017, in Proceedings of
  the 26th International Joint Conference on Artificial Intelligence,
  \href{https://ui.adsabs.harvard.edu/abs/2016arXiv161105148J}{1965--1972}

\bibitem[{{Kingma} \& {Ba}(2015)}]{kingmaBa15}
{Kingma}, D.~P. \& {Ba}, J. 2015, in International Conference on Learning
  Representations (ICLR),
  \href{https://ui.adsabs.harvard.edu/abs/2014arXiv1412.6980K}{arXiv:1412.6980}

\bibitem[{{Kingma} \& {Welling}(2014)}]{kingma2014}
{Kingma}, D.~P. \& {Welling}, M. 2014, in International Conference on Learning
  Representations (ICLR),
  \href{https://ui.adsabs.harvard.edu/abs/2013arXiv1312.6114K}{arXiv:1312.6114}

\bibitem[{{Laroche} \& {Speagle}(2023)}]{laroche2023closing}
{Laroche}, A. \& {Speagle}, J.~S. 2023, in International Conference in Machine
  Learning, Workshop on Machine Learning for Astrophysics,
  \href{https://ui.adsabs.harvard.edu/abs/2023arXiv230706378L}{arXiv:2307.06378}

\bibitem[{{Lebzelter} {et~al.}(2023){Lebzelter}, {Mowlavi}, {Lecoeur-Taibi},
  {Trabucchi}, {Audard}, {Garc{\'\i}a-Lario}, {Gavras}, {Holl}, {Jevardat de
  Fombelle}, {Nienartowicz}, {Rimoldini}, \& {Eyer}}]{Lebzelter2023gaia_lpv}
{Lebzelter}, T., {Mowlavi}, N., {Lecoeur-Taibi}, I., {et~al.} 2023,
  \href{http://dx.doi.org/10.1051/0004-6361/202244241}{\color{magenta}\aap},
  \href{https://ui.adsabs.harvard.edu/abs/2023A&A...674A..15L}{674, A15}

\bibitem[{{Lei Ba} {et~al.}(2016){Lei Ba}, {Kiros}, \& {Hinton}}]{lei2016layer}
{Lei Ba}, J., {Kiros}, J.~R., \& {Hinton}, G.~E. 2016,
  \href{https://ui.adsabs.harvard.edu/abs/2016arXiv160706450L}{\href{http://dx.doi.org/10.48550/arXiv.1607.06450}{\color{magenta}arXiv
  e-prints}, arXiv:1607.06450}

\bibitem[{Liu {et~al.}(2021)Liu, Zhang, Hou, Mian, Wang, Zhang, \&
  Tang}]{liu2021self}
Liu, X., Zhang, F., Hou, Z., {et~al.} 2021,
  \href{http://dx.doi.org/10.48550/arXiv.2006.08218}{\color{magenta}IEEE
  transactions on knowledge and data engineering},
  \href{https://ui.adsabs.harvard.edu/abs/2020arXiv200608218L}{35, 857}

\bibitem[{{Lomb}(1976)}]{lomb1976least}
{Lomb}, N.~R. 1976,
  \href{http://dx.doi.org/10.1007/BF00648343}{\color{magenta}\apss},
  \href{https://ui.adsabs.harvard.edu/abs/1976Ap&SS..39..447L}{39, 447}

\bibitem[{{Luongo} {et~al.}(2024){Luongo}, {Ripepi}, {Marconi}, {Prudil},
  {Rejkuba}, {Clementini}, \& {Longo}}]{Luongo24}
{Luongo}, E., {Ripepi}, V., {Marconi}, M., {et~al.} 2024,
  \href{http://dx.doi.org/10.1051/0004-6361/202451596}{\color{magenta}\aap},
  \href{https://ui.adsabs.harvard.edu/abs/2024A&A...690L..17L}{690, L17}

\bibitem[{Mahabal {et~al.}(2017)Mahabal, Sheth, Gieseke, Pai, Djorgovski,
  Drake, \& Graham}]{mahabal2017deep}
Mahabal, A., Sheth, K., Gieseke, F., {et~al.} 2017, in 2017 IEEE symposium
  series on computational intelligence (SSCI), IEEE,
  \href{https://ui.adsabs.harvard.edu/abs/2017arXiv170906257M}{1--8}

\bibitem[{{Mart{\'\i}nez-Palomera} {et~al.}(2022){Mart{\'\i}nez-Palomera},
  {Bloom}, \& {Abrahams}}]{martinez2022deep}
{Mart{\'\i}nez-Palomera}, J., {Bloom}, J.~S., \& {Abrahams}, E.~S. 2022,
  \href{http://dx.doi.org/10.3847/1538-3881/ac9b3f}{\color{magenta}\aj},
  \href{https://ui.adsabs.harvard.edu/abs/2022AJ....164..263M}{164, 263}

\bibitem[{{Mombarg} {et~al.}(2024){Mombarg}, {Aerts}, {Van Reeth}, \&
  {Hey}}]{Mombarg2024}
{Mombarg}, J. S.~G., {Aerts}, C., {Van Reeth}, T., \& {Hey}, D. 2024,
  \href{http://dx.doi.org/10.1051/0004-6361/202451651}{\color{magenta}\aap},
  \href{https://ui.adsabs.harvard.edu/abs/2024A&A...691A.131M}{691, A131}

\bibitem[{{Mowlavi} {et~al.}(2023){Mowlavi}, {Holl}, {Lecoeur-Ta{\"\i}bi},
  {Barblan}, {Kochoska}, {Pr{\v{s}}a}, {Mazeh}, {Rimoldini}, {Gavras},
  {Audard}, {Jevardat de Fombelle}, {Nienartowicz}, {Garc{\'\i}a-Lario}, \&
  {Eyer}}]{Mowlavi2023gaia}
{Mowlavi}, N., {Holl}, B., {Lecoeur-Ta{\"\i}bi}, I., {et~al.} 2023,
  \href{http://dx.doi.org/10.1051/0004-6361/202245330}{\color{magenta}\aap},
  \href{https://ui.adsabs.harvard.edu/abs/2023A&A...674A..16M}{674, A16}

\bibitem[{Nair \& Hinton(2010)}]{nair2010rectified}
Nair, V. \& Hinton, G.~E. 2010, in Proceedings of the 27th international
  conference on machine learning (ICML-10),
  \href{https://www.cs.toronto.edu/~fritz/absps/reluICML.pdf}{807--814}

\bibitem[{{Naul} {et~al.}(2018){Naul}, {Bloom}, {P{\'e}rez}, \& {van der
  Walt}}]{naul2018recurrent}
{Naul}, B., {Bloom}, J.~S., {P{\'e}rez}, F., \& {van der Walt}, S. 2018,
  \href{http://dx.doi.org/10.1038/s41550-017-0321-z}{\color{magenta}Nature
  Astronomy}, \href{https://ui.adsabs.harvard.edu/abs/2018NatAs...2..151N}{2,
  151}

\bibitem[{{Oosterhoff}(1939)}]{Oosterhoff39}
{Oosterhoff}, P.~T. 1939, The Observatory,
  \href{https://ui.adsabs.harvard.edu/abs/1939Obs....62..104O}{62, 104}

\bibitem[{{Oosterhoff}(1944)}]{Oosterhoff44}
{Oosterhoff}, P.~T. 1944, \bain,
  \href{https://ui.adsabs.harvard.edu/abs/1944BAN....10...55O}{10, 55}

\bibitem[{{Pang} {et~al.}(2021){Pang}, {Shen}, {Cao}, \& {van den
  Hengel}}]{pang2020anomaly}
{Pang}, G., {Shen}, C., {Cao}, L., \& {van den Hengel}, A. 2021,
  \href{http://dx.doi.org/10.1145/3439950}{\color{magenta}ACM Comput. Surv.},
  \href{https://ui.adsabs.harvard.edu/abs/2020arXiv200702500P}{54}

\bibitem[{{Parker} {et~al.}(2024){Parker}, {Lanusse}, {Golkar}, {Sarra},
  {Cranmer}, {Bietti}, {Eickenberg}, {Krawezik}, {McCabe}, {Morel}, {Ohana},
  {Pettee}, {R{\'e}galdo-Saint Blancard}, {Cho}, {Ho}, \& {Polymathic AI
  Collaboration}}]{parker2024astroclip}
{Parker}, L., {Lanusse}, F., {Golkar}, S., {et~al.} 2024,
  \href{http://dx.doi.org/10.1093/mnras/stae1450}{\color{magenta}\mnras},
  \href{https://ui.adsabs.harvard.edu/abs/2024MNRAS.531.4990P}{531, 4990}

\bibitem[{Paszke {et~al.}(2019)Paszke, Gross, Massa, Lerer, Bradbury, Chanan,
  Killeen, Lin, Gimelshein, Antiga, Desmaison, Kopf, Yang, DeVito, Raison,
  Tejani, Chilamkurthy, Steiner, Fang, Bai, \& Chintala}]{paszke2019}
Paszke, A., Gross, S., Massa, F., {et~al.} 2019, in Advances in Neural
  Information Processing Systems {(NIPS)} 32 (Curran Associates, Inc.),
  \href{https://ui.adsabs.harvard.edu/abs/2019arXiv191201703P}{8024--8035}

\bibitem[{{Pedregosa} {et~al.}(2011){Pedregosa}, {Varoquaux}, {Gramfort},
  {Michel}, {Thirion}, {Grisel}, {Blondel}, {M{\"u}ller}, {Nothman}, {Louppe},
  {Prettenhofer}, {Weiss}, {Dubourg}, {Vanderplas}, {Passos}, {Cournapeau},
  {Brucher}, {Perrot}, \& {Duchesnay}}]{pedregosa2011sklearn}
{Pedregosa}, F., {Varoquaux}, G., {Gramfort}, A., {et~al.} 2011,
  \href{http://dx.doi.org/10.48550/arXiv.1201.0490}{\color{magenta}Journal of
  Machine Learning Research},
  \href{https://ui.adsabs.harvard.edu/abs/2011JMLR...12.2825P}{12, 2825}

\bibitem[{{Prudil} {et~al.}(2019){Prudil}, {D{\'e}k{\'a}ny}, {Catelan},
  {Smolec}, {Grebel}, \& {Skarka}}]{Prudil2019rrlyr}
{Prudil}, Z., {D{\'e}k{\'a}ny}, I., {Catelan}, M., {et~al.} 2019,
  \href{http://dx.doi.org/10.1093/mnras/stz311}{\color{magenta}\mnras},
  \href{https://ui.adsabs.harvard.edu/abs/2019MNRAS.484.4833P}{484, 4833}

\bibitem[{Radford {et~al.}(2021)Radford, Kim, Hallacy, Ramesh, Goh, Agarwal,
  Sastry, Askell, Mishkin, Clark, {et~al.}}]{radford2021clip}
Radford, A., Kim, J.~W., Hallacy, C., {et~al.} 2021, in International
  conference on machine learning, PMLR,
  \href{https://ui.adsabs.harvard.edu/abs/2021arXiv210300020R}{8748--8763}

\bibitem[{{Rimoldini} {et~al.}(2022){Rimoldini}, {Eyer}, {Audard}, {Barblan},
  {Carnerero}, {Clementini}, {De Ridder}, {Distefano}, {Faigler}, {Garofalo},
  {Gavras}, {Gomel}, {Holl}, {Jevardat de Fombelle}, {Kruszy{\'n}ska},
  {Lanzafame}, {Lebzelter}, {Leccia}, {Lecoeur-Ta{\"\i}bi}, {Mazeh},
  {Molinaro}, {Mowlavi}, {Muraveva}, {Nienartowicz}, {Panahi}, {Raiteri},
  {Ripepi}, {Rybicki}, {Trabucchi}, {Wyrzykowski}, \&
  {Zucker}}]{rimoldini2022gaiadr3manual}
{Rimoldini}, L., {Eyer}, L., {Audard}, M., {et~al.} 2022, {Gaia DR3
  documentation Chapter 10: Variability}, Gaia DR3 documentation, European
  Space Agency; Gaia Data Processing and Analysis Consortium. Online at
  \url{https://gea.esac.esa.int/archive/documentation/GDR3/index.html}, id. 10

\bibitem[{{Rimoldini} {et~al.}(2023){Rimoldini}, {Holl}, {Gavras}, {Audard},
  {De Ridder}, {Mowlavi}, {Nienartowicz}, {Jevardat de Fombelle},
  {Lecoeur-Ta{\"\i}bi}, {Karbevska}, {Evans}, {{\'A}brah{\'a}m}, {Carnerero},
  {Clementini}, {Distefano}, {Garofalo}, {Garc{\'\i}a-Lario}, {Gomel},
  {Klioner}, {Kruszy{\'n}ska}, {Lanzafame}, {Lebzelter}, {Marton}, {Mazeh},
  {Molinaro}, {Panahi}, {Raiteri}, {Ripepi}, {Szabados}, {Teyssier},
  {Trabucchi}, {Wyrzykowski}, {Zucker}, \& {Eyer}}]{rimoldini2023gaia}
{Rimoldini}, L., {Holl}, B., {Gavras}, P., {et~al.} 2023,
  \href{http://dx.doi.org/10.1051/0004-6361/202245591}{\color{magenta}\aap},
  \href{https://ui.adsabs.harvard.edu/abs/2023A&A...674A..14R}{674, A14}

\bibitem[{{Rizhko} \& {Bloom}(2025)}]{rizhko2024self}
{Rizhko}, M. \& {Bloom}, J.~S. 2025,
  \href{http://dx.doi.org/10.3847/1538-3881/adcbad}{\color{magenta}\aj},
  \href{https://ui.adsabs.harvard.edu/abs/2024arXiv241108842R}{170, 28}

\bibitem[{{Ruz-Mieres}(2024)}]{ruz-mieres2024}
{Ruz-Mieres}, D. 2024, {gaia-dpci/GaiaXPy: GaiaXPy v2.1.2}, Available online
  at: \url{https://gaia-dpci.github.io/GaiaXPy-website/}

\bibitem[{{S{\'a}nchez-S{\'a}ez} {et~al.}(2021){S{\'a}nchez-S{\'a}ez}, {Lira},
  {Mart{\'\i}}, {S{\'a}nchez-Pi}, {Arredondo}, {Bauer}, {Bayo},
  {Cabrera-Vives}, {Donoso-Oliva}, {Est{\'e}vez}, {Eyheramendy}, {F{\"o}rster},
  {Hern{\'a}ndez-Garc{\'\i}a}, {Arancibia}, {P{\'e}rez-Carrasco},
  {Sep{\'u}lveda}, \& {Vergara}}]{sanchez2021agns}
{S{\'a}nchez-S{\'a}ez}, P., {Lira}, H., {Mart{\'\i}}, L., {et~al.} 2021,
  \href{http://dx.doi.org/10.3847/1538-3881/ac1426}{\color{magenta}\aj},
  \href{https://ui.adsabs.harvard.edu/abs/2021AJ....162..206S}{162, 206}

\bibitem[{{Sartoretti} {et~al.}(2023){Sartoretti}, {Marchal}, {Babusiaux},
  {Jordi}, {Guerrier}, {Panuzzo}, {Katz}, {Seabroke}, {Th{\'e}venin},
  {Cropper}, {Benson}, {Blomme}, {Haigron}, {Smith}, {Baker}, {Chemin},
  {David}, {Dolding}, {Fr{\'e}mat}, {Jan{\ss}en}, {Jasniewicz}, {Lobel},
  {Plum}, {Samaras}, {Snaith}, {Soubiran}, {Vanel}, {Zwitter}, {Brouillet},
  {Caffau}, {Crifo}, {Fabre}, {Fragkoudi}, {Jean-Antoine Piccolo}, {Huckle},
  {Lasne}, {Leclerc}, {Mastrobuono-Battisti}, {Royer}, {Viala}, \&
  {Zorec}}]{sartoretti2023rvs}
{Sartoretti}, P., {Marchal}, O., {Babusiaux}, C., {et~al.} 2023,
  \href{http://dx.doi.org/10.1051/0004-6361/202243615}{\color{magenta}\aap},
  \href{https://ui.adsabs.harvard.edu/abs/2023A&A...674A...6S}{674, A6}

\bibitem[{{Scargle}(1982)}]{scargle1982studies}
{Scargle}, J.~D. 1982,
  \href{http://dx.doi.org/10.1086/160554}{\color{magenta}\apj},
  \href{https://ui.adsabs.harvard.edu/abs/1982ApJ...263..835S}{263, 835}

\bibitem[{{Skarka} {et~al.}(2020){Skarka}, {Prudil}, \&
  {Jurcsik}}]{Skarka2020rrlyr}
{Skarka}, M., {Prudil}, Z., \& {Jurcsik}, J. 2020,
  \href{http://dx.doi.org/10.1093/mnras/staa673}{\color{magenta}\mnras},
  \href{https://ui.adsabs.harvard.edu/abs/2020MNRAS.494.1237S}{494, 1237}

\bibitem[{{Smith}(1995)}]{Smith95}
{Smith}, H.~A. 1995, Cambridge Astrophysics Series,
  \href{https://ui.adsabs.harvard.edu/abs/1995CAS....27.....S}{27}

\bibitem[{{Soraisam} {et~al.}(2020){Soraisam}, {Saha}, {Matheson}, {Lee},
  {Narayan}, {Vivas}, {Scheidegger}, {Oppermann}, {Olszewski}, {Sinha},
  {Desantis}, \& {ANTARES Collaboration}}]{soraisam2020classification}
{Soraisam}, M.~D., {Saha}, A., {Matheson}, T., {et~al.} 2020,
  \href{http://dx.doi.org/10.3847/1538-4357/ab7b61}{\color{magenta}\apj},
  \href{https://ui.adsabs.harvard.edu/abs/2020ApJ...892..112S}{892, 112}

\bibitem[{{The Multimodal Universe Collaboration} {et~al.}(2024){The Multimodal
  Universe Collaboration}, {Audenaert}, {Bowles}, {Boyd}, {Chemaly},
  {Cherinka}, {Ciuc{\u{a}}}, {Cranmer}, {Do}, {Grayling}, {Hayes}, {Hehir},
  {Ho}, {Huertas-Company}, {Iyer}, {Jablonska}, {Lanusse}, {Leung}, {Mandel},
  {Mart{\'\i}nez-Galarza}, {Melchior}, {Meyer}, {Parker}, {Qu}, {Shen},
  {Smith}, {Stone}, {Walmsley}, \& {Wu}}]{angeloudimultimodal}
{The Multimodal Universe Collaboration}, {Audenaert}, J., {Bowles}, M.,
  {et~al.} 2024, in The Thirty-eight Conference on Neural Information
  Processing Systems {(NIPS)} Datasets and Benchmarks Track,
  \href{https://ui.adsabs.harvard.edu/abs/2024arXiv241202527T}{arXiv:2412.02527}

\bibitem[{{Udalski} {et~al.}(2008){Udalski}, {Szymanski}, {Soszynski}, \&
  {Poleski}}]{udalski2008ogleiii}
{Udalski}, A., {Szymanski}, M.~K., {Soszynski}, I., \& {Poleski}, R. 2008,
  \href{http://dx.doi.org/10.48550/arXiv.0807.3884}{\color{magenta}\actaa},
  \href{https://ui.adsabs.harvard.edu/abs/2008AcA....58...69U}{58, 69}

\bibitem[{{Zhang} {et~al.}(2024){Zhang}, {Helfer}, {Gagliano}, {Mishra-Sharma},
  \& {Ashley Villar}}]{zhang2024maven}
{Zhang}, G., {Helfer}, T., {Gagliano}, A.~T., {Mishra-Sharma}, S., \& {Ashley
  Villar}, V. 2024,
  \href{http://dx.doi.org/10.1088/2632-2153/ad990d}{\color{magenta}Machine
  Learning: Science and Technology},
  \href{https://ui.adsabs.harvard.edu/abs/2024MLS&T...5d5069Z}{5, 045069}

\end{thebibliography}

\begin{appendix}

\section{Supporting figures and tables}
\label{sec:supporting_material}

\subsubsection*{Section 3}

The ADQL query to retrieve the source identifiers for the training set:

\begin{verbatim}
SELECT vari.source_id
FROM gaiadr3.vari_summary AS vari
JOIN gaiadr3.gaia_source AS src 
USING (source_id)
WHERE src.has_xp_continuous='t' 
AND vari.num_selected_g_fov >= 40 
AND vari.num_selected_rp >= 40 
AND vari.num_selected_bp >= 40
\end{verbatim}

\subsubsection*{Section 4}

\begin{figure*}[t]
     \centering     
     \includegraphics[width=0.93\textwidth]{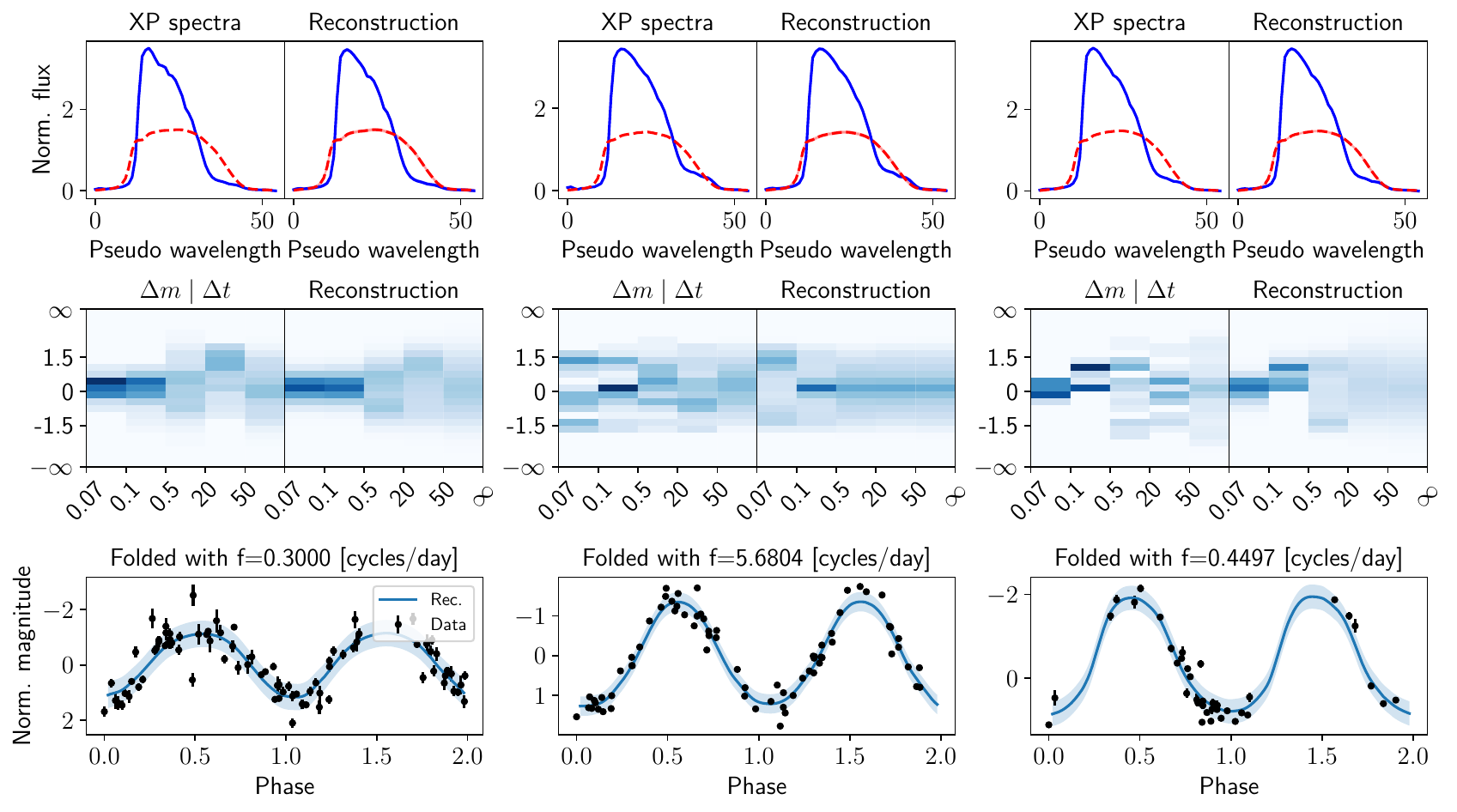}
     \caption{Input data and VAE reconstructions for three \gaia DR3 sources belonging to the ACV, \bcep and SPB classes, respectively. The description of the rows and columns follows that of Fig.~\ref{fig:example_modalities}. 
     } 
     \label{fig:example_modalities_a1}
\end{figure*}

\begin{figure*}[t]
     \centering     
     \includegraphics[width=0.93\textwidth]{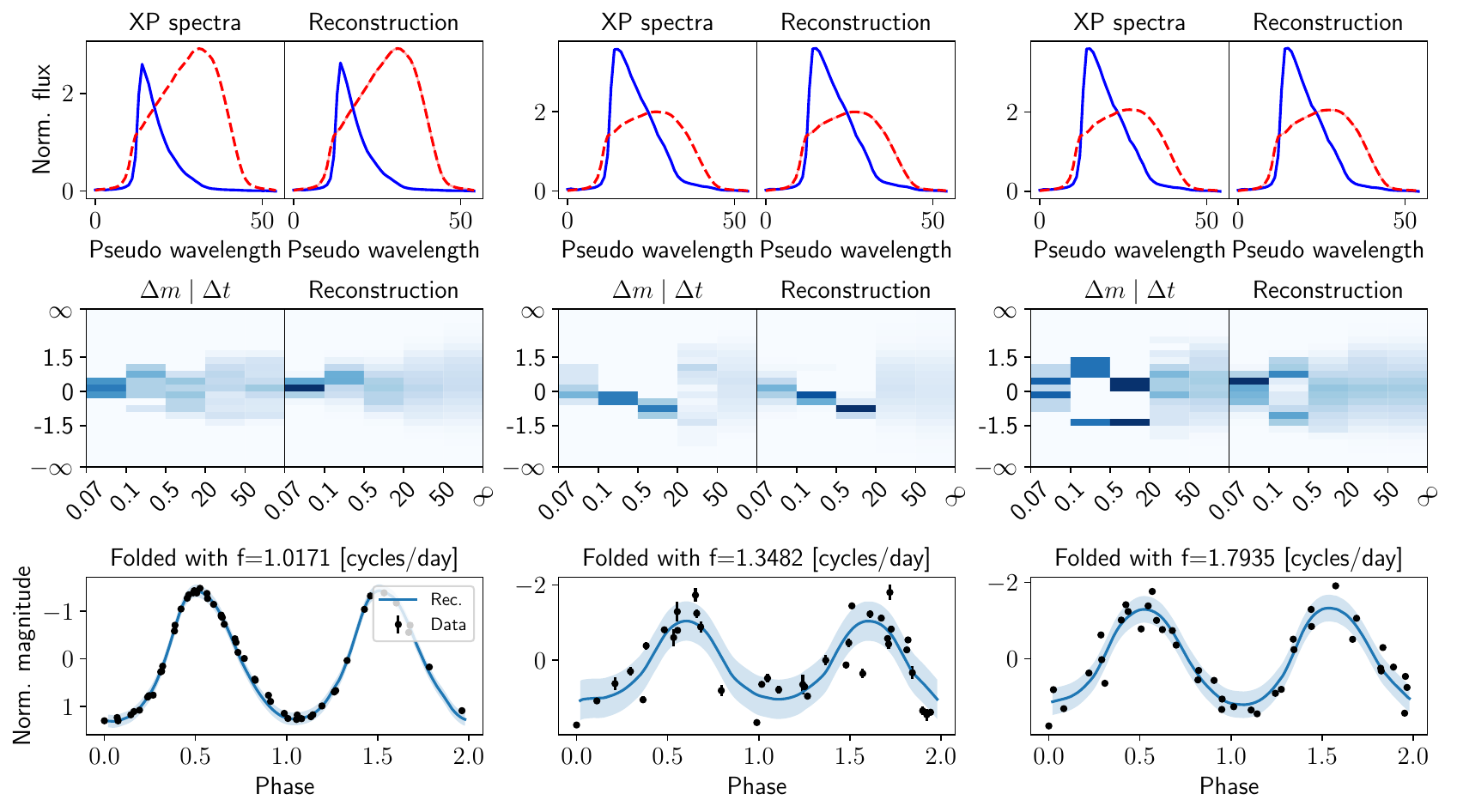}
     \caption{Input data and VAE reconstructions for three \gaia DR3 sources belonging to the \dcep, \dsct and \gdor star classes, respectively. The description of the rows and columns follows that of Fig.~\ref{fig:example_modalities}. 
     } 
     \label{fig:example_modalities_a2}
\end{figure*}

\begin{figure*}[t]
     \centering     
     \includegraphics[width=0.93\textwidth]{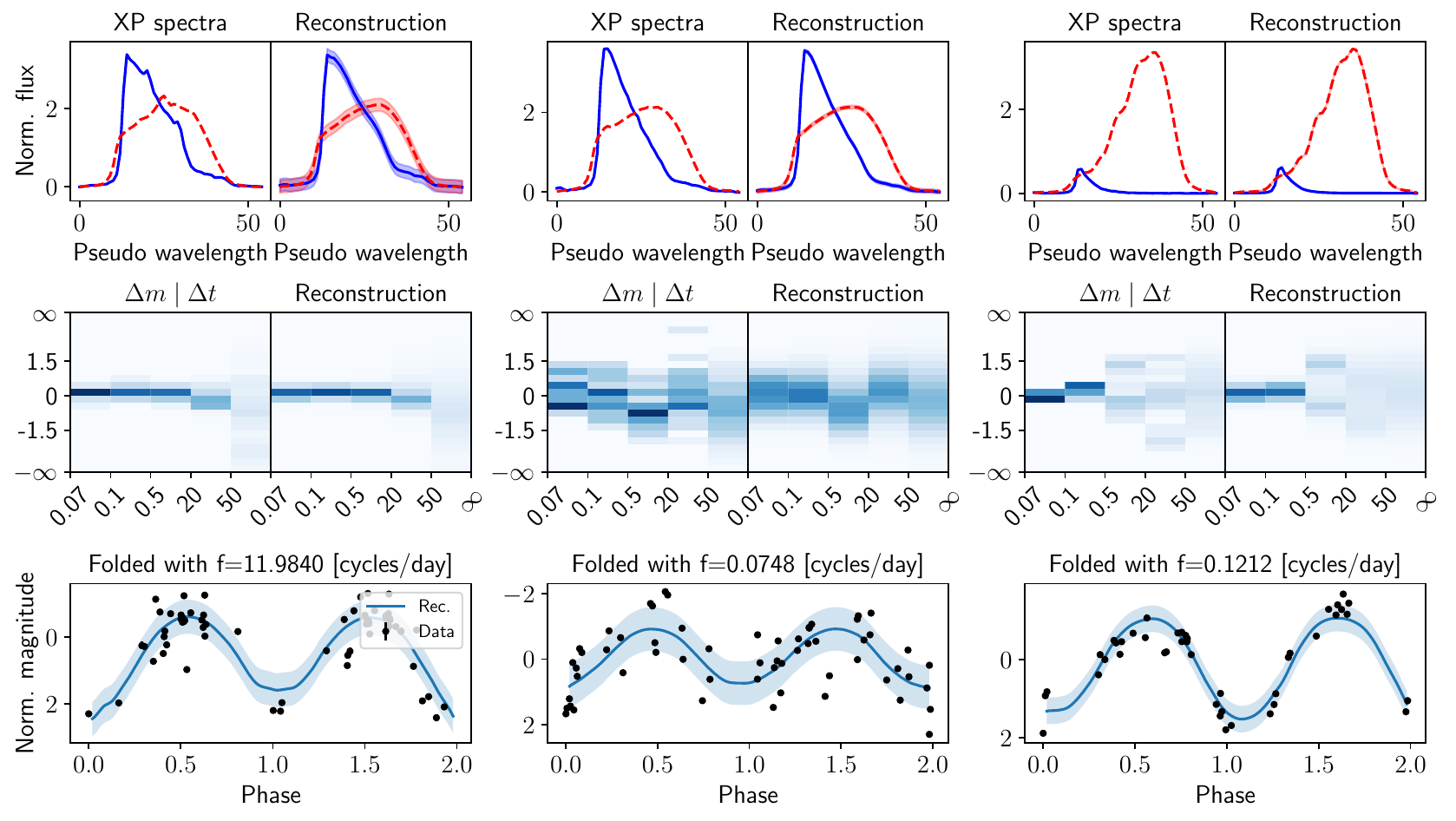}
     \caption{Input data and VAE reconstructions for three \gaia DR3 sources belonging to the AGN, CV and YSO classes, respectively. The description of the rows and columns follows that of Fig.~\ref{fig:example_modalities}. 
     } 
     \label{fig:example_modalities_a3}
\end{figure*}

\begin{figure*}[t]
     \centering     
     \includegraphics[width=0.93\textwidth]{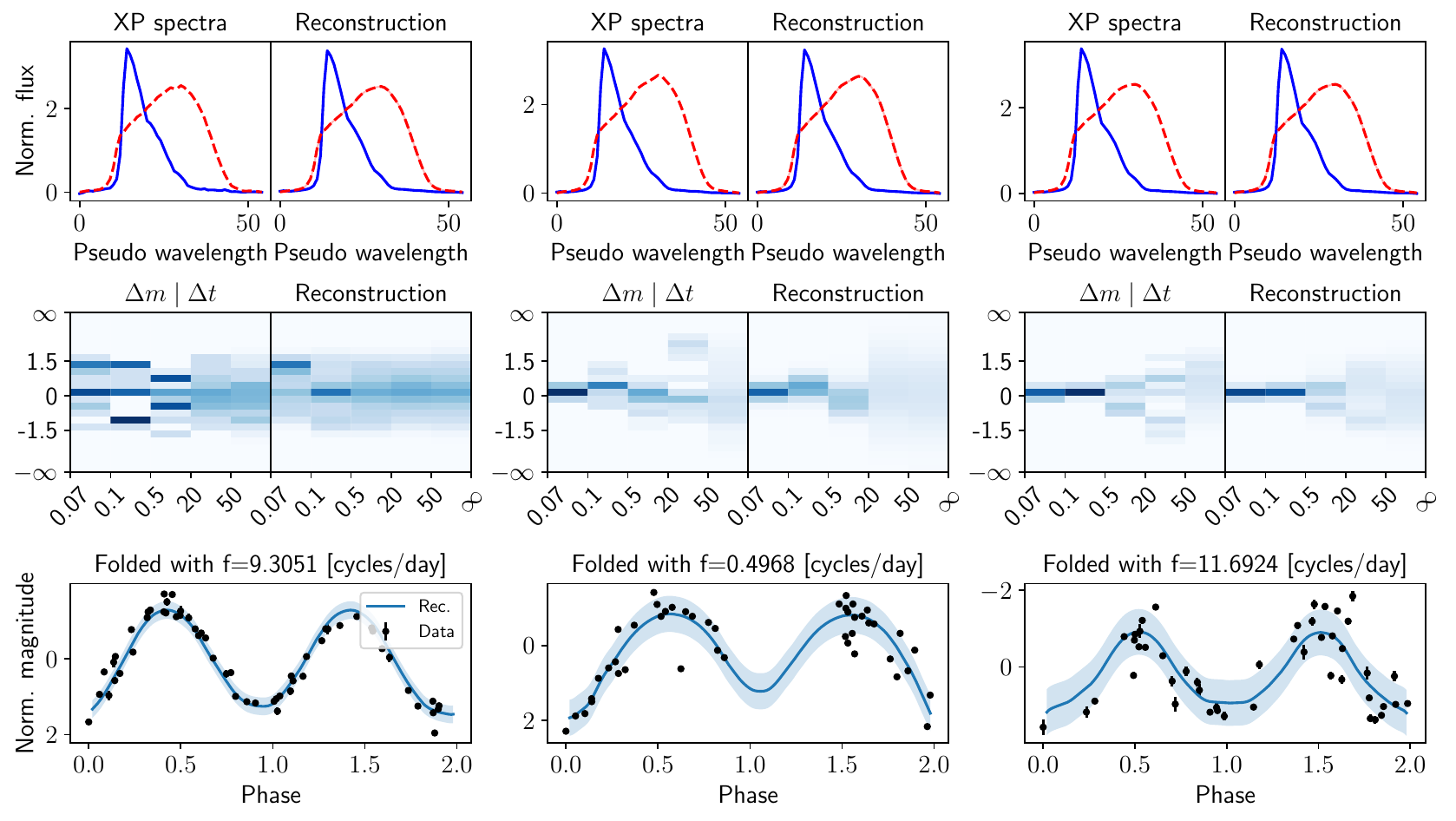}
     \caption{Input data and VAE reconstructions for three \gaia DR3 sources belonging to the RS, ELL and SOLAR classes, respectively.  The description of the rows and columns follows that of Fig.~\ref{fig:example_modalities}. 
     } 
     \label{fig:example_modalities_a4}
\end{figure*}

Figures~\ref{fig:example_modalities_a1}--\ref{fig:example_modalities_a4} show examples of the preprocessed input data and VAE reconstructions for 12 of the variability classes present in the labelled subset, extending what was shown in Fig.~\ref{fig:example_modalities}.

\begin{figure}[t]
     \centering     
     \includegraphics[width=0.4\textwidth]{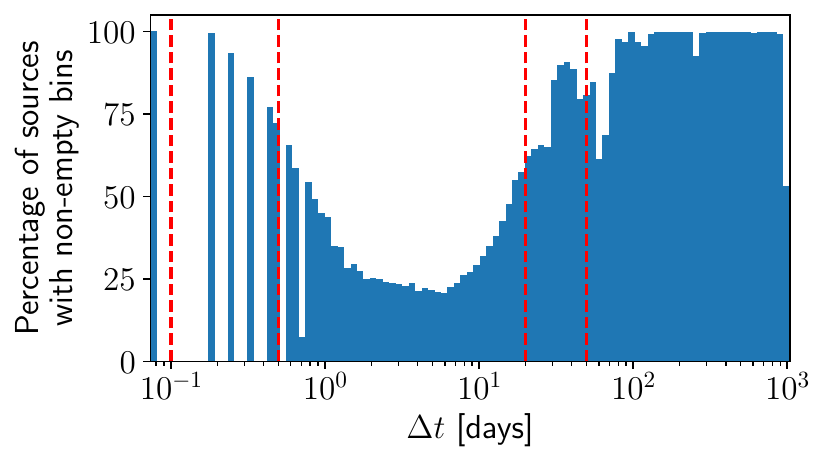}
     \caption{
     Percentage of sources with at least one pair of observations whose time difference falls in a given $\Delta t$ bin. For each source the time differences are extracted and then binned in 100 equally spaced logarithmic bins. The structure is a result of \gaia's sampling. The red dashed lines mark the edges of the low-resolution $\Delta t$ bins used in this work. All sources have at least one observation in each of these low-resolution bins. } 
     \label{fig:dt_histogram}
\end{figure}

\begin{figure}[t]
     \centering     
     \includegraphics[width=0.43\textwidth]{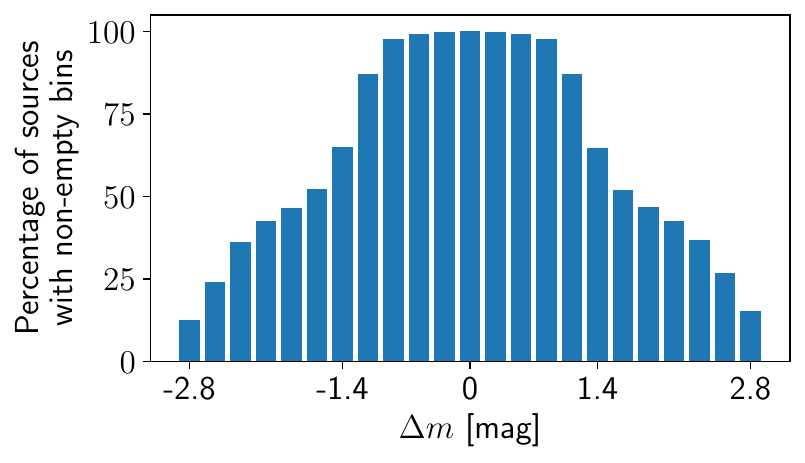}
     \caption{Percentage of sources with at least one pair of observations whose magnitude difference falls in a given $\Delta m$ bin. The normalisation applied to the light curves ensures that the $\Delta m$ distribution is centred around zero.} 
     \label{fig:dm_histogram}
\end{figure}

\begin{table}[b]
\caption{Average computational time required to process a \gaia source with 100 observations. GPU times are measured for a batch of size $256$. Hardware: Intel Xeon 5117 CPU, NVIDIA A4000 GPU.}
\label{tab:time}
\centering
\begin{tabular}{r|cc}
\toprule
Inference step & CPU time [ms] & GPU time [$\mu$s]\\
\midrule
LS periodogram & $41.21 \pm 2.343$ & NA \\
XP VAE & $0.482 \pm 0.063$ & $3.518 \pm 0.599$\\
Folded LC VAE & $14.72 \pm 1.277$ & $19.23 \pm 0.716$\\
$\Delta m| \Delta t$ VAE & $8.561 \pm 0.457$ & $98.96 \pm 0.875$\\
2-dim VAE & $0.144 \pm 0.015$ & $1.064 \pm 0.036$ \\
\bottomrule
\end{tabular}
\end{table}

\subsubsection*{Section 7}

Table~\ref{tab:pre_rec_f1} shows the precision, recall and $F_1$-scores per class for the LR classifier trained using the latent variables from the XP, $\Delta m| \Delta t$ and folded light curve VAEs. The confusion matrix for this model is shown in Fig.~\ref{fig:cm_all}.

\begin{table}[t]
\caption{Supervised classification performance metrics per class using all latent variables as input}
\label{tab:pre_rec_f1}
\begin{tabular}{r|ccc}
\toprule
\toprule
Class & Precision [\%]& Recall [\%]& $F_1$-score [\%]\\
\midrule
\dcep & 77.1(0.9) & 70.8(0.5) & 73.8(0.5) \\
\tiicep & 54.7(1.7) & 78.1(2.0) & 64.3(1.7) \\
\rrab & 87.4(0.4) & 84.5(0.8) & 85.9(0.4) \\
\rrc & 78.0(1.3) & 83.1(0.4) & 80.4(0.6) \\
\rrd & 48.7(1.7) & 54.6(1.3) & 51.4(0.9) \\
ECLa & 70.5(0.3) & 79.2(0.6) & 74.6(0.4) \\
ECLb & 71.3(1.1) & 51.4(0.6) & 59.7(0.7) \\
ECLw & 67.4(0.3) & 70.6(1.4) & 69.0(0.8) \\
LPV & 98.5(0.0) & 93.1(0.7) & 95.7(0.4) \\
\bcep & 56.4(2.6) & 70.6(0.0) & 62.6(1.6) \\
\dsct & 96.9(0.4) & 83.7(0.6) & 89.8(0.2) \\
\gdor & 41.6(1.4) & 94.4(0.7) & 57.8(1.5) \\
SPB & 19.7(1.6) & 81.0(3.3) & 31.6(2.0) \\
WD & 90.4(4.5) & 97.0(4.2) & 93.5(3.1) \\
RS & 28.8(0.7) & 46.5(1.5) & 35.5(0.8) \\
ELL & 14.6(0.5) & 60.4(1.2) & 23.6(0.6) \\
SOLAR & 88.6(0.3) & 72.1(0.7) & 79.5(0.3) \\
ACV$\mid$CP & 75.6(1.8) & 67.1(3.6) & 71.1(2.9) \\
YSO & 59.9(2.3) & 64.1(0.6) & 61.9(1.4) \\
BE$\mid$GCAS & 72.3(1.8) & 74.0(0.8) & 73.1(1.0) \\
CV & 64.7(4.9) & 93.0(1.2) & 76.2(3.1) \\
AGN & 90.9(2.8) & 86.3(3.1) & 88.5(1.9) \\
\midrule
macro & 66.1(0.5) & 75.2(0.3) & 68.2(0.4) \\
weighted & 82.6(0.2) & 78.6(0.4) & 79.9(0.4) \\
\bottomrule
\end{tabular}
\end{table}

\subsubsection*{Section 8}

\begin{figure}[t]
     \centering     
     \includegraphics[width=0.4\textwidth]{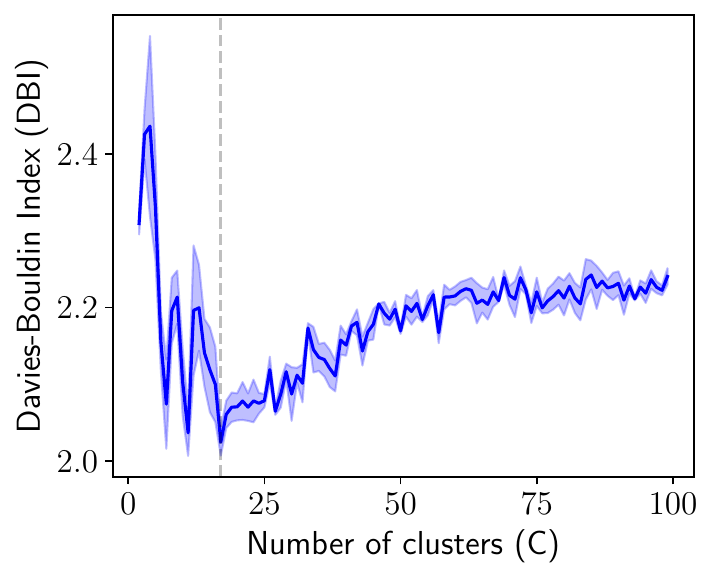}
     \caption{Davies-Bouldin Index as a function of the number of clusters (lower is better). The minimum (C=17) is marked with a black dashed line.} 
     \label{fig:dbi}
\end{figure}

The clustering capability of the latent variables is evaluated using the $k$-means and GMM methods, both of which require the user to predefine the number of clusters, $C$. 
Fig.~\ref{fig:dbi} presents the Davies-Bouldin Index \citep[DBI;][]{Davies1979} as a function of $C$. The DBI measures the ratio between within-cluster distances (compactness) and between-cluster distances (separation), with lower values indicating better clustering performance. In this case, the DBI reaches its minimum at $C=17$, suggesting this is the optimal number of clusters.

\begin{figure}[t]
\centering
    \begin{subfigure}[t]{0.23\textwidth}
    \includegraphics[width=\textwidth]{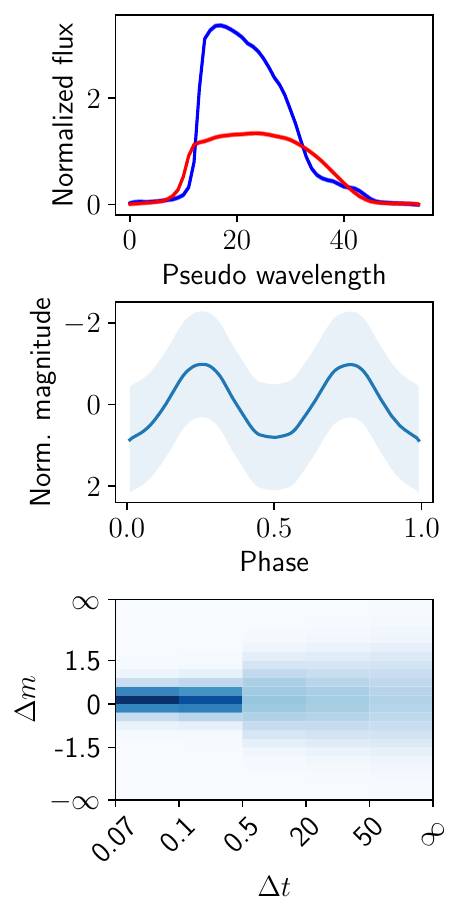}
    \caption{Cluster 2 mean.}
    \label{fig:cluster2}
    \end{subfigure}    
    \hfill
    \begin{subfigure}[t]{0.23\textwidth}
    \includegraphics[width=\textwidth]{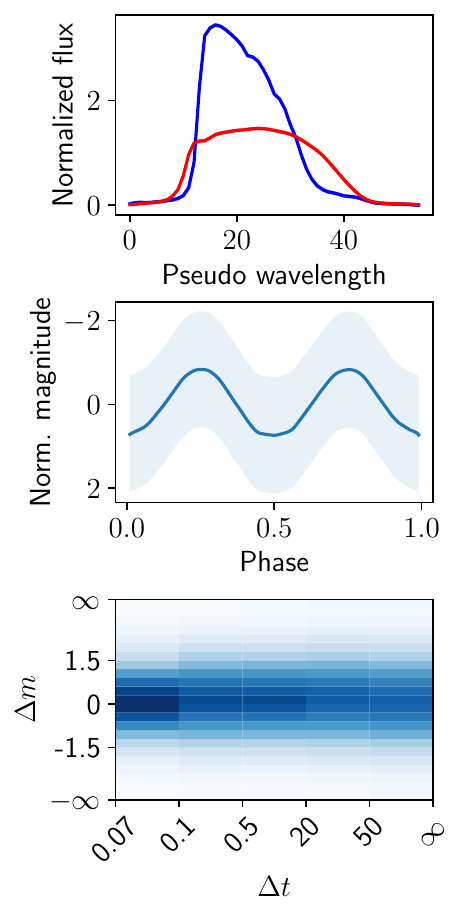}
    \caption{Cluster 5 mean.}
    \label{fig:cluster5}
    \end{subfigure}
    \vfill
    \begin{subfigure}[t]{0.23\textwidth}    \includegraphics[width=\textwidth]{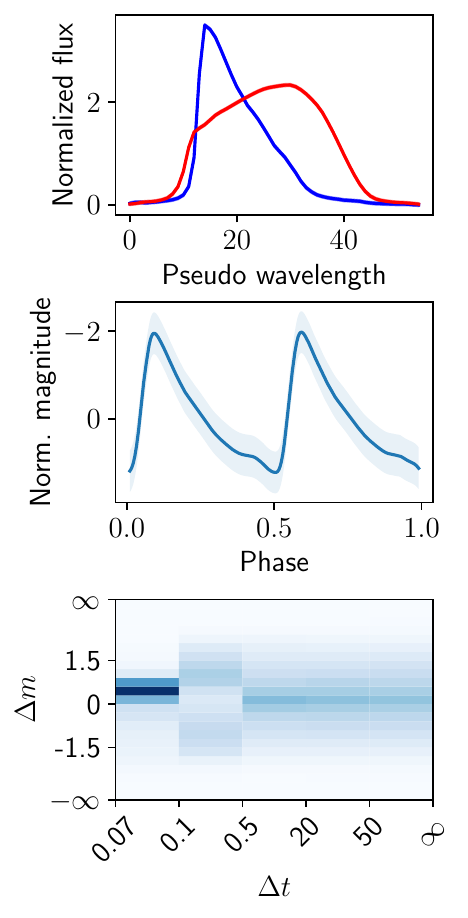}
    \caption{Cluster 13 mean.}
    \label{fig:cluster13}
    \end{subfigure}
    \hfill
    \begin{subfigure}[t]{0.23\textwidth}
    \includegraphics[width=\textwidth]{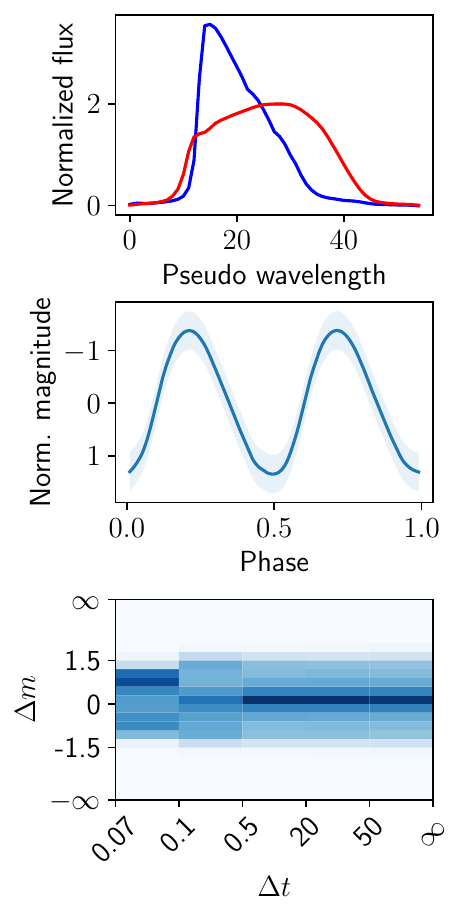}
    \caption{Cluster 14 mean.}
    \label{fig:cluster14}
    \end{subfigure}
    
\caption{Using the decoder of the VAEs, we reconstruct the means of the clusters found by GMM to visualise the prototypical XP spectra, folded time series and $\Delta m | \Delta t$ that characterise these clusters.}
\label{fig:centroids}
\end{figure}

After fitting $k$-means or GMM, in addition to the cluster assignment for each source, one obtains the means or centres of the clusters. Using the decoders of the VAEs, we reconstruct the centres to obtain the prototypical XP spectra, folded time series and $\Delta m|\Delta t$ that are associated with each of the clusters. This is shown for four of the GMM clusters in Fig.~\ref{fig:centroids}, allowing us to further understand the features that characterise these clusters. Panels (a) and (b) show the reconstructions from clusters 2 and 5, respectively. These clusters differ mainly in their time scale, as shown by their prototypical $\Delta m|\Delta t$. Panels (c) and (d) show the reconstructions from clusters 13 and 14, respectively, both of which are predominantly populated by \rr stars. However, the reconstructed folded light curve and $\Delta m|\Delta t$ suggest that cluster 13 corresponds to \rrab, while cluster 14 corresponds to \rrc stars.

\subsubsection*{Section 9}
\label{app:astrophysical_interpretation}
\begin{figure}
\begin{center}
\includegraphics[width=0.45\textwidth,keepaspectratio]{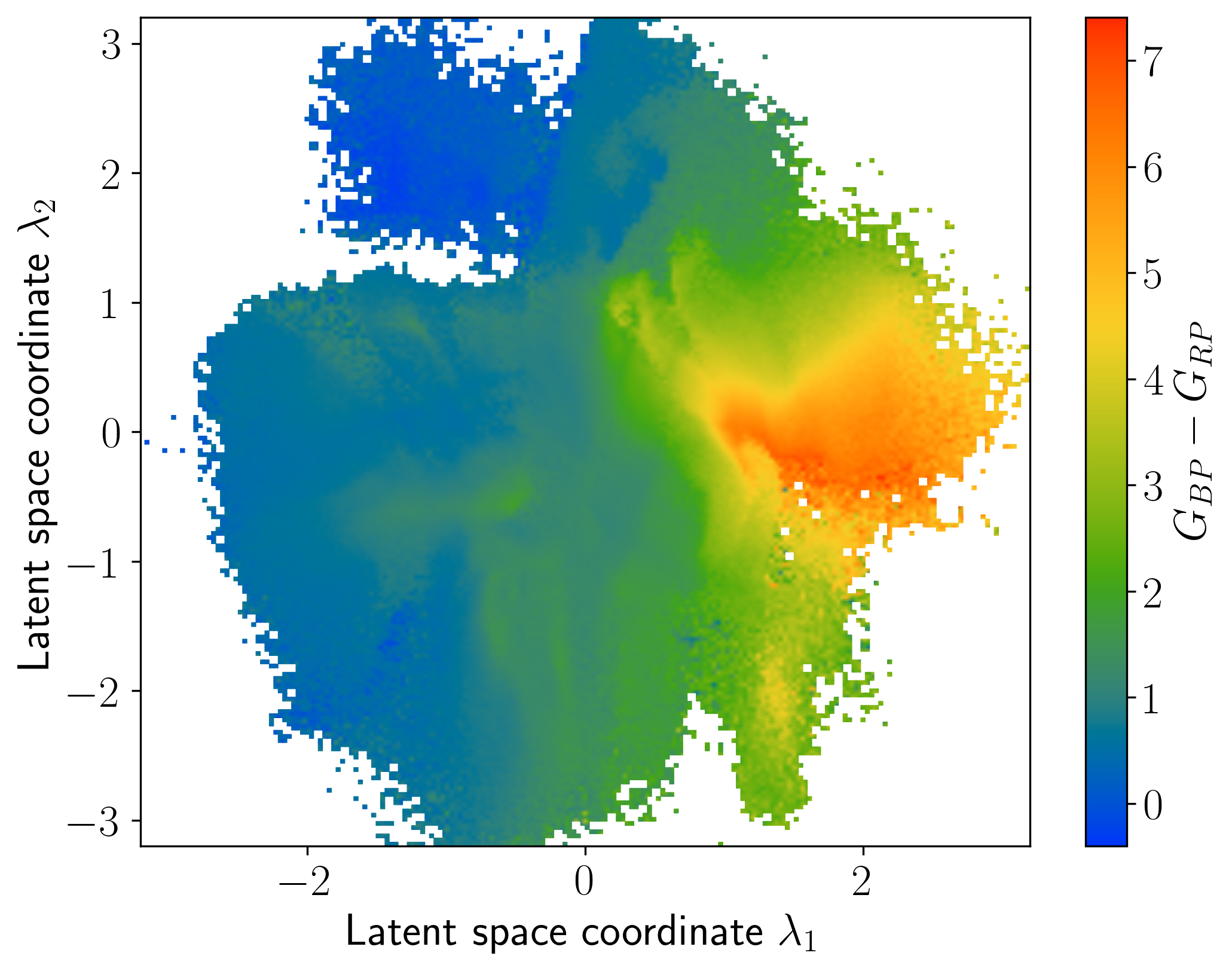}
\caption{\label{fig:bp_rp_heatmap}The distribution of the (potentially reddened) mean colour \bprp in the latent space.}
\end{center}
\end{figure}

\begin{figure}
\begin{center}
\includegraphics[width=0.45\textwidth,keepaspectratio]{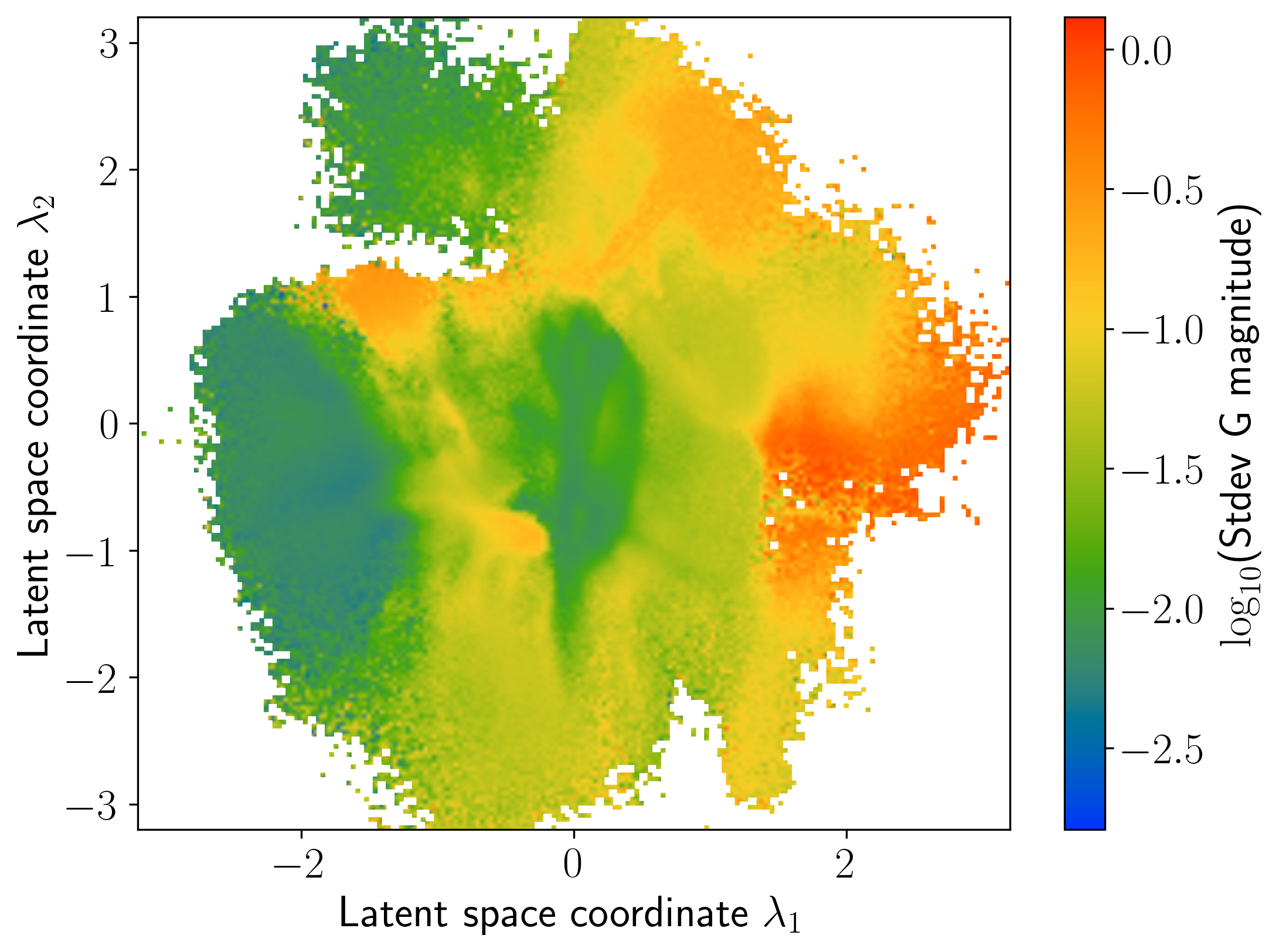}
\caption{\label{fig:stdev_mag_heatmap}The distribution of the standard deviation of the observed \gaia \g magnitude in the latent space.}
\end{center}
\end{figure}

\begin{figure}
\begin{center}
\includegraphics[width=0.45\textwidth,keepaspectratio]{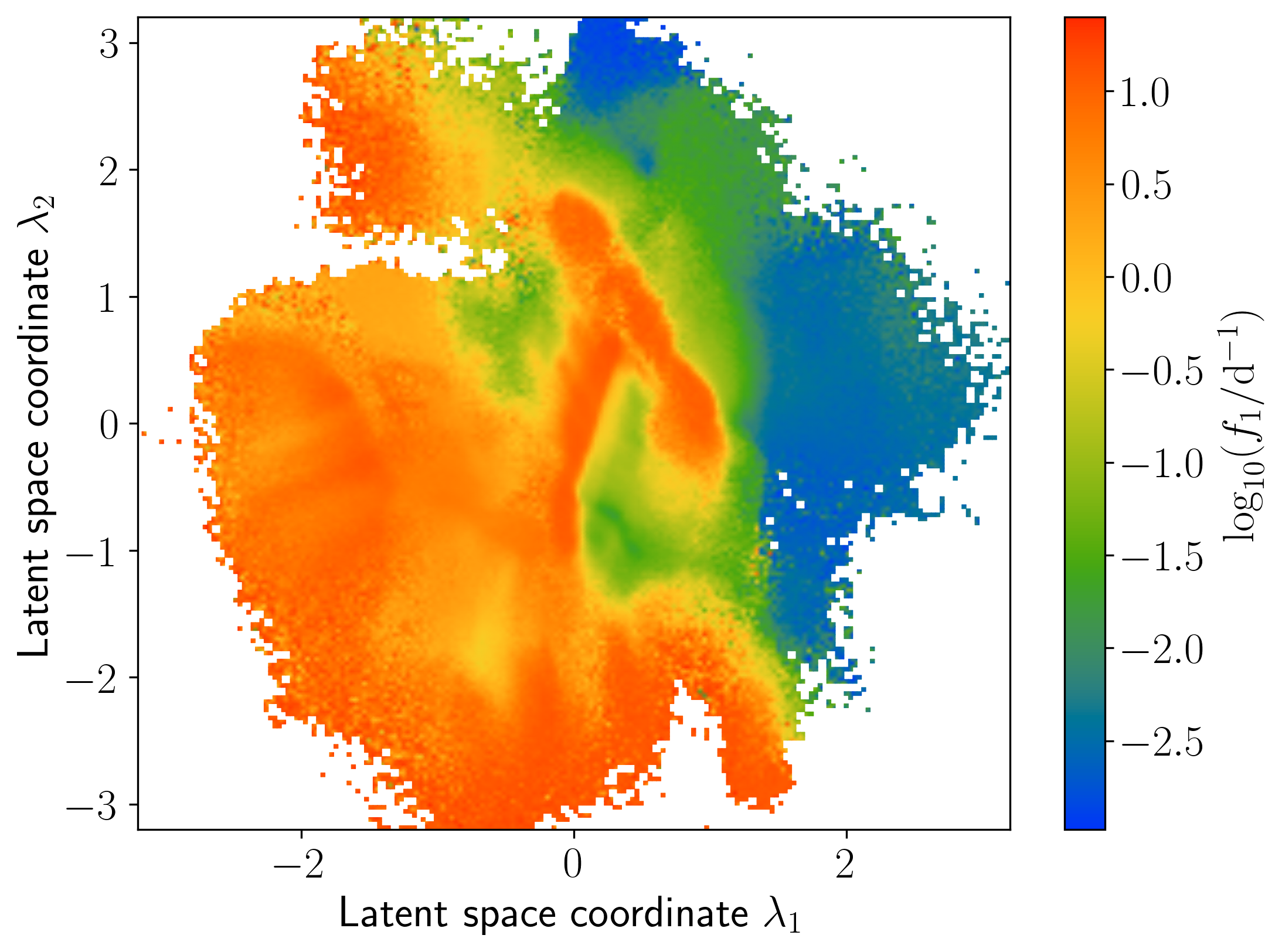}
\caption{\label{fig:main_freq_heatmap}The distribution of the main frequency $f_1$.}
\end{center}
\end{figure}

To interpret the distribution of variability classes in the 2D latent space (Figs.\ref{fig:density_heatmap}-\ref{fig:purity_class_heatmap}), it is useful to learn how the ANN projects some of the observables that would be relevant for a human classifier into the latent space. In Figs.~\ref{fig:bp_rp_heatmap}--\ref{fig:main_freq_heatmap} we divided the latent space in $200\times 200$ bins, assigned each source to the relevant bin, and computed then for each bin respectively the median colour $\overline{\bprp}$, the median standard deviation $\sigma_G$ in the \gaia \g passband as a proxy for the variability amplitude, and the median main frequency $f_1$ derived from a Lomb-Scargle frequencygram. The ANN does not have direct access to these quantities when constructing the $3K = 15$-dimensional latent space, but it does receive them as additional input when constructing the 2-dimensional latent space as shown in Fig.~\ref{fig:ann_2D_latent_architecture}. The distributions in Figs.~\ref{fig:bp_rp_heatmap}--\ref{fig:main_freq_heatmap} vary fairly smoothly locally, but also show that the variation on a global scale can be very non-smooth. Although such distribution would be atypical for a human classifier, it is not unexpected for an ANN given its highly non-linear way of operating.

\begin{figure}
\begin{center}
\includegraphics[width=0.45\textwidth,keepaspectratio]{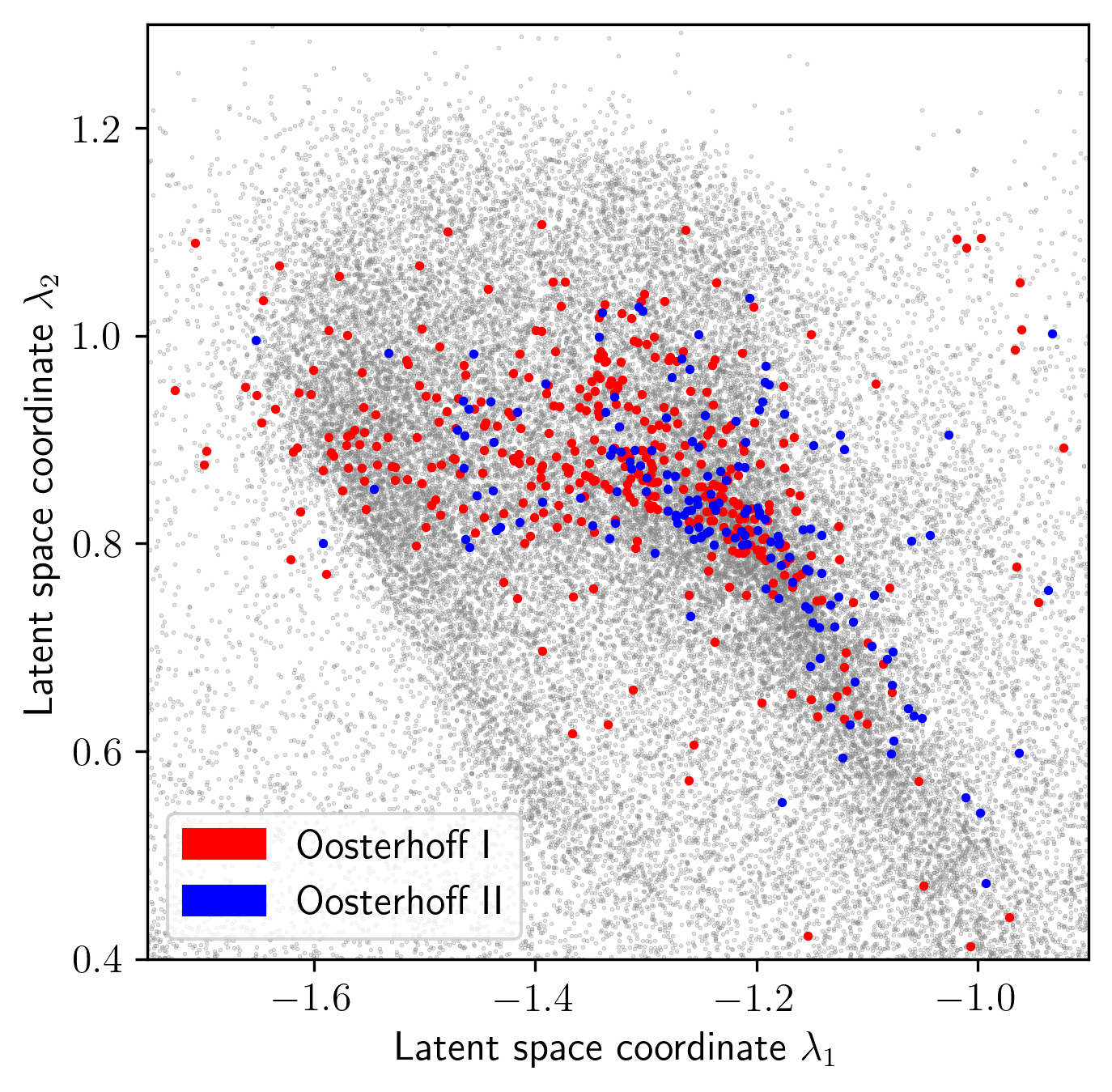}
\caption{\label{fig:rrlyr_oosterhoff}The region of the latent space containing the \rrab variables. The sample of \rr stars plotted to show the Oosterhoff classification was taken from \citet{Prudil2019rrlyr}.}
\end{center}
\end{figure}

Concerning the bimodal distribution of the \rrab variables in Fig.~\ref{fig:rrlyr_heatmap}, we investigated several possible causes. None of the observables published in the SOS variability catalogue in \gaia DR3 correlates with this dichotomy. Over-plotting the sample of \citet{Skarka2020rrlyr} containing \rr stars that exhibit the Blazhko effect \citep{Blazko1907},  
does not show any evidence that the latter is causing this bimodal distribution (figure not shown for the sake of brevity). Over-plotting in Fig.~\ref{fig:rrlyr_oosterhoff} the sample of \rr stars in the Galactic bulge for which an Oosterhoff (Oo) classification \citep{Oosterhoff39,Oosterhoff44} is published in \citet{Prudil2019rrlyr}. Oosterhoff I (OoI) systems typically (predominantly) contain \rr stars with fundamental mode period distribution peaking around $P=0.55$ d and have a metallicity distribution that peaks around [Fe/H] $\sim$ -1.5 dex. The \rr stars in Oosterhoff II (OoII) populations instead have $P\sim 0.65$ d and a metal abundance [Fe/H]$< -1.5$ dex and typically around [Fe/H] $\sim$ -2.0 dex \citep[][and references therein]{Smith95,Catelan15}. The Oo dichotomy is best seen in a period-amplitude diagram, and since these quantities correlate strongly with the latent coordinates, it is plausible that the ANN can detect these subtypes. Figure~\ref{fig:rrlyr_oosterhoff} shows that OoII \rr stars occur indeed predominantly in one of the overdensities, but the contrast between the two types in the latent space is not very high. This does not mean that we can discard the Oo dichotomy as a possible explanation for the bimodal distribution. It is known that \rr stars in the Magellan Clouds and in dwarf spheroidal galaxies orbiting our Milky Way have periods intermediate between OoI and OoII types \citep[e.g.][and references therein]{Catelan09,Luongo24}, and even within our own Milky Way there are clusters (e.g.~NGC6388 and NGC6441) whose \rr population do not conform with the Oo classification \citep[][and references therein]{Catelan09}. We therefore can expect the Oo dichotomy to be noisy in the latent space.

\begin{figure}[t]
\begin{center}
\includegraphics[width=0.42\textwidth,keepaspectratio]{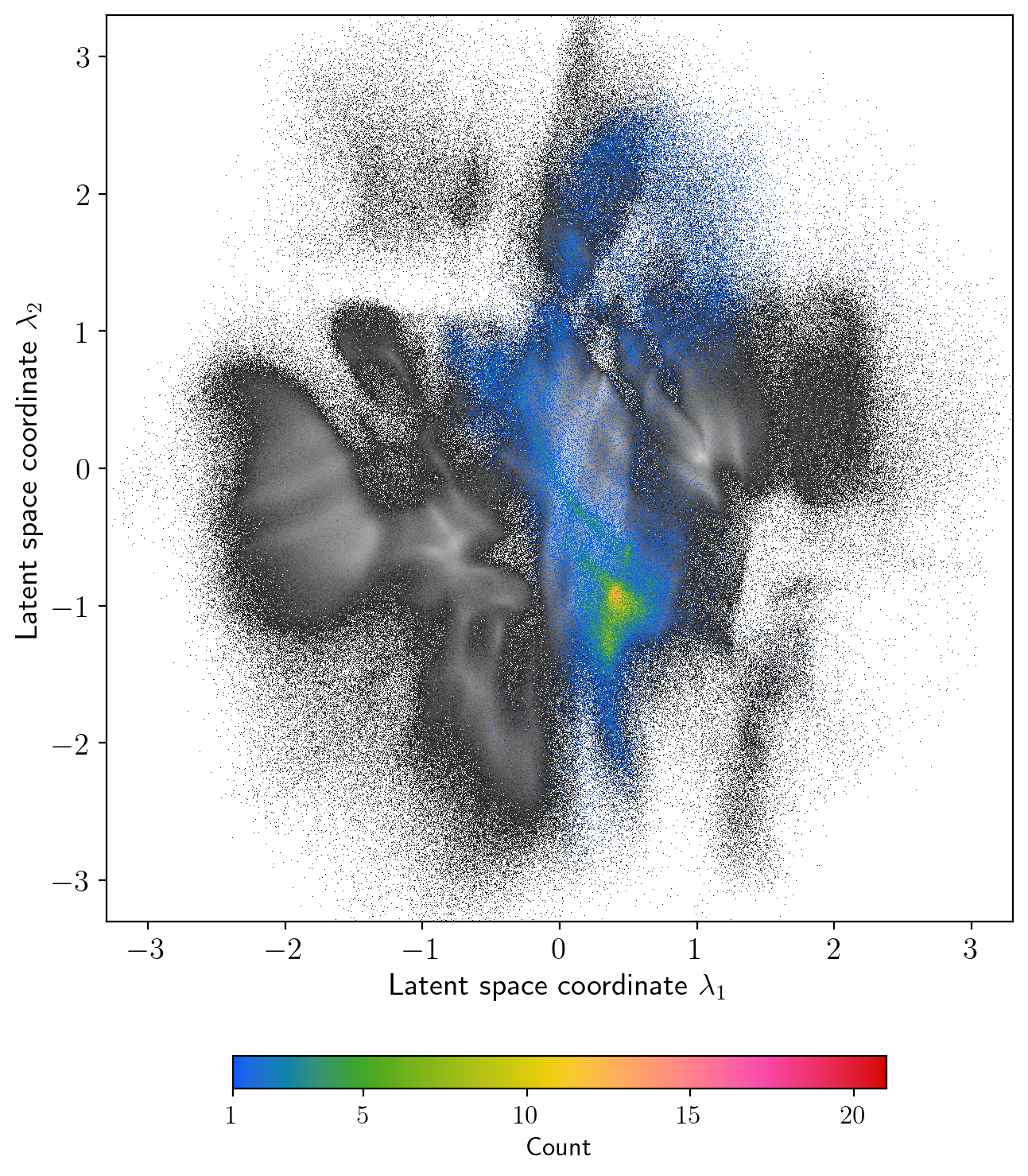}
\caption{\label{fig:spurious_heatmap}Location of \gaia DR3 sources with significant scan-angle dependent signal that induces a spurious variability signal. The sample was taken from \citet{holl2023gaia} with thresholds \texttt{spearman\_corr\_ipd\_g\_fov} > 0.9 and \texttt{scan\_angle\_model\_ampl\_sig\_g\_fov} > 10.}
\end{center}
\end{figure}
\begin{figure}[t]
\begin{center}
\includegraphics[width=0.38\textwidth,keepaspectratio]{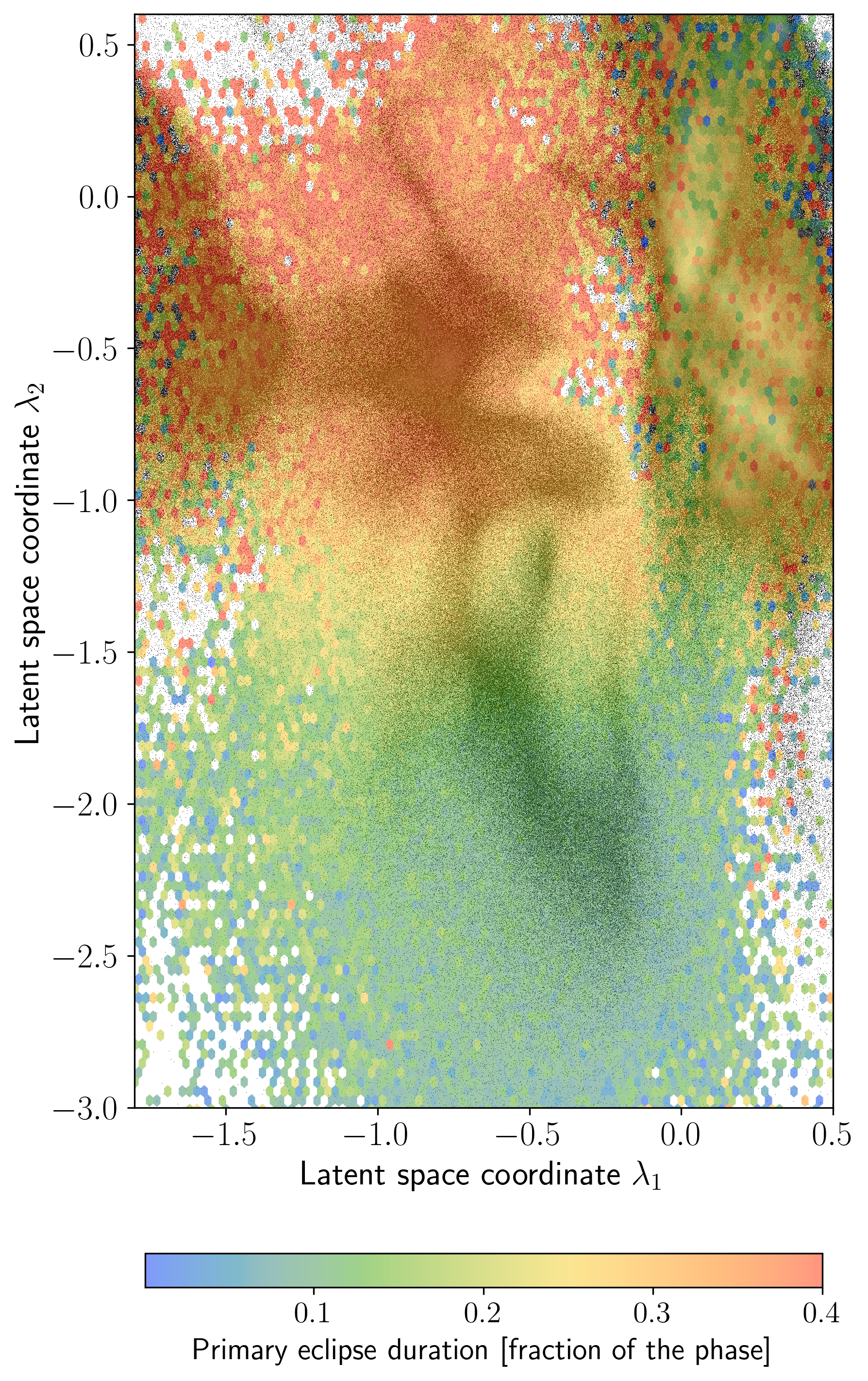}
\caption{\label{fig:eclipsing_duration_heatmap}The same as Fig.~\ref{fig:eclipsing_depth_diff_heatmap}, but colour represents the duration of the primary eclipse expressed as the fraction of the phase.}
\end{center}
\end{figure}
\begin{figure}[t]
\begin{center}
\includegraphics[width=0.38\textwidth,keepaspectratio]{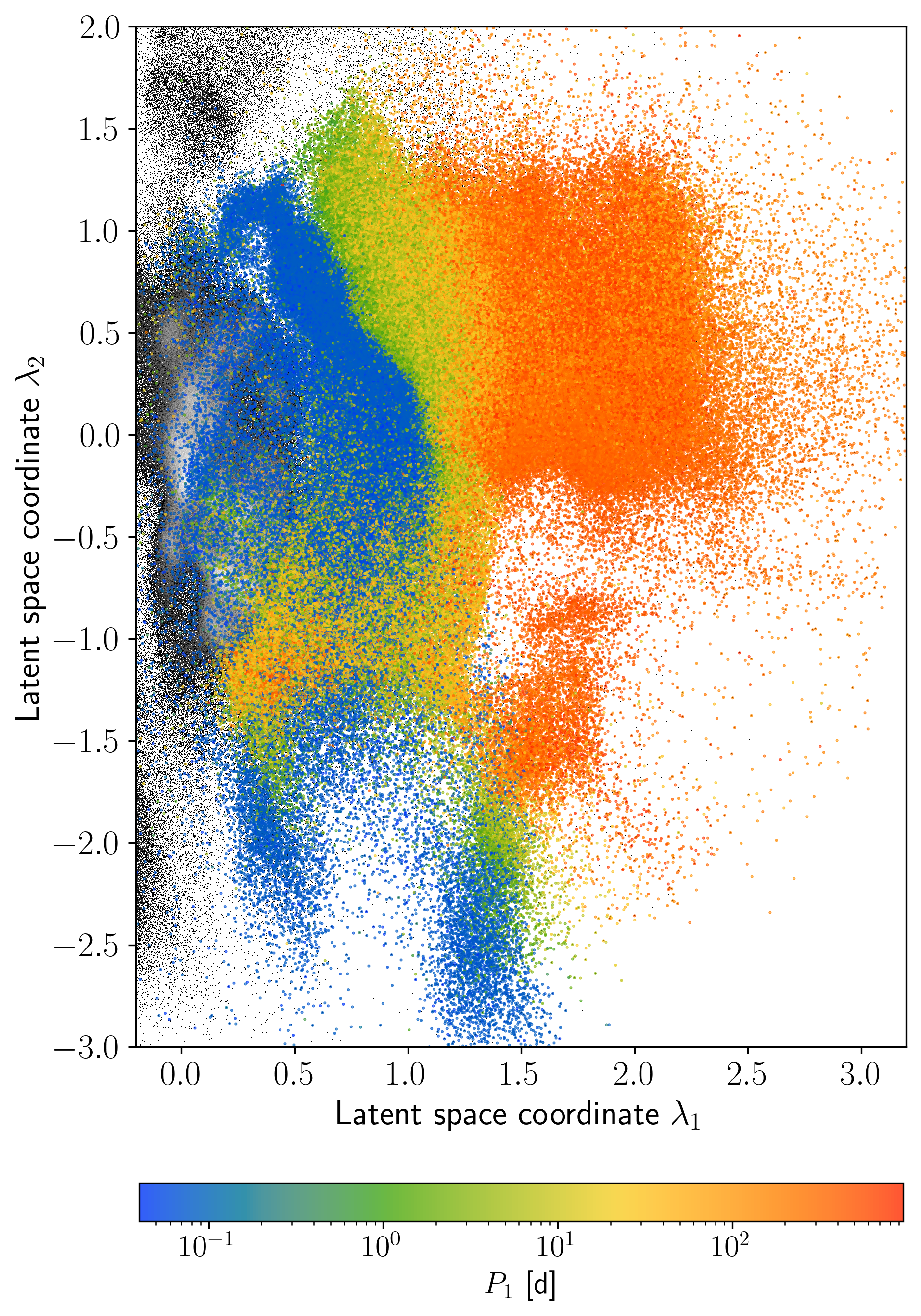}
\caption{\label{fig:lpv_f1_heatmap} The same as for Fig.~\ref{fig:lpv_xp_heatmap}, but the Lomb-Scargle dominant period $P_1$ is used for the colour coding.}
\end{center}
\end{figure}
\begin{figure}[t]
\begin{center}
\includegraphics[width=0.38\textwidth,keepaspectratio]{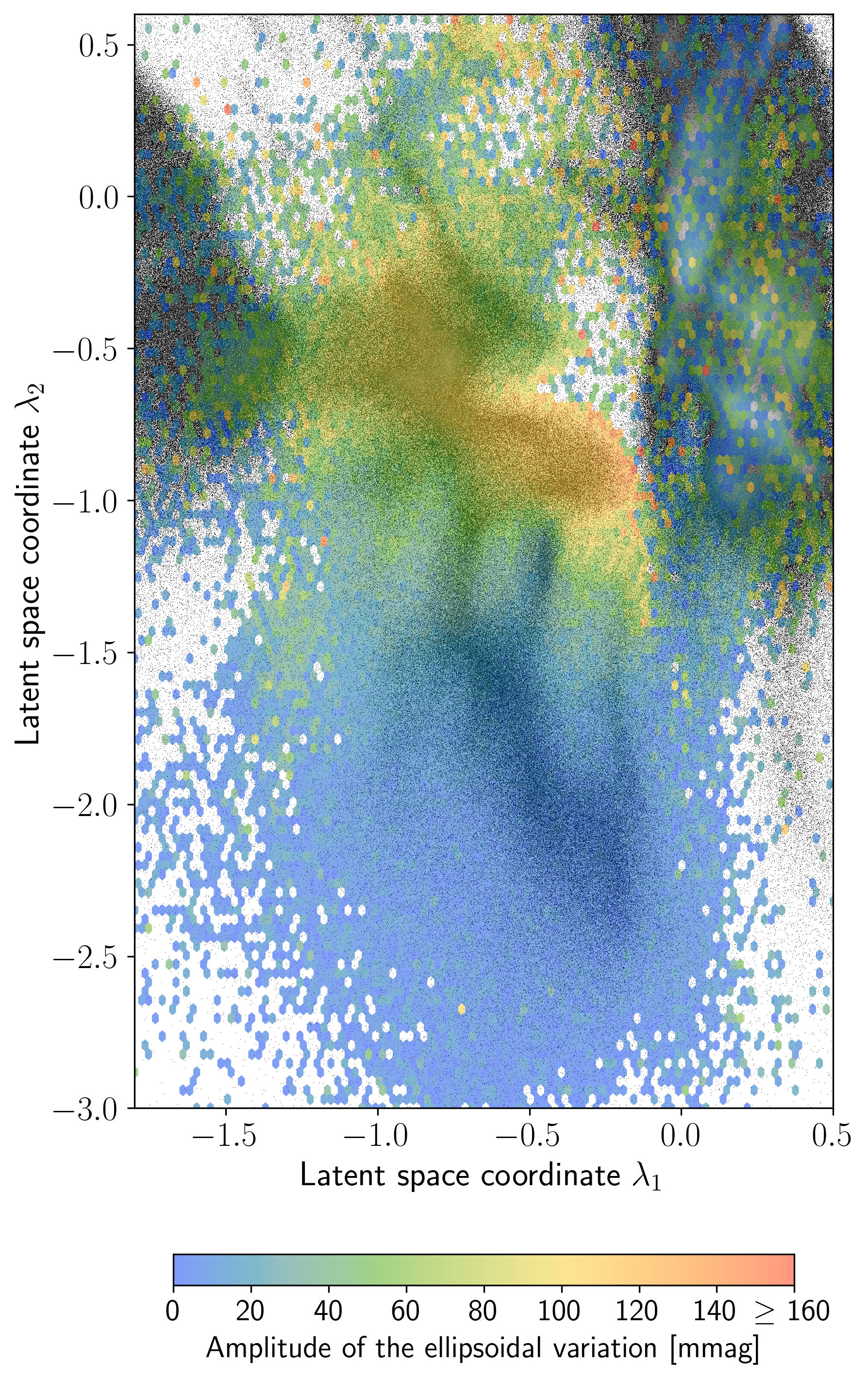}
\caption{\label{fig:ellipsoidal_amplitude_heatmap}The same as Fig.~\ref{fig:eclipsing_depth_diff_heatmap}, but colour represents the amplitude of the ellipsoidal variation (if any) in mmag.}
\end{center}
\end{figure}

\FloatBarrier

\section{Neural network architecture details}
\label{sec:architectures}

The architecture of the encoder of the VAE for sampled XP low-resolution mean spectra is described as follows:
\begin{enumerate}
\item An FC layer with 110 inputs and 128 outputs, followed by layer normalisation (LN) and a GELU activation function.
\item An FC layer with 128 inputs and 128 outputs with LN and GELU activation.
\item An FC layer with 128 inputs and $2K$ outputs. The output is then split into two $K$ arrays.
\end{enumerate}
The output of the last layer represents the mean and the logarithm of the standard deviation, from which the latent variable is sampled using Eq.~\eqref{eq:reparam_trick}. After this, the data is reconstructed by a decoder ANN with the following architecture:
\begin{enumerate}
\item An FC layer with $K$ inputs and 128 outputs with LN and GELU activation.
\item An FC layer with 128 inputs and 128 outputs with LN and GELU activation.
\item An FC layer with 128 inputs and 110 outputs with a ReLU activation to ensure non-negativity.
\end{enumerate}

The log of the variance of the reconstructed spectra $\log \sigma_{\rm XP}^2$, is predicted from the latent variables using a smaller network with the following architecture:
\begin{enumerate}
\item An FC layer with $K$ inputs and 128 outputs with LN and GELU activation.
\item An FC layer with 128 inputs and 1 output.
\end{enumerate}

The $\Delta m|\Delta t$ representation is compressed using an encoder composed of FC layers. The architecture is described as follows:
\begin{enumerate}
\item An FC layer with 115 ($23\times5$) inputs and 128 outputs, followed by LN and a GELU activation function.
\item An FC layer with 128 inputs and 128 outputs with LN and GELU activation.
\item An FC layer with 128 inputs and 128 outputs with LN and GELU activation.
\item An FC layer with 128 inputs and $2K$ outputs. The output is then split into two $K$ arrays.
\end{enumerate}
The latent variables are sampled using  Eq.~\eqref{eq:reparam_trick} and given as input to a decoder with the following architecture:
\begin{enumerate}
\item An FC layer with $K$ inputs and 128 outputs with LN and GELU activation.
\item An FC layer with 128 inputs and 128 outputs with LN and GELU activation.
\item An FC layer with 128 inputs and 128 outputs with LN and GELU activation.
\item An FC layer with 128 inputs and 115 outputs. The output is then reshaped into a $23\times5$ matrix and followed by a Softmax activation function to ensure that all values are non-negative and that they add up to one column-wise.
\end{enumerate}

The architecture of the VAE for the folded \gaia \g band light curves is described as follows:
\begin{enumerate}
\item The dimensionality of the phases $\phi_j$ is expanded using trigonometric functions as $[\cos(2\pi h\phi_j), \sin(2\pi h \phi_j)]_{h=1}^H \in \mathbb{R}^{2 H}$  and concatenated with the magnitudes forming an real-valued array $E_1$ with dimensionality $L \times (1+2H)$.
\item A GRU\footnote{Preliminary experiments showed that the difference in performance between GRU and the more complex LSTM \citep{hochreiter1997lstm} was negligible, hence the former being more efficient was preferred.} with input size $(1+2H)$, hidden state size of $128$ and dropout probability of $0.2$, receives $E_1$ and produces a real-valued array $E_2$ with dimensionality $L \times 128$. 
\item A second GRU with input and hidden state size of $128$, receives $E_2$ and produces a real-valued array $E_3$ with dimensionality $L \times 128$.
\item An FC layer with 128 inputs and 1 output is used to obtain an attention score $A=\exp(\text{FC}(E_3)) \in \mathbb{R}^{L}$ \citep{Bahdanau2014attention}. The score is multiplied by the binary mask $B$ to set the attention of padded values to zero. Then, the attention scores are normalised so that they add to unity. Note that this is equivalent to applying the Softmax activation function to the attention scores, but compensating for the masked values. The normalised scores are used as weights to generate $E_4$, a 128-dimensional vector that is the weighted average of $E_3$ along the sequence dimension.
\item An FC layer with 128 inputs and $2K$ outputs reduces the dimensionality of $E_4$. The output is then split into two $K$-dimensional arrays.
\end{enumerate}

After sampling from the parameters using Eq.~\eqref{eq:reparam_trick}, the latent variable and the phases are given as input to the decoder. The architecture uses FC layers and is described as follows:
\begin{enumerate}
\item The phases are expanded using the same procedure described for the encoder, and concatenated with the latent variables forming an array $D_1$ with dimensionality $L \times (K+2H)$. 
\item An FC layer with $K+2H$ inputs and 128 outputs with LN and GELU activation transforms $D_1$ into $D_2$.
\item An FC layer with 128 inputs and 128 outputs, with LN and GELU activation, transforms $D_2$ into $D_3$.
\item An FC layer with 128 inputs and 1 output transforms $D_3$ into $D_4 \in \mathbb{R}^{L}$. This corresponds to the reconstructed magnitudes.
\end{enumerate}

During training, the phases given to the decoder have to match the ones used to encode the latent variables. However, at inference time, the decoder can reconstruct into a finer phase grid for visualisation purposes. This can be seen in the last row of Fig.~\ref{fig:example_modalities}.

The architecture of the VAE to build the 2-dimensional representation that combines the latent variables from previous VAEs is described as follows:
\begin{enumerate}
\item An FC layer with $3K + 3$ inputs and 128 outputs, followed by LN and a hyperbolic tangent (tanh) activation function.
\item An FC layer with 128 inputs and 128 outputs with LN and tanh activation.
\item An FC layer with 128 inputs and 4 outputs. The output is then split into two arrays with two variables each, corresponding to the mean and standard deviation of the latent variables.
\end{enumerate}
The latent variables are sampled using  Eq.~\eqref{eq:reparam_trick} and given as input to a decoder ANN with the following architecture:
\begin{enumerate}
\item An FC layer with 2 inputs and 128 outputs with LN and tanh activation.
\item An FC layer with 128 inputs and 128 outputs with LN and tanh activation.
\item An FC layer with 128 inputs and $3K + 3$ outputs. 
\end{enumerate}

\end{appendix}

\end{document}